\documentclass{aa}  

\usepackage{graphicx}
\usepackage{txfonts}
\usepackage{xcolor}
\usepackage{mathrsfs}
\usepackage[colorlinks=true, citecolor=blue]{hyperref}
\usepackage{physics}
\usepackage{nicefrac}
\usepackage[normalem]{ulem}
\usepackage{soul}
\usepackage[colorlinks=true, citecolor=blue]{hyperref}

\newcommand{\vr}{\vec{r}}

\newcommand{\reynolds}{\bar{\alpha}_{\rm R}}
\newcommand{\grav}{\bar{\alpha}_{\rm grav}}

\newcommand{\FargoCPT}{\textsc{FargoCPT}\xspace}
\newcommand{\Fargo}{\textsc{Fargo}\xspace}

\begin{document} 
\title{Self-gravity in thin protoplanetary discs:}
\subtitle{2. Numerical convergence solved and revealing the overestimation in mass of formed planets with softening}

\author{S. Rendon Restrepo \inst{1}
\thanks{\email{srendon@aip.de}}
}
\institute{Leibniz-Institut für Astrophysik Potsdam (AIP), Potsdam, Germany}


 
\abstract
{ 
The Gravitational instability (GI) is a leading theory proposed to explain both angular momentum transport and early planet formation in young, massive discs. In the early 2010s, 3D smoothed particle hydrodynamics (SPH) simulations investigating GI failed to achieve convergence. Although this issue was initially addressed and attributed to resolution-dependent viscosity, it subsequently emerged in 2D SPH and 2D grid-based simulations -- suggesting a numerical artifact inherent to the 2D approximation of gravity.
}
{
Recently, we derived from first principles a much improved prescription for gravity in 2D discs (via a Bessel kernel). 
This prescription introduces a characteristic length, $H_{\rm rms}$, below which gravity smoothly transitions from a 3D to a 2D scaling.
This cannot be captured by standard smoothing length approaches, widely used in 2D simulations. 
Our objective is to employ this new prescription to resolve the convergence issue of GI in 2D, and compare the outcomes of the instability in runs using the Bessel kernel with those obtained using softening prescriptions at high resolution. 
}
{
I conducted numerical simulations with the \FargoCPT{} code, where the Bessel prescription was implemented.
}
{
The 2D Bessel formalism of gravity  effectively resolves the convergence issues encountered in 2D simulations.
When compared to simulations employing softened or unsoftened potentials, I observe that a softening parameter $\epsilon/H\leq 0.3$ tends to overestimate gravitational effects.
This may result in an artificially high number of fragments, potentially leading to final fragment masses that are overestimated by a factor of 2–3.
Conversely, employing a softening parameter exceeding the scale height inhibits gravitational effects.
Although our analysis initially suggests that a softening parameter of $0.6 H$ might offer the best compromise, in reality, the resulting fragments fail to remain gravitationally bound—a limitation not observed when using the Bessel kernel.
}
{
Our findings on the Bessel kernel, coupled with its high numerical efficiency, strongly suggest its adoption in future studies to ensure a consistent and accurate treatment of gravity in thin discs.
}


\maketitle
%

\section{Introduction}

Gravitational Instability (GI) is a key physical process in which a disc becomes unstable to its own gravity, driving outward angular momentum transport and enabling mass accretion onto the star. 
While GI primarily operates in young protoplanetary discs, it may also play a role at later stages, such as in Class II discs, where late infall could contribute \citep{2025_longarini}.
Additionally, it is a key theory in planet formation, describing how a disc can fragment into clumps for efficient cooling \citep{2001_gammie, 2003_rice, 2016_Kratter_Lodato}. 
In this context, a disc is gravitationally unstable provided that its Toomre's parameter,
\begin{equation}
Q = \frac{c_s \kappa}{\pi G \Sigma},
\end{equation}
falls below unity.
This indicates that the gravity of a density perturbation can overcome both the gas pressure support and the tidal forces exerted by the central object \citep{1960_safronov, 1964_toomre, 1964_lin_shu}.
While GI can be driven by viscosity through the reduction of rotational support \citep{1974_lynden,1992_willerding,1996_gammie}, numerical simulations commonly satisfy the instability condition using a cooling prescription, the most widely used of which is $\beta$-cooling.
This prescription places the disc in an unstable regime, where the final outcome depends on the efficiency of radiative cooling \citep{2001_gammie}.
For slow cooling, $\beta \gtrsim 3$, the initial perturbations evolve into spiral structures, generating heat through shocks.
This process stabilizes the disc in a marginally stable state, eventually quenching the instability—a regime commonly referred to as gravito-turbulence.
Conversely, for efficient cooling, the heating mechanism becomes insufficient to counteract cooling, resulting to disc fragmentation and the formation of objects bound by gravity.
The initial mass of these GI-induced fragments is typically of few Jupiter masses, but ongoing accretion onto the newborn objects likely results in final masses characteristic of brown dwarfs \citep{1997_boss, 2019_stamatellos, 2010_kratter, 2013_forgan, 2015_rice}. 
Furthermore, there is substantial numerical evidence that GI-induced spirals can capture dust and facilitate the formation of planetesimals \citep{2004_rice,2006_rice,2014_gibbons}.
Recently, \citet{2023a_longarini} and \citet{2023b_longarini} demonstrated that a mixture of gas and dust can trigger a two-fluid gravitational instability, potentially explaining the early formation of planetesimals or Earth-mass planets.

In numerical simulations, convergence tests are typically conducted by increasing the resolution until the results become independent of resolution, ensuring that all relevant physical length scales are adequately resolved.
In the context of GI, this involves identifying a clear separation, dependent on cooling, between the regimes of gravito-turbulence and fragmentation as resolution increases. 
However, achieving numerical convergence in GI simulations emerged quickly as a significant challenge, first observed in 3D smoothed particle hydrodynamics (SPH) studies \citep{2011a_meru,2011b_meru}.
This behavior was primarily attributed to the implementation of radiative cooling \citep{2012_rice} and, more critically, to the resolution-dependent artificial viscosity \citep{2011_lodato_clarke,2014_rice, 2017_deng}, which is typically used to handle shocks and prevent particle interpenetration.
Although this issue was linked to viscosity and addressed in 3D, it even arose in 2D SPH \citep{2015_young_clarke} and 2D grid-based simulations using the \Fargo code \citep{2012_meru}. 
\citet{2015_young_clarke} attributed the problem to the absence of smoothing in gravity, which amplifies gravitational forces with increasing resolution. 
They mitigated this by softening gravity on the scale $H$, but acknowledged that this approach suppresses the gravitational interaction between fluid elements separated by less than one scale height, thereby inhibiting the collapse of pressure-supported clumps.
Consequently, they proposed that a "more sophisticated model for approximating gravity in two dimensions [...] would reduce the fragmentation-suppressing effects of gravitational softening", a challenge that remains unresolved.

It is important to note that, within the thin-disc approximation (2D), gravity is not well-defined, necessitating to approximate the gravitational potential.
This is typically achieved using either a Plummer potential with a smoothing length prescription \citep{muller_kley_2012} or by solving a 2D Poisson equation \citep{2012_paardekooper}.
However, introducing a finite softening inherently suppresses the Newtonian nature of gravity \citep{1989_adams,hockney2021computer, 2015_young_clarke}, potentially inhibiting gravitational collapse at small scales. 
Conversely, reducing the smoothing length to zero or solving the Poisson equation in two dimensions artificially amplifies gravitational effects, likely contributing to the convergence issues discussed earlier. 
The effects of softening have been extensively investigated in N-body simulations of disc galaxies, particularly its impact on GI \citep{1994_romeo, 1997_romeo, 1998_romeo}.
\citet{2025c_rendon} analytically derived the correct gravity prescription for the 2D approximation of discs, which necessitates the use of a Bessel kernel. 
This prescription enables gravity to transition smoothly from a purely 3D behavior at long ranges to a purely 2D behavior at short ranges—a feature that cannot be replicated by approaches relying on finite smoothing lengths.
Furthermore, this prescription strictly adheres to Newton's third law, unlike the smoothing length approach which necessitates correction terms to account for self-accelerations \citep{2008_baruteau}.
When directly compared to 3D simulations, the Bessel prescription demonstrated a maximum deviation of only 5\%, whereas the smoothing length approach either underestimates or overestimates self-gravity, with errors reaching up to 129\%. 
Consequently, by restoring the Newtonian character of gravity at scales smaller than the scale height—without introducing artificial overestimations—the Bessel prescription emerges as an ideal solution for addressing the numerical convergence challenges encountered in 2D simulations of GI.
It is worth noting that the Plummer potential formulation is also commonly employed in 2D simulations of planet-disc interactions in order to model the planet gravitational field, or more accurately, the vertical average of the gradient of its potential. 
In such contexts, the torques and planet-induced flow structures are highly sensitive to the choice of softening length, which is typically selected based on the specific aspect of the interaction under investigation.
However, recent studies have highlighted that these inherent limitations can be effectively addressed by adopting a Bessel potential \citep{2024_brown_joshua, 2025_cordwell}.

In this second paper of a two-part series, I address the convergence of 2D grid-based simulations of GI through the use of the correct self-gravity prescription, that is, the Bessel kernel. 
Additionally, I compare the outcomes of fragmentation vs. gravito-turbulence when employing the Bessel prescription in place of the Plummer potential approach with varying smoothing lengths.
I begin by outlining the numerical setup, cooling prescription, gravity prescriptions, and simulation framework in Section~\ref{sec: numerical setup}. 
In Section~\ref{sec: numerical convergence of GI}, I demonstrate the numerical convergence of 2D GI simulations using the Bessel kernel. 
Finally, in Section~\ref{sec: characterisation of GI as a function of gravity prescription}, I characterize the outcomes of GI as a function of the gravity prescription.
Section~\ref{sec: discussion and perspectives} discusses the implications of using more accurate potentials for modeling planet-disc interactions, the initial mass of giant gas planets, numerical considerations, and future research directions.
Finally, Section~\ref{sec: conclusion} provides concluding remarks.

\section{Numerical setup}\label{sec: numerical setup}

\subsection{Equations and code}

For the purpose of this investigation, the vertically-integrated hydrodynamical equations ---namely, the continuity, momentum and energy equation--- were solved numerically in their differential form:
\begin{equation}
\begin{array}{ll}
\displaystyle \frac{\partial \Sigma}{\partial t} + \vec{\nabla} \cdot (\Sigma \vec{u})   &  \displaystyle = 0 \\ [8pt]
\displaystyle \frac{\partial \vec{u}}{\partial t} + (\vec{u} \cdot \vec{\nabla}) \, \vec{u}     & \displaystyle= - \frac{\vec{\nabla}P}{\Sigma} - \vec{\nabla} \Phi_{\odot} + \vec{f}_{SG} \\ [8pt]
\displaystyle \frac{\partial e}{\partial t} + \vec{\nabla} \cdot \left( e \vec{u} \right) & \displaystyle = - P \nabla \cdot \vec{u} - \Gamma 
\end{array}
\end{equation}
Here, $\Sigma$ represents the surface density, $P$ the vertically integrated pressure, $\Phi_{\odot}=-{G M_{\odot}}/{r}$ the gravitational potential of the central object, and $e$ the internal energy density.
The cooling term, $\Gamma$, and the self-gravity forces exerted by the gas, $\vec{f}_{SG}$, will be elaborated upon in subsequent sections.
The numerical simulations were conducted using \FargoCPT, which is a 2D finite difference code with a staggered mesh, employing an advection scheme akin to finite volume methods \citep{2024_rometsch}. 
It solves the hydrodynamics equations using operator splitting and a second-order upwind scheme.
\FargoCPT and its variants, such as \textsc{FARGOCA}\xspace \citep{2014_lega}, \textsc{FARGOADSG}\xspace \citep{2008_baruteau}, and \textsc{FARGO3D}\xspace \citep{2016_llambay}, are built upon the Fargo code presented in \citep{2000_masset}.
It is formally accurate up to second-order in space and first-order in time, and relies on artificial viscosity to handle shocks. 

\subsection{Cooling prescription and viscosity precautions}

To induce gravitational instability in the disc, it is standard practice to implement a cooling mechanism, optionally supplemented by a heating term, using the $\beta$-cooling prescription as introduced by \citet{2001_gammie}:
\begin{equation}
\Gamma = \frac{\Omega_K(r)}{\beta} e
\end{equation}
In this framework, $\beta$ remains spatially constant throughout the disc. 
However, due to the dependence of the cooling rate on $\Omega_K$, the inner regions of the disc cool more rapidly. 
While alternative approaches—such as incorporating a heating term or adopting more realistic cooling models, including radiative cooling \citep{2015_baehr, 2016_takahashi}—are feasible, I have chosen not to pursue them in this study.
This decision is motivated by my objective to ensure a rigorous and direct comparison with the extensive body of literature addressing the convergence challenges in 2D simulations of GI.
Once the convergence issue is resolved—the aim of the present work—future investigations may explore the integration of more sophisticated cooling prescriptions.

GI can also be driven by viscosity, which reduces rotational support \citep{2016_lin_kratter}.
Consequently, precise control of viscosity is essential to prevent numerical artifacts, as highlighted in the introduction. 
In finite difference schemes, such as those used in \FargoCPT, shocks are managed through the \citet{1950_vonneuman_richtmyer} artificial viscosity. 
However, this approach can introduce artificial pressure effects that I addressed adopting the \citet{1979_tscharnuter_winkler} artificial viscosity prescription with a parameter $q=1.41$.

\subsection{Self-gravity computation}

The 2D SG force per unit volume exerted by the disc on a volume element is given by:
\begin{equation}
\vec{f}_{2D}(\vr) = -\Sigma(\vr) \iint\limits_{disc} \Sigma(\vr') \mbox{K}(\vr, \vr') \, \vec{e}_s \, d^2 \vr'
\end{equation}
where $\mbox{K}$ is the SG kernel, corresponding to the vertical average of the Green's function used to compute the 3D SG force.
In the general case, the two-dimensional SG force is not inherently conservative, as the vertical averaging procedure does not necessarily preserve this property.

For this study, we employ two distinct gravitational prescriptions for the SG kernel. 
The first approach uses an approximation of the SG kernel in the form of the standard Plummer potential,
\begin{equation}
\mbox{K}_{\rm Plumm} = \displaystyle \frac{1}{\sqrt{ ||\vr-\vr'||^2 + \epsilon(r)^2}} 
,\end{equation}
which requires the introduction of a softening length, $\epsilon(r)$, typically ranging from 0 to $1.2 H(r)$.
It is important to note that the purpose of this softening is not to avoid numerical singularities, but rather to account for the vertical structure of the disc, thereby aiming for more physically realistic configuration. 

The second prescription, which is exact for a Gaussian stratified disc and analytically derived by \citet{2025c_rendon}, employs a Bessel kernel,
\begin{equation}\label{Eq: SG force kernel exact}
\begin{array}{lll}
\text{K}_{\rm B}
      & = & \displaystyle \frac{1}{\sqrt{\pi}}\left( H_{rms} \right)^{-2} \frac{d}{8} 
            \displaystyle \exp\left(\frac{d^2}{8} \right) \left[ K_1\left(\frac{d^2}{8} \right) - K_0\left(\frac{d^2}{8} \right) \right]
\end{array}
,\end{equation}
where $K_\alpha$ are modified Bessel functions of the second kind and order $\alpha$.
Here $d=||\vr-\vr'||/H_{rms}$ denotes the normalised distance between two fluid elements.
The root mean square scale height is defined as:
\begin{equation}\label{Eq: mean root square height}
H_{rms}(r,r') = \sqrt{\frac{{H_{sg}(r)}^2 + {H_{sg}(r')}^2}{2}}
\,.\end{equation}
with $H_{sg}$ being the scale height of gas in presence of SG. 
This characteristic length, $H_{\rm rms}$, marks the transition scale below which gravity smoothly transitions from a 3D behavior $(\propto 1/||\vr-\vr'||^2)$  to a 2D behavior $(\propto 1/||\vr-\vr'||)$.
It is crucial to distinguish $H_{\rm rms}$ from the smoothing length used in the Plummer potential.
It is important to emphasize that this formulation remains valid as long as the vertical profile is Gaussian, a condition that remains reasonable for the isothermal and polytropic cases in presence of SG, provided the scale height is appropriately adapted \citep[see Fig. 2 and Eq. 28]{2025_rendon_restrepo_et_al} and \citep[see Fig. 1 and Appendix B]{2025_gordon}.
Additionally, the Bessel formalism accommodates variations in disc thickness both temporally and spatially, which is particularly relevant for spirals generated during GI or during the formation of massive clumps.
In these cases, the scale height can locally increase or decrease, respectively.
However, the current numerical method, which is based on Fast Fourier Transforms, is constrained to a constant disc aspect ratio and thus cannot capture these variations. 
Consequently, in spirals and clumps, the SG tends to be slightly overestimated and underestimated, respectively. 
This limitation is expected to be mitigated by developing a numerical method based on Hankel transforms.

Both SG prescriptions were efficiently computed using full fast Fourier transform (FFT) methods.
This approach necessitates expressing SG forces in polar coordinates as a convolution in both the radial and azimuthal directions.
While the azimuthal condition is inherently satisfied with a linear grid, the radial condition is more complex.
It requires the use of a logarithmic radial grid and the enforcement of $H_{sg}/r=const.$
This condition imposes stringent conditions on the Toomre parameter, which directly influences the density profile (for a detailed discussion, see Sect. 5.2 of \citep{2025c_rendon}).
To simplify the analysis, I chose to neglect the effect of SG on the vertical stratification, thereby adopting the more conventional and computationally convenient condition of $H/r=const.$
It is also important to highlight that the Bessel prescription strictly adheres to Newton's third law, as the force between two column densities is symmetric.
In contrast, the current Plummer potential formulation does not inherently satisfy this symmetry and introduces a spurious radial acceleration, which is typically compensated for in practice \citep{2008_baruteau}.
To mitigate this issue in the Plummer potential formulation, the smoothing length was chosen to be proportional to the mean square scale height, $H_{\rm rms}(r,r')$, rather than the local scale height, $H(r)$, thereby ensuring symmetry between any two fluid elements.
Additionally, zero-padding of the density field was employed to enable aperiodic convolution in the radial direction.

The presence of asymmetries in the disc can induce an offset between the barycenter of mass and the reference frame centered at the star’s position. 
This offset is typically accounted for through an indirect term in the potential, expressed as $\Phi_{ind}(\vr)=- \vec{a}_{\star}\cdot \vec{r}$, where $\Vec{a}_{\star}$ represents the acceleration of the central object \citep{2016_zhu, regaly_vorobyov_2017, rendon_2022}. 
For the purposes of this initial study, I intentionally neglected the effect of the indirect term to ensure a direct and consistent comparison with previous results in the literature, which also do not include this effect \citep{2003_rice, 2011_paardekooper, 2012_meru, 2015_young_clarke, 2018_vorobyov, 2021_bethune}.
However, it is essential to recognize that the indirect term and SG are intrinsically linked \citep{2025a_crida, 2025b_crida}.
As such, both components are critical for realistic simulations, particularly those involving non-axisymmetric features—such as those arising from GI. 
In this regard, \citet[Sect. 5.4]{2025c_rendon} derived a new expression for the indirect term that is compatible with a Gaussian-stratified disc, which may be incorporated in future work to enhance the physical fidelity of such simulations.

\subsection{Initial conditions and simulations outline}

Prior to executing the production runs, I performed preliminary tests using a fixed density profile and a radial domain extending from 1 to 10 AU.
In some of these test simulations, I observed the rapid onset of the Rossby Wave Instability (RWI) \citep{1999_lovelace, lovelace_2013}, as well as the occasional propagation of a strong wave from the inner radii toward the outer disc.
I identified that the development of GI, which typically initiates in the inner region of the disc and leads to accretion, is the potential cause for this phenomenon.
When combined with outflow boundary conditions at the inner radius, this accretion process generates a steep density gradient, which is Rossby wave unstable.
This outcome is enhanced at low resolution, as numerical viscosity provides an additional contribution to radial accretion.
A possible solution would have been to uniformly reduce the disc density. 
However, this approach would have necessitated a simultaneous decrease in temperature to maintain a Toomre parameter of unity. 
Instead, I opted for an alternative solution: extending the radial domain to span from 10 to 100 AU. 
This adjustment preserved the Toomre parameter while reducing the density at the inner boundary by two orders of magnitude compared to the case where the inner radius was set at 1 AU.

The numerical window and initial conditions of the simulations undertaken in this study are:
\begin{equation}
\begin{array}{ll}
[r_{in}, r_{out}]  & = [10, 100] \, \text{AU} \\
\Sigma(r)     & = 200000 \left( r/1 \text{AU} \right)^{-2} \text{g/cm}^2 \\
H/r           & = 0.1 
\end{array}
\end{equation}
This setup results in an initial Toomre parameter equal to ~${Q=c_s \kappa / (\pi G \Sigma) = 1.8}$ for the whole disc, where $c_s=\sqrt{\gamma P / \Sigma}$ is the sound speed and a disc mass of $0.32 M_{\odot}$, where $M_{\odot}$ is the solar mass.
The adiabatic index is $\gamma=1.667$.
I use the \textsc{FARGO}\xspace algorithm for orbital advection and a CFL parameter of 0.5.
I introduced a random seed perturbation in density of an amplitude of 1\%.
I employed outflow boundary conditions in the inner and outer radii of the disc and disregarded using damping zones since it resulted in matter accumulation in the inner and outer boundaries.
Finally, the azimuthal velocity at the inner and outer boundary is set in such a way that the disc remains in centrifugal equilibrium. 
As it will be detailed in next paragraph, for a given resolution, all production runs start from the same gravito-turbulent state reached at $t=3.5$ kyr and continue for 4.5 kyr.

To avoid artificial fragmentation, careful consideration must be given to the initial setup of the disc. 
As the disc cools and its Toomre parameter drops below unity, initial perturbations are amplified. 
The disc responds to perturbations by developing shocks, which heat the gas via compressional work ($P \vec{\nabla} \cdot \vec{v}$) and dissipation, leading to a gravito-turbulent state.
However, if the initial cooling occurs too rapidly compared to the disc’s response time, artificial fragmentation may arise because gravito-turbulence has insufficient time to develop.
As demonstrated by \citet{2012_paardekooper} in shearing box simulations, the disc requires approximately 10 orbital periods to generate enough heat via shocks to balance the cooling. To address this, I adopt a time-dependent $\beta$ parameter, defined as:
\begin{equation}
\beta(t) = \beta_{ini} + \frac{1}{2} \left(1+\tanh{\left( \frac{t-t_0}{t_{ramp}} \right)} \right) (\beta_{f} - \beta_{ini})
\end{equation}
where $\beta_{ini}=30$, $t_0=3.5$ kyr, $t_{ramp}=0.1$ kyr and $\beta_f$ the targeted beta cooling.
The choice of $\beta_{ini}$ and $t_0$ ensures that the simulation covers 110 inner orbits and 3.5 outer orbits, allowing sufficient time for the gravito-turbulent state to establish.
This setup closely mirrors the approach used by \citep{2015_young_clarke}. 
This approach simultaneously mitigates the propagation of strong inner shock wave fronts outward, which can otherwise lead to spurious fragmentation in the inner regions of the disc \citep{2011_paardekooper}.

It is important to note that, in my simulations, the evolution is continued even after fragmentation occurs. 
However, this approach is no strictly appropriate.
Previous studies \citep{1997_truelove} have established that the Jeans length,
\begin{equation}
\lambda_J = \pi (2 \pi)^{1/4}  \sqrt{Q} H \, ,
\end{equation}
in self-gravitating hydrodynamic simulations must be resolved by at least four grid cells in order to avoid artificial fragmentation. 
In simulations where the local Jeans length becomes under-resolved, sink particles \citep{2010_federrath} are commonly employed to overcome this limitation by removing gas from cells in which the Jeans length can no longer be resolved by the computational grid. 
Since a sink particle algorithm is not implemented in the present study, the simulation results obtained after the Jeans length becomes unresolved may not be quantitatively reliable.
Nonetheless, it permits a qualitative and empirical comparison of the outcomes arising from the different gravity prescriptions, which I anticipate to remain valid.
In my simulations for the initial setup, I have $\lambda_J \sim 2 H$, and later, as gravito-turbulence develops at $Q\sim0.5$, the Jeans length decreases to $\lambda_J \sim H$.
Thus the Jeans criterion is satisfied in my simulations across all resolutions (see Table \ref{tab: resolution numerical convergence}).
Nevertheless, it is important to acknowledge that, under conditions of efficient cooling, both the Toomre parameter and the scale height could reach sufficiently small values such that the Jeans length may no longer be adequately resolved by my grid.
An approach using adaptive mesh refinement would be more appropriated, but it is out of the scope of this study.

\section{Numerical convergence of 2D global simulations with the Bessel kernel}\label{sec: numerical convergence of GI}

\begin{table}
\caption{Resolutions used for the numerical convergence} 
\label{tab: resolution numerical convergence}            
\centering                                          
\begin{tabular}{c | c}                            
\hline                                              
 $N_r \times N_\varphi$ & cells per scale-height       \\ 
\hline                                      
$128 \times 364$        & $5.5 \times 6$  \\
$256 \times 768$        & $11  \times 12$ \\ 
$512 \times 1536$       & $22  \times 24$ \\ 
$768 \times 2304$       & $33  \times 36$ \\
$1024 \times 3072$      & $44  \times 49$ \\ 
$2048 \times 6144$      & $89  \times 98$ \\
\hline
\end{tabular}
\end{table}

\begin{figure}
\centering
\includegraphics[width=\hsize]{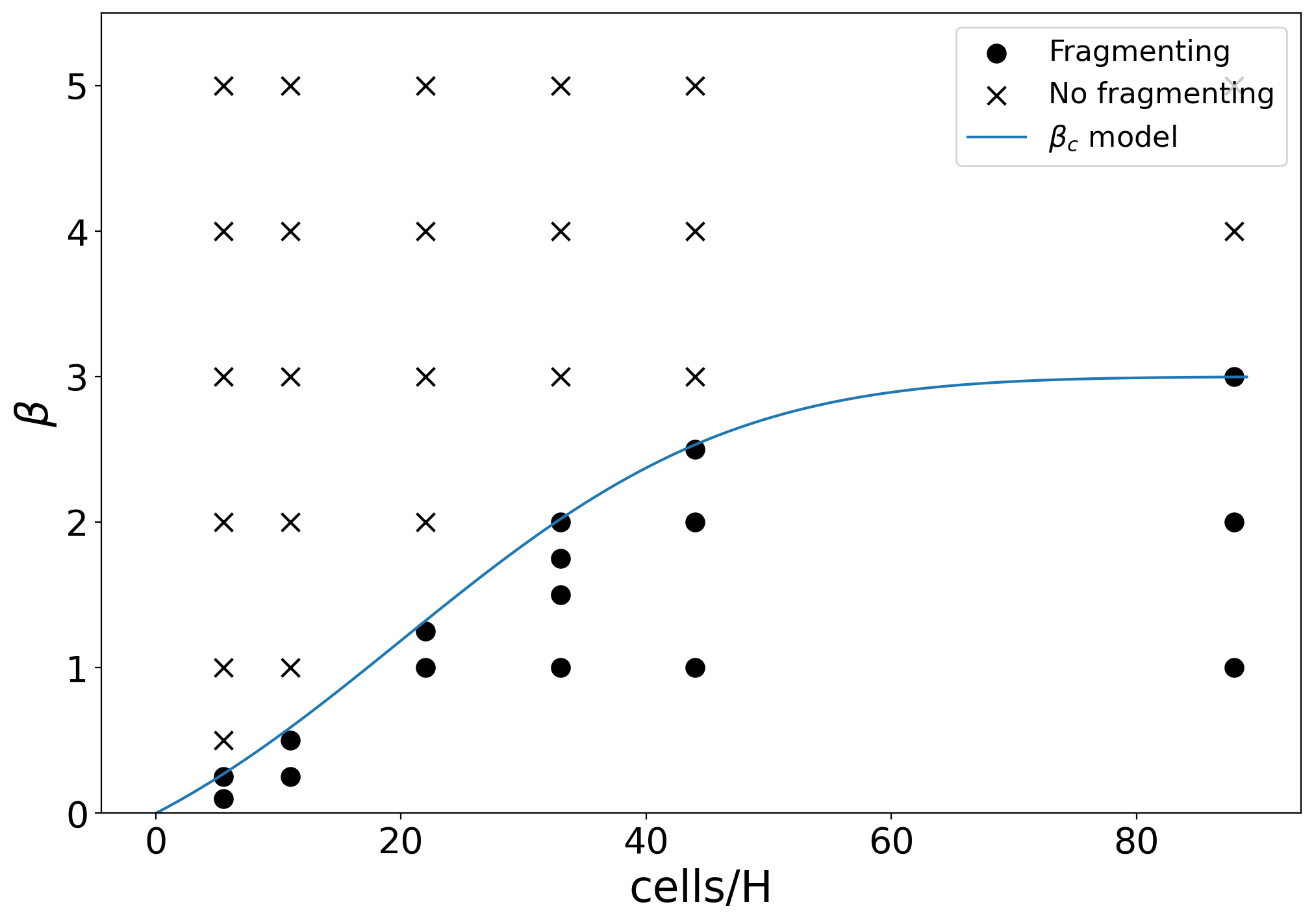}
\caption{Numerical convergence of 2D Gravitational instability simulations using the gravitational Bessel Kernel.
The solid line represents a sigmoid fit to the critical cooling threshold that separates fragmentation from gravito-turbulence. 
Above $\beta=5$ none of my simulations fragments.
} 
\label{fig: numerical convergence}
\end{figure}

As it is understood from the literature, the numerical convergence of GI simulations consists on finding a clear separation, as a function of cooling, between a regime of gravito-turbulence and fragmentation when the numerical resolution increases.
This requires nonetheless to define what is fragmentation, which is a delicate task.
In particular, fragmentation should be distinguished from clumping, that is the generation of transient overdensity regions which can be destroyed by shear or encounters with spirals.
Fragments are also overdensity regions but they cannot be destroyed and will further contract under the influence of their own gravity, with the ultimate fate to form an object bound by gravity, i.e. a planet, brown dwarf or star.

While fragments are often easily identifiable by eye in simulations, most studies in the literature surprisingly omit the definition they used.
Fortunately, some works provide their criteria.
For instance, \citet{2011_paardekooper} define a fragment in 2D shearing box simulations as a region where the density ratio $\Sigma/\Sigma(t=0)$ exceeds 100, a definition also adopted by \citet{2011_baruteau} and  \citet{2012_paardekooper}. 
Alternatively, \citet{2016_Kratter_Lodato} propose a fragmentation criterion based on the fragment's radius being smaller than its Hill radius. 
However, this approach introduces ambiguity, as calculating the clump's mass—and thus its Hill radius—requires an arbitrary choice of clump radius, making the fragmentation criterion dependent on this selection.
In contrast, \citet{2015_baehr} and \citet{2017_baehr} suggest a physically grounded criterion: fragmentation occurs when the density exceeds the Roche density. 
This definition is advantageous because it avoids ad hoc parameters and is rooted in physical principles.
Recently, \citet{2023_zier} reconciled these definitions in non-dimensionalised 2D and 3D shearing box simulations.
Given the use of dimensional units in this study, the fragmentation criterion is based on physical considerations. 
Therefore, a fragment is identified as a region where the density exceeds two times the Roche density,
\begin{equation} \label{Eq: fragmentation criterion}
\Sigma > 2 \Sigma_{\rm Roche} 
\end{equation}
where $\Sigma_{\rm Roche} = \frac{7}{2} \sqrt{2 \pi} \frac{H}{r} \frac{M_\odot}{r^2} \simeq 7 \frac{c_s^2}{H G}$ \citep{2015_baehr,2020_klahr_schreiber}.
The factor of two serves as a conservative threshold to ensure that the system has definitively entered the fragmentation regime.
For numerical peak detection, the \textsc{scikit-image} library is employed \citep{scikit_image}, with the requirement that two peaks must be separated by at least $20 N_r/512$ cells to be considered distinct. 
If this condition is not satisfied, the peaks are treated as part of a single fragment. 
Fragmentation is deemed to have occurred if a fragment forms between 3.5 kyr and 8 kyr.

I conducted a series of numerical simulations with progressively increasing radial and azimuthal resolution per scale height, as detailed in Table \ref{tab: resolution numerical convergence}. 
In these simulations, only the Bessel kernel was employed. 
It is worth noting that, due to the substantial computational cost, the highest resolution simulations were not conducted until 8 kyr.
The numerical convergence of these simulations, as a function of $\beta$ and resolution per scale height, is illustrated in Figure \ref{fig: numerical convergence}. 
Filled circles and crosses denote simulations that did and did not result in fragmentation, respectively.
At low resolutions of 5 and 11 cells per scale height (cells/H), no fragmentation was observed, likely due to the effects of grid diffusion. 
At a resolution of 24 cells/H, fragmentation occurred only for $\beta=1$. 
For higher resolutions of 33 and 44 cells/H, the disc fragmented for both $\beta \leq 2$. 
Notably, at 33 cells/H and $\beta=2$, a fragment formed but was immediately disrupted; this case is therefore considered as the upper bound for fragmentation at this resolution. 
Finally, at 89 cells/H, fragmentation was observed for $\beta \leq 3$.
Based on these findings, the critical fragmentation threshold, $\beta_c$, is modeled as a function of resolution using a sigmoid function:
\begin{equation}
\beta_c(N) = \left.3 \left[ \erf\left( \frac{N-a}{b} \right)  + \erf\left( \frac{a}{b} \right) \right] \, \middle/ \, \left[ 1 + \erf\left( \frac{a}{b} \right) \right]\right.
\end{equation}
where $N$ is the number of cells per scale height and $(a,b)=(10, 41)$.
In this modelling, the critical parameter $\beta_c$ is set to 0 at 0 resolution, as convergence is not expected in this hypothetical scenario, and to 3 for infinite resolution, consistent with theoretical estimates \citep{2001_gammie}. 
Additionally, the curve is constrained to pass through the point (33, 2).
While this modeling can certainly be questioned, it is important to acknowledge that I cannot afford to perform simulations with infinite resolution to check its validity at extremely high resolution. 
However, what I can confidently assert is that a linear fit would not align with the data; otherwise, the disc would necessarily fragment at $\beta=4, 5, 6$ for a resolution of 89 cells/H, which is not the case.
This suggests that my model is more realistic than a linear fit.
Our modeling suggests that, at a resolution of 40 cells/H, the results are converged at 93\%.
Thus, if convergence is defined as the distinction between the gravito-turbulent and fragmentation regimes, my results confirm that my simulations have achieved convergence at 44 cells/H.

\begin{figure*}
\centering
\begin{tabular}{ccc}
$\beta=1$ &
$\beta=2$ &
$\beta=3$ \\
\includegraphics[width=0.315\hsize]{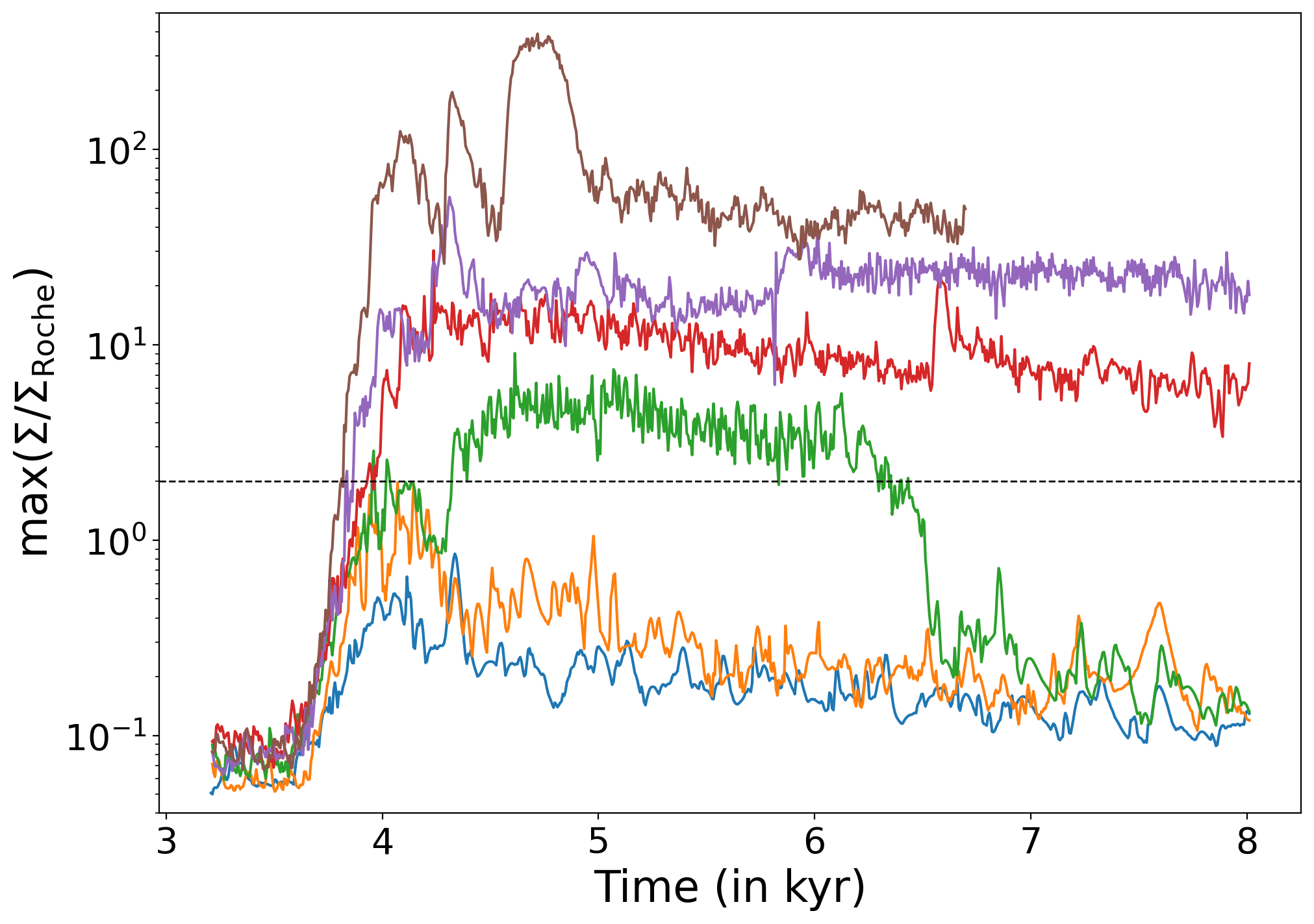} &
\includegraphics[width=0.315\hsize]{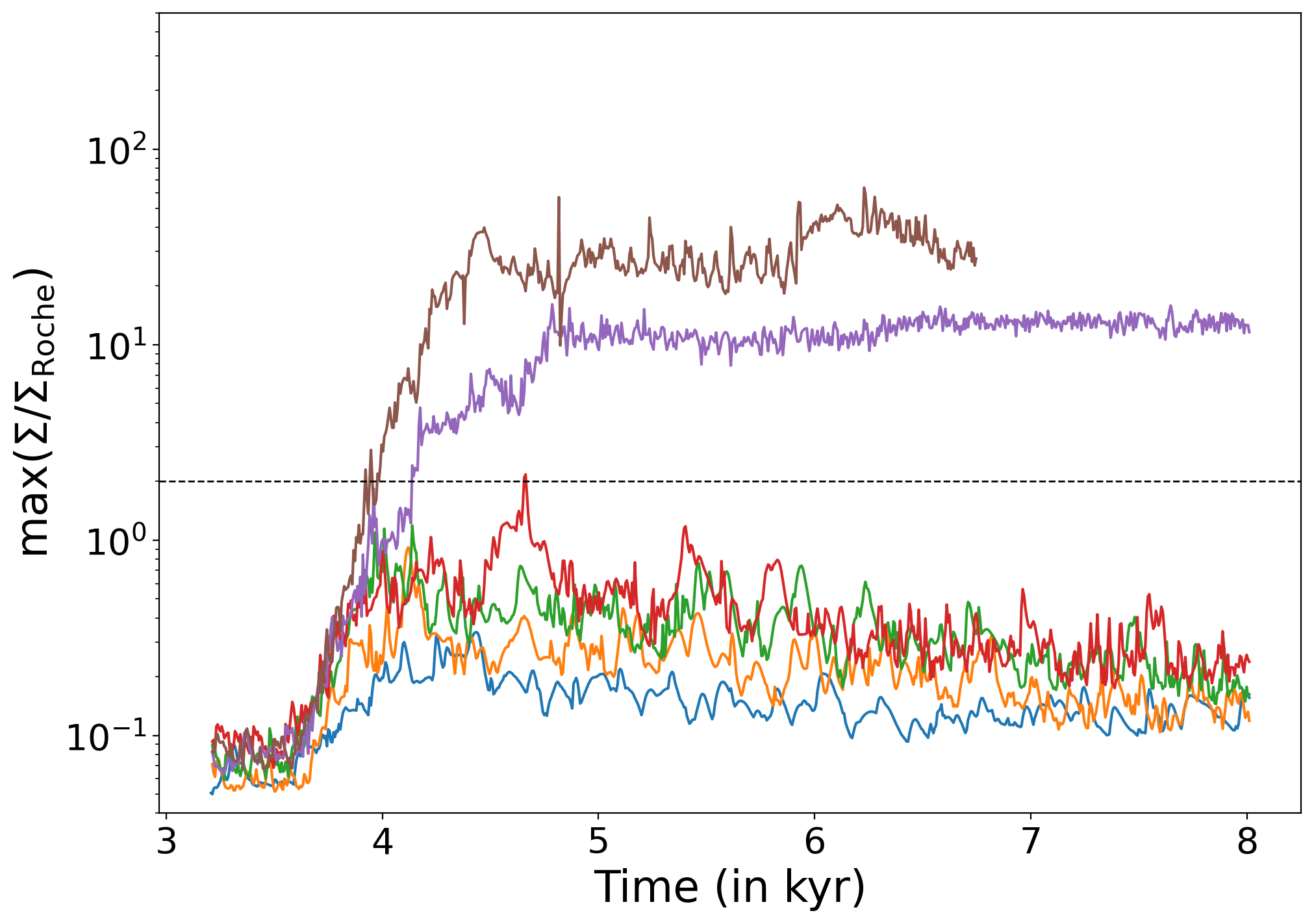} &
\includegraphics[width=0.315\hsize]{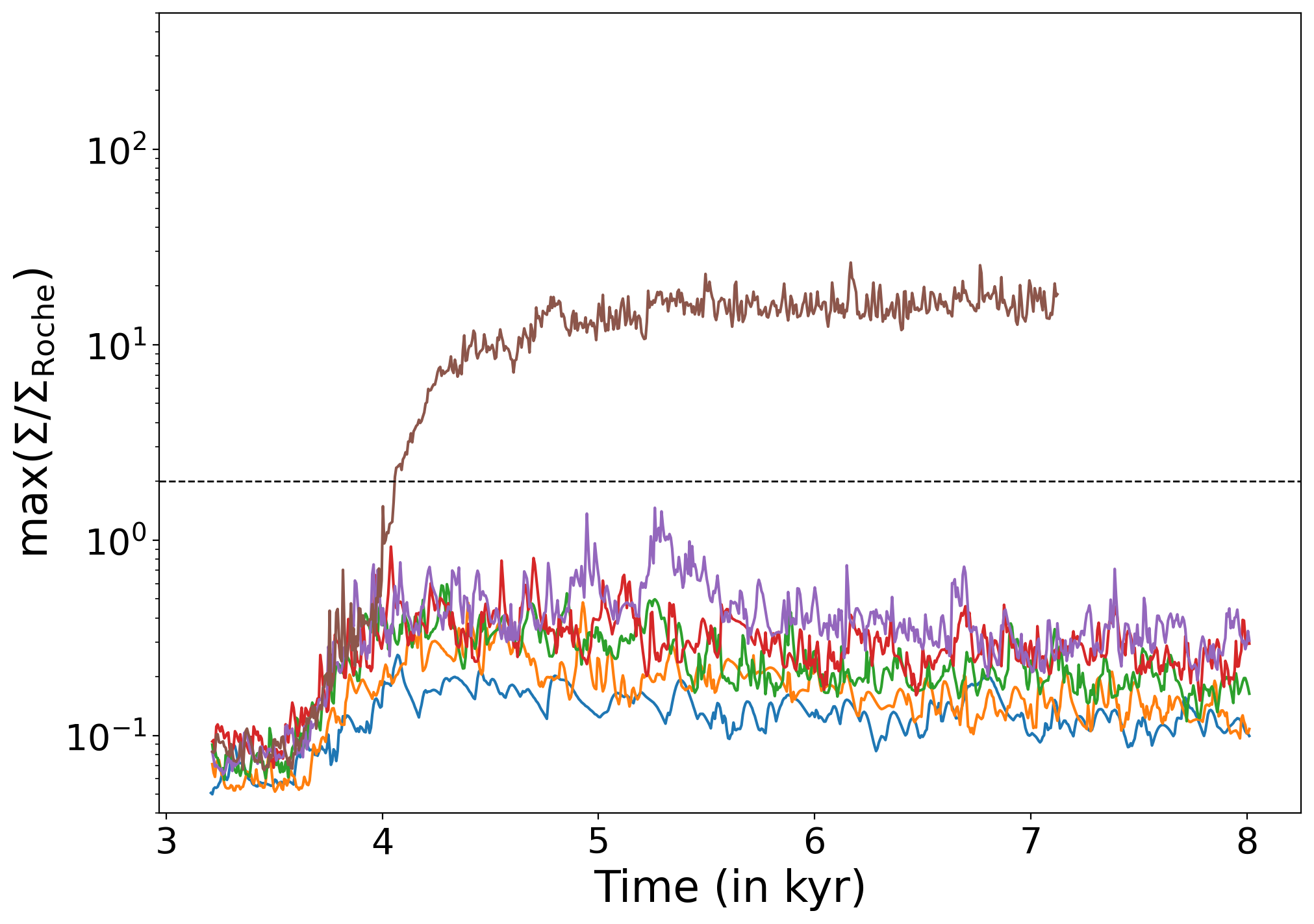} \\
\includegraphics[width=0.315\hsize]{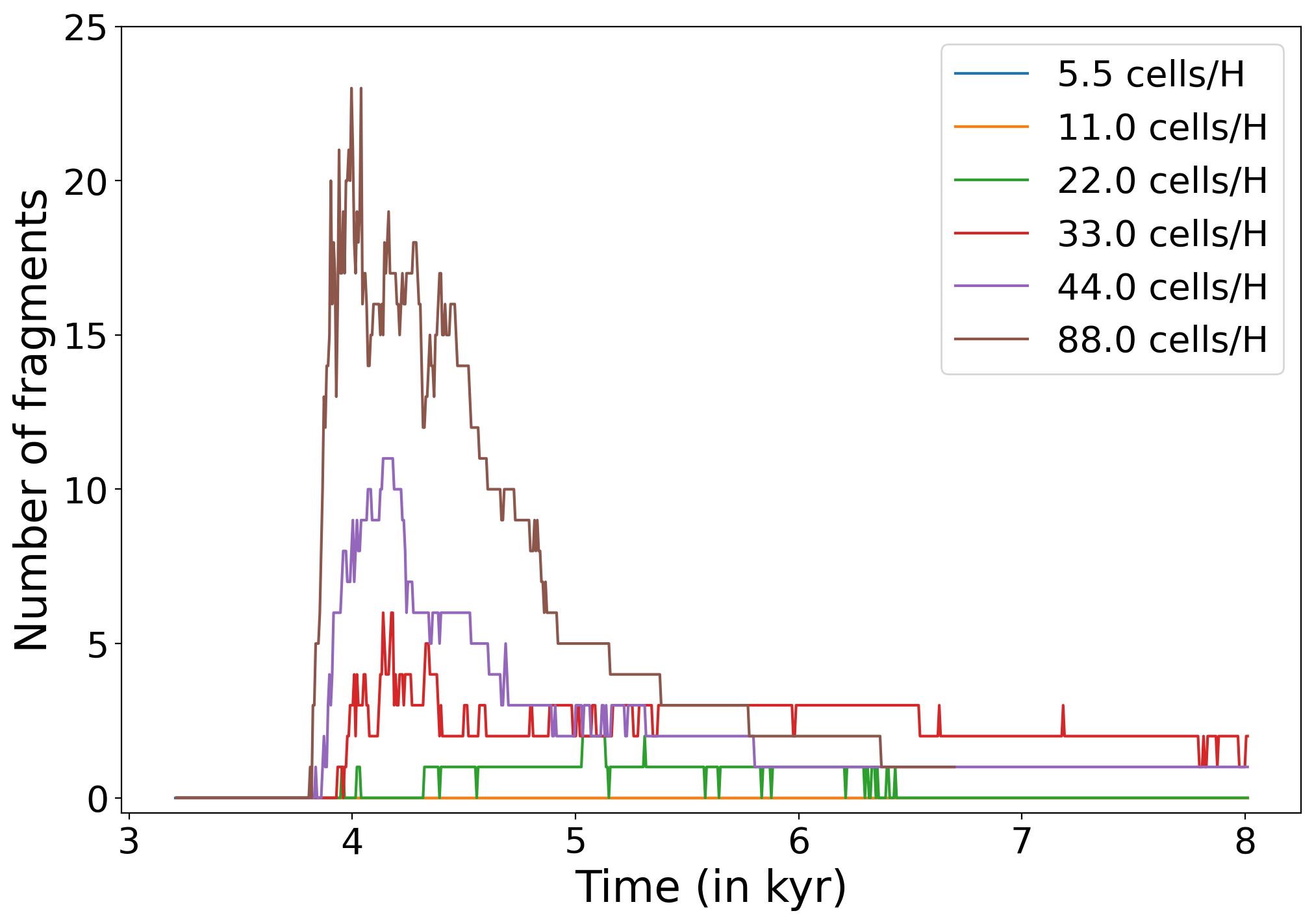} &
\includegraphics[width=0.315\hsize]{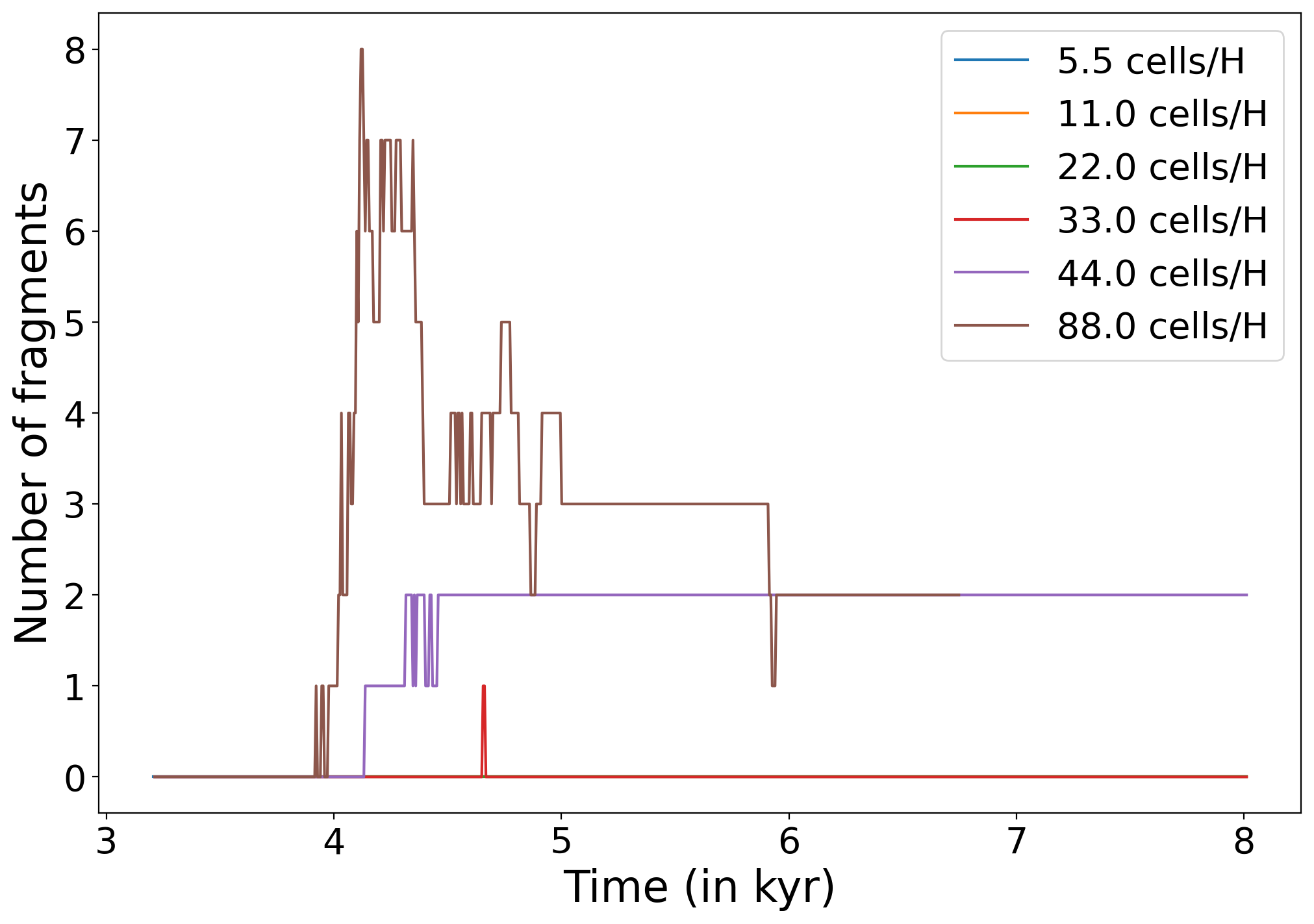} &
\includegraphics[width=0.315\hsize]{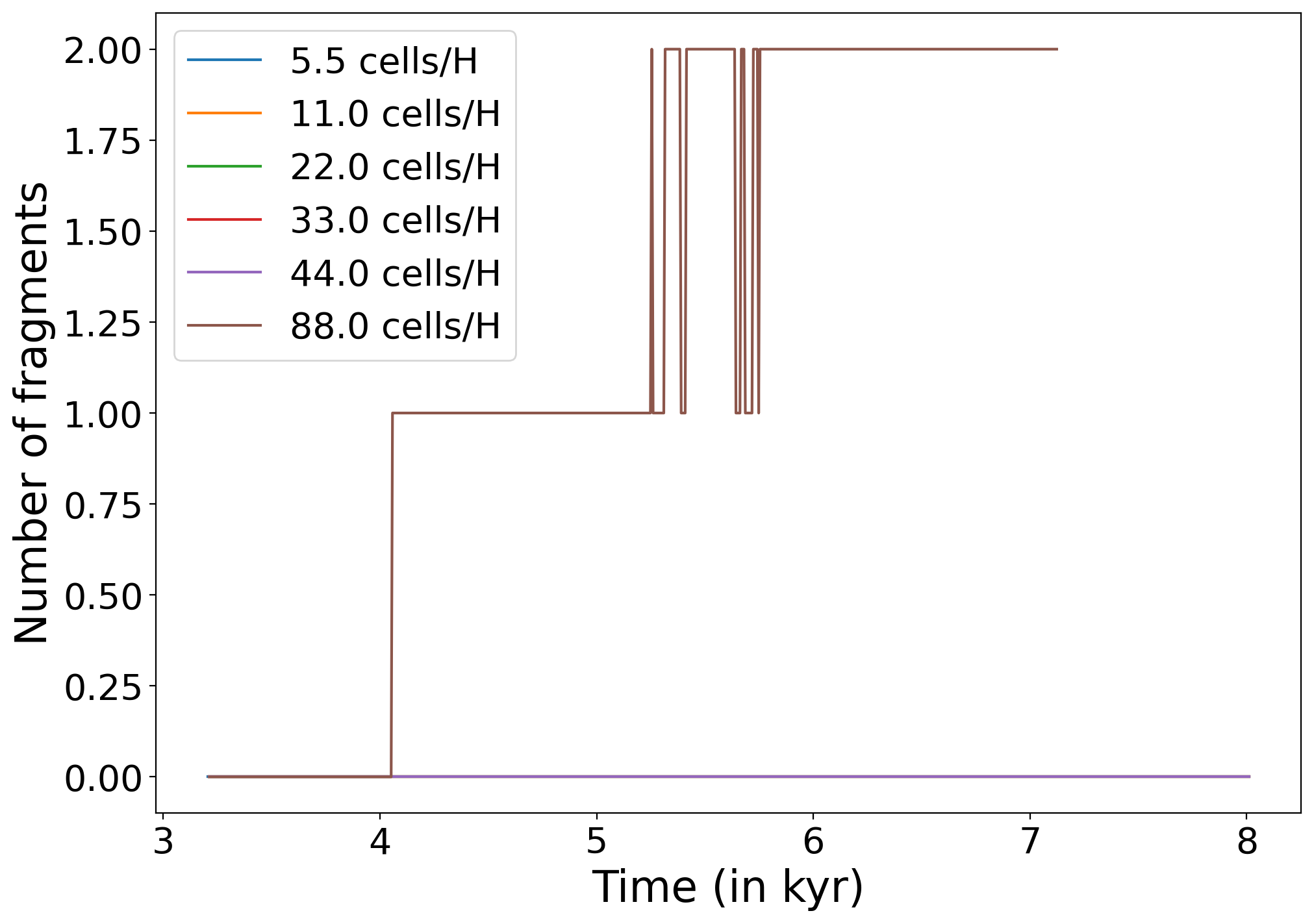} \\
\end{tabular}

\caption{Numerical convergence of the properties of the fragmentation regime when using the Bessel kernel for $\beta=1,2,3$.
The black dash line indicates the fragmentation threshold.
}
\label{fig: numerical convergence beta=1 and beta=2}
\end{figure*}

One may question whether my simulations also achieve convergence in terms of the physical properties characteristic of the gravito-turbulent and fragmentation regimes.
In the gravito-turbulent regime, this includes the estimation of stresses, whereas in the fragmentation regime, it involves assessing the number of fragments and the maximum density reached.

To investigate this, Figure \ref{fig: numerical convergence beta=1 and beta=2} shows the maximum density, normalized by the Roche density, as a function of time for $\beta=[1,2,3]$. 
For $\beta=1$, fragments form at resolutions higher than 22 cells/H, with densities exceeding the Roche density by factors ranging from 7 (for 22 cells/H) to 300 (for the highest resolution).
Additionally, the number of initial fragments increases with resolution.
For $\beta=2$, fragments only form at resolutions exceeding 44 cells/H, with densities surpassing the Roche density by a factor of 16. 
Specifically, for 44 cells/H and $\beta=2$, only two fragments emerge and persist throughout the simulation—from the onset of fragmentation to its conclusion—appearing to settle into a stable configuration.
Finally, for $\beta=3$, only the highest resolution simulation exhibits fragmentation, suggesting proximity to the fragmentation threshold.
In both cooling regimes, the high densities reached confirm the gravitationally bound nature of the fragments.
At the lowest resolutions (less than 22 cells/H) and with very efficient cooling ($\beta\leq1$), the disc may begin to fragment at surface density ratios as high as $\Sigma/\Sigma_{\rm Roche} \sim 4$, but the fragments do not survive.
In contrast, at high resolutions (greater than 44 cells/H), once the threshold defined by Eq.~\ref{Eq: fragmentation criterion} is reached, runaway accretion begins, as evidenced by the significant increase in density.
These results confirm two key points: first, that disruption at low resolution is likely a grid diffusion effect; and second, that the fragmentation criterion employed (Eq.~\ref{Eq: fragmentation criterion}) is appropriate.
Regarding the properties of fragmentation—namely, peak density and number of fragments—it appears that no numerical convergence is reached.
This could be explained by the fact that, for efficient cooling, the Jeans length might not be resolved, which enhances artificial fragmentation.

\begin{figure}
\centering
\includegraphics[width=\hsize]{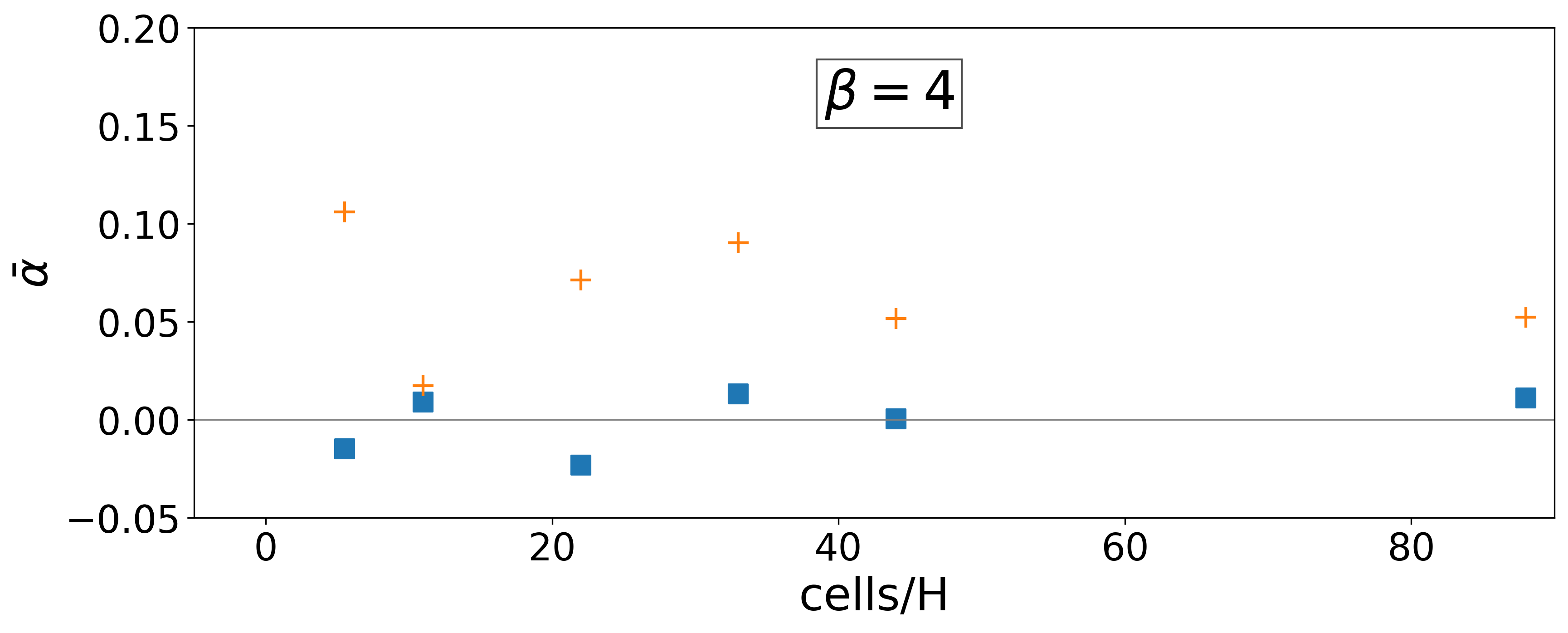}
\includegraphics[width=\hsize]{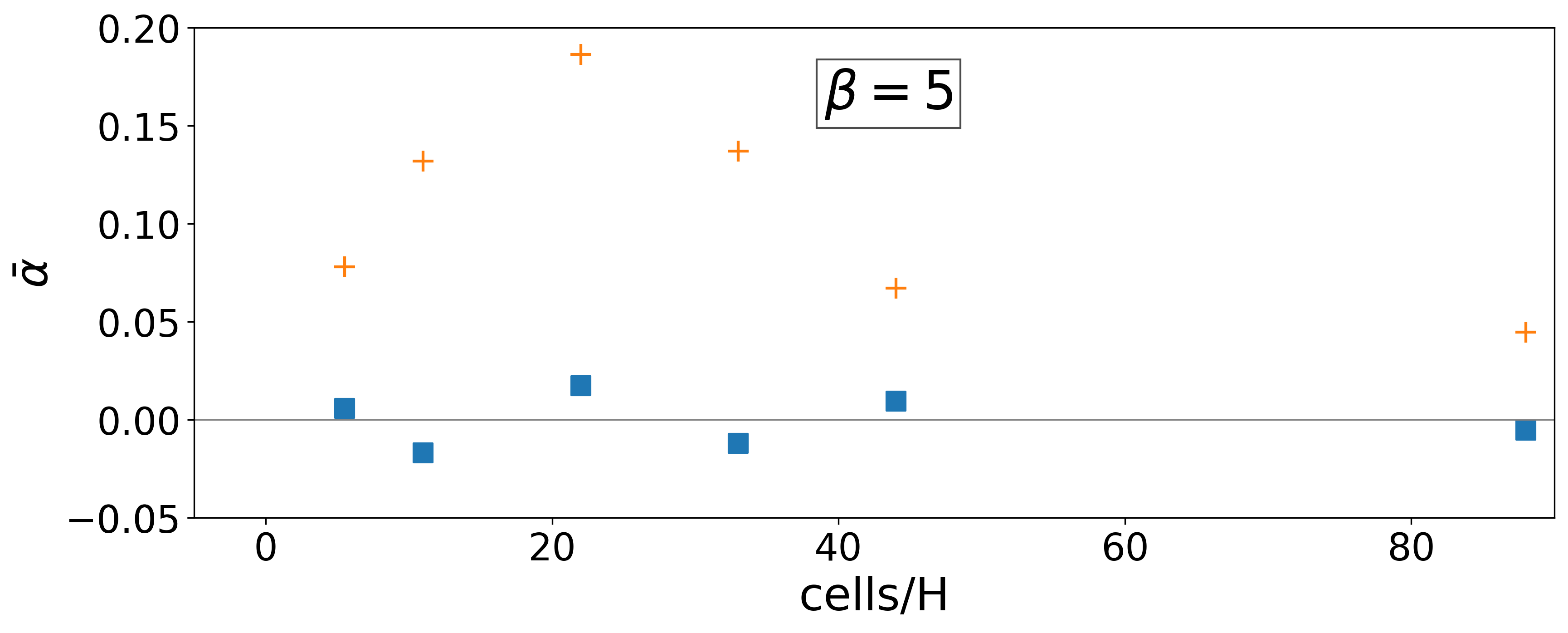}
\includegraphics[width=\hsize]{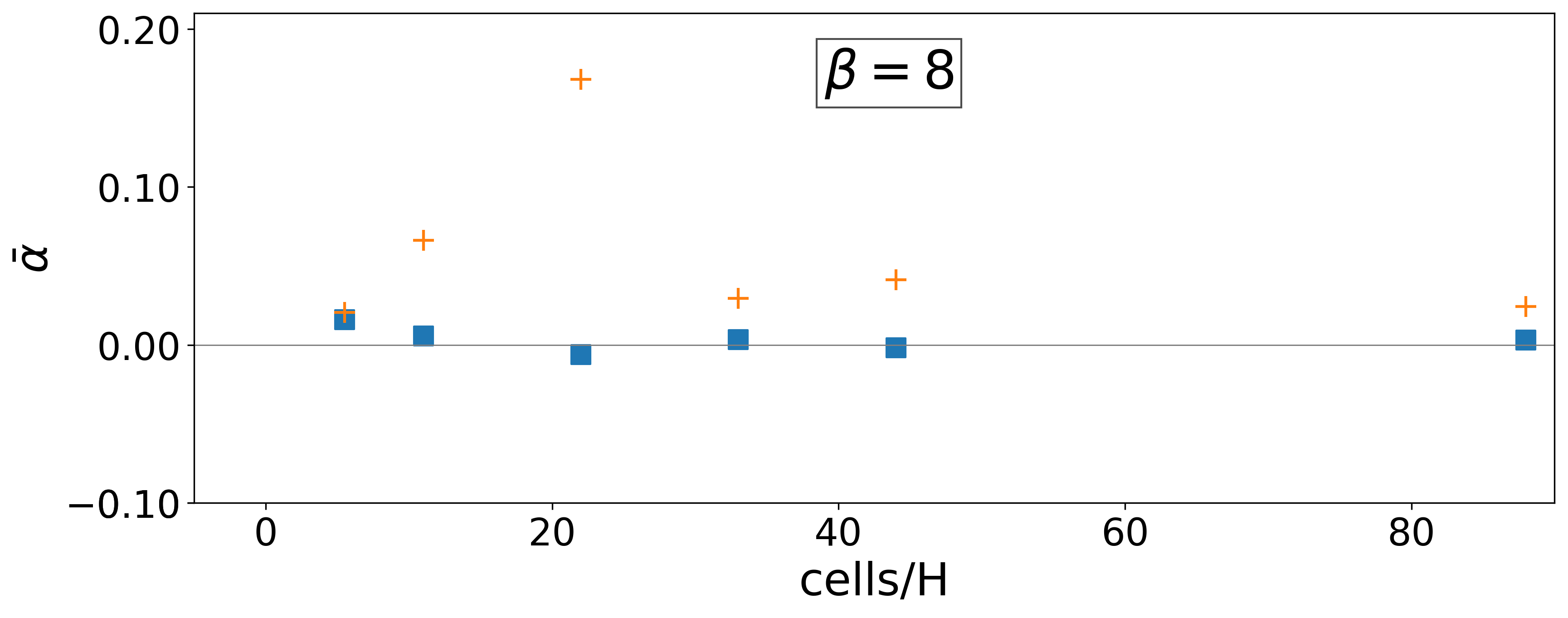}
\includegraphics[width=\hsize]{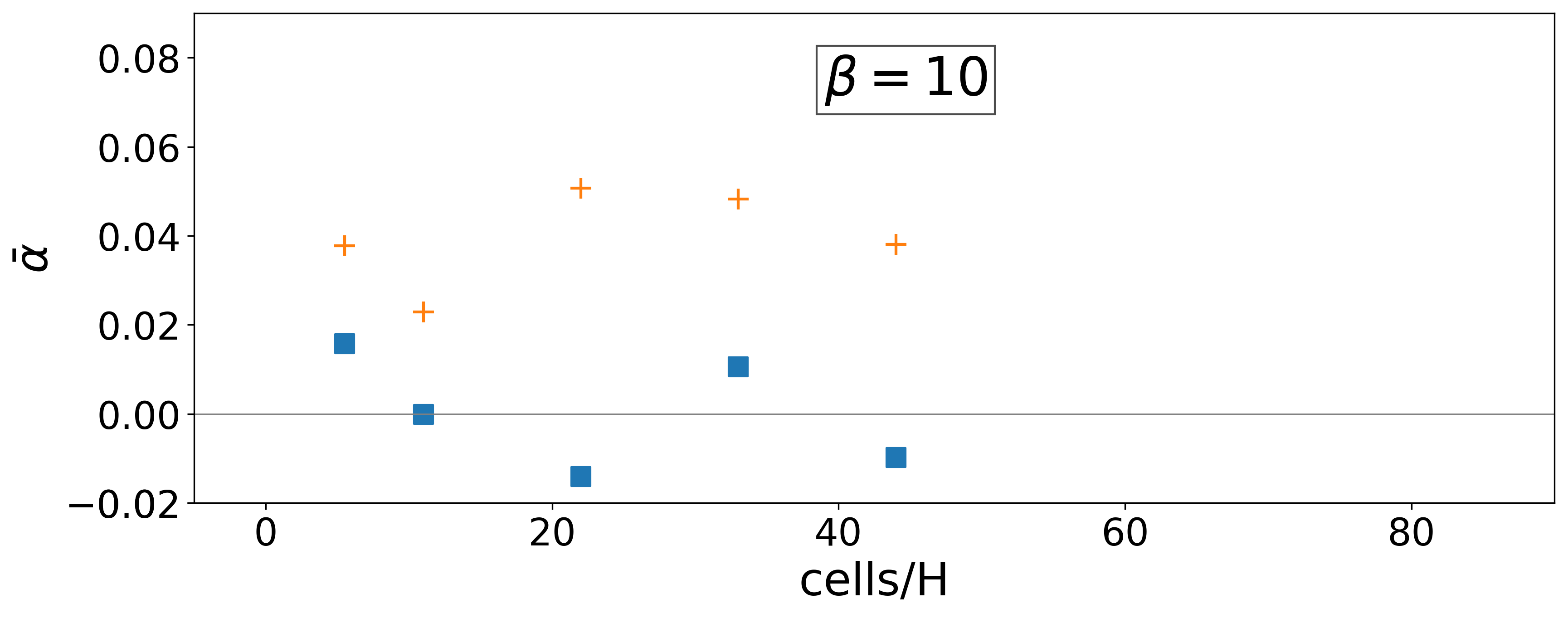}
\includegraphics[width=\hsize]{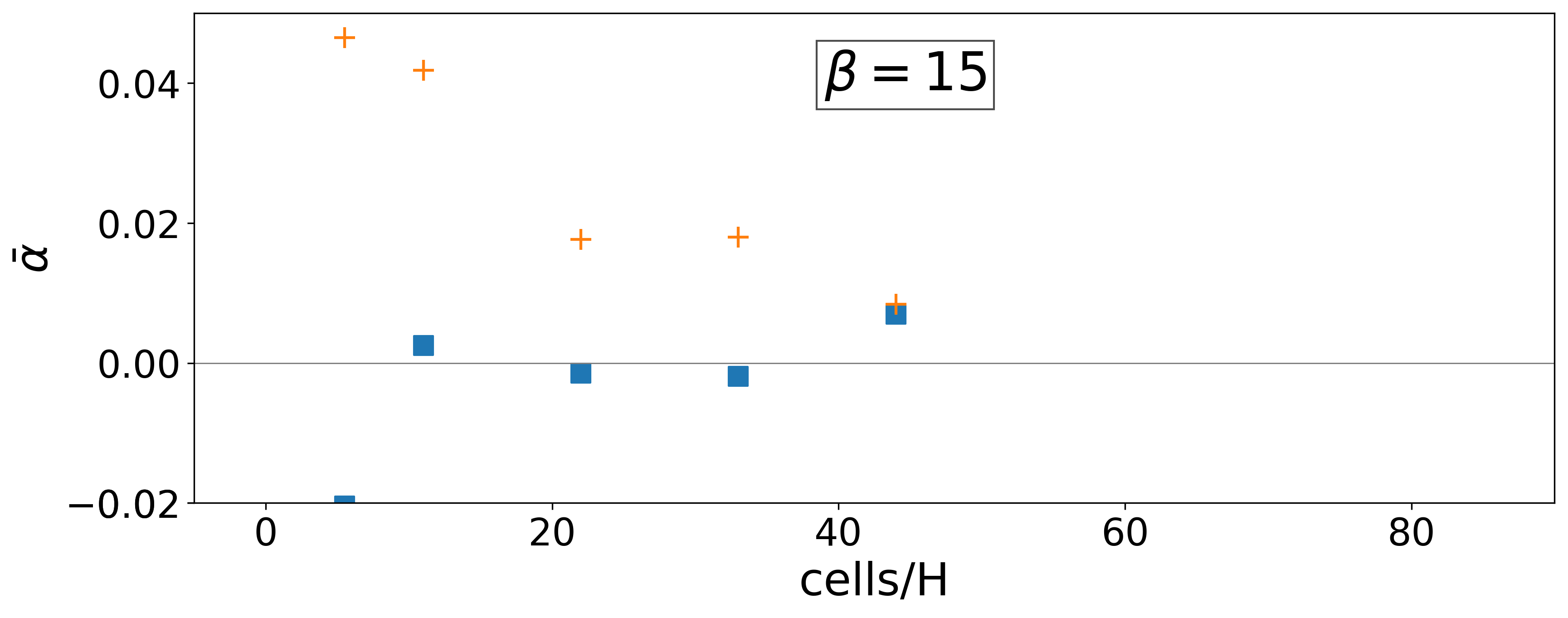}
\includegraphics[width=0.35\hsize]{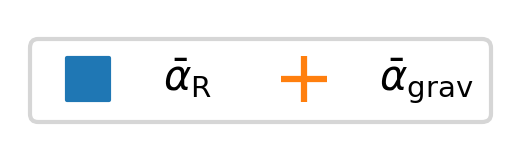}

\caption{Numerical convergence of the time and space averaged Reynolds and gravitational stresses in the regime of gravito-turbulence for different cooling rates.
} 
\label{fig: numerical convergence reynolds and grav stresses}
\end{figure}

In Figure \ref{fig: numerical convergence reynolds and grav stresses}, I present the time- and space-averaged Reynolds stress parameter, $\reynolds$, and the gravitational stress parameter, $\grav$, for various resolutions and cooling parameters. 
To conserve numerical resources, simulations were not conducted at the highest resolution for $\beta=[10,15]$. 
To mitigate boundary effects and avoid contamination from the initial relaxation phase at $\beta_{ini}=30$, these stresses were spatially averaged between $R_{\rm min}=25$ AU and $R_{\rm max}=85$ AU, and temporally averaged between $T_{min}=3.2$ kyr and $T_{\rm max}=7.4$~kyr.
For all cooling parameters, the gravitational stress dominates the Reynolds stress, with the latter eventually taking negative values, although the total stress remains positive. 
As expected, the smaller the $\beta$ parameter, the higher the stresses.
The key result is that the resolution at which numerical convergence is reached in the gravito-turbulent regime is inversely proportional to the cooling efficiency: for $\beta=[4,5,8]$, convergence appears to be achieved at 44 cells/H, whereas for $\beta=[10,15]$, it is reached at 22 cells/H.

In this section, I investigated how the use of the Bessel kernel affects numerical convergence according to three distinct criteria.
The first criterion is the separation between gravito-turbulence and fragmentation, which converges at 40 cells/H.
 In the fragmentation regime, convergence is not achieved. 
This is likely due to the intrinsic nature of fragmentation, which requires increasingly higher resolutions at the fragment location—something that can only be addressed using AMR techniques combined with the insertion of sink particles.
Finally, in the gravito-turbulent regime, the stress values converge differently depending on the $\beta$ parameter.

\section{Characterisation of GI for different gravity prescriptions}\label{sec: characterisation of GI as a function of gravity prescription}

Now that I have demonstrated the convergence of GI simulations using the Bessel kernel, my next objective is to compare the results obtained with this approach to those using the Plummer potential prescription for various smoothing lengths. 
Specifically, I have selected ~$\epsilon/H_{\rm rms}=[0, 0.3, 0.6, 1.2]$, which are standard values in PPD studies involving self-gravity \citep{2001_gammie, 2011_paardekooper, 2015_young_clarke, 2016_zhu, 2016_baruteau, 2018_vorobyov, 2026_nayakshin}.
I remind that setting the smoothing length to zero is equivalent to solving the two-dimensional Poisson equation, a common practice in shearing box simulations \citep{2012_paardekooper, 2015_baehr, 2017_klee, 2019_klee}.
While the theoretical work of \citet{muller_kley_2012} recommends a general value of $\epsilon/H_{\rm rms}=1.2$, in practice simulations often adopt different values depending on the focus of the study. 
For instance, simulations investigating fragmentation typically use $\epsilon/H_{\rm rms}=0.3$, whereas those examining the gravito-turbulent state commonly employ $\epsilon/H_{\rm rms}=0.6$.

For this comparative study, I use a grid resolution of $(N_r, N_\varphi)=(1024, 3072)$ cells, corresponding to approximately 44 cells per scale height. 
Additionally, I have chosen to conduct this study with two values of the $\beta$ parameter, $\beta=[2, 8]$, representing the fragmentation regime and the gravito-turbulent regime, respectively.
The value $\beta=8$ was specifically selected because, as demonstrated in the previous section, the time- and space-averaged Reynolds and gravitational stresses have converged for this parameter.
The value $\beta=2$ was chosen because, although fragmentation simulations did not converge, the outcomes for resolutions of 44 cells/H and 89 cells/H are very similar

\subsection{Fragmentation: $\beta=2$}\label{subsec:characterisation of GI different kernels - beta 2}

\begin{figure*}
\centering

\begin{tabular}{cp{0.3\hsize}p{0.3\hsize}p{0.3\hsize}}
& \centering{\textbf{time = $3.9$ kyr}} & \centering{\textbf{time = $4.3$ kyr}} & \textbf{time = $6.1$ kyr} \\
\rotatebox{90}{\hspace{6.4em} \textbf{Bessel}} &
\includegraphics[width=\linewidth]{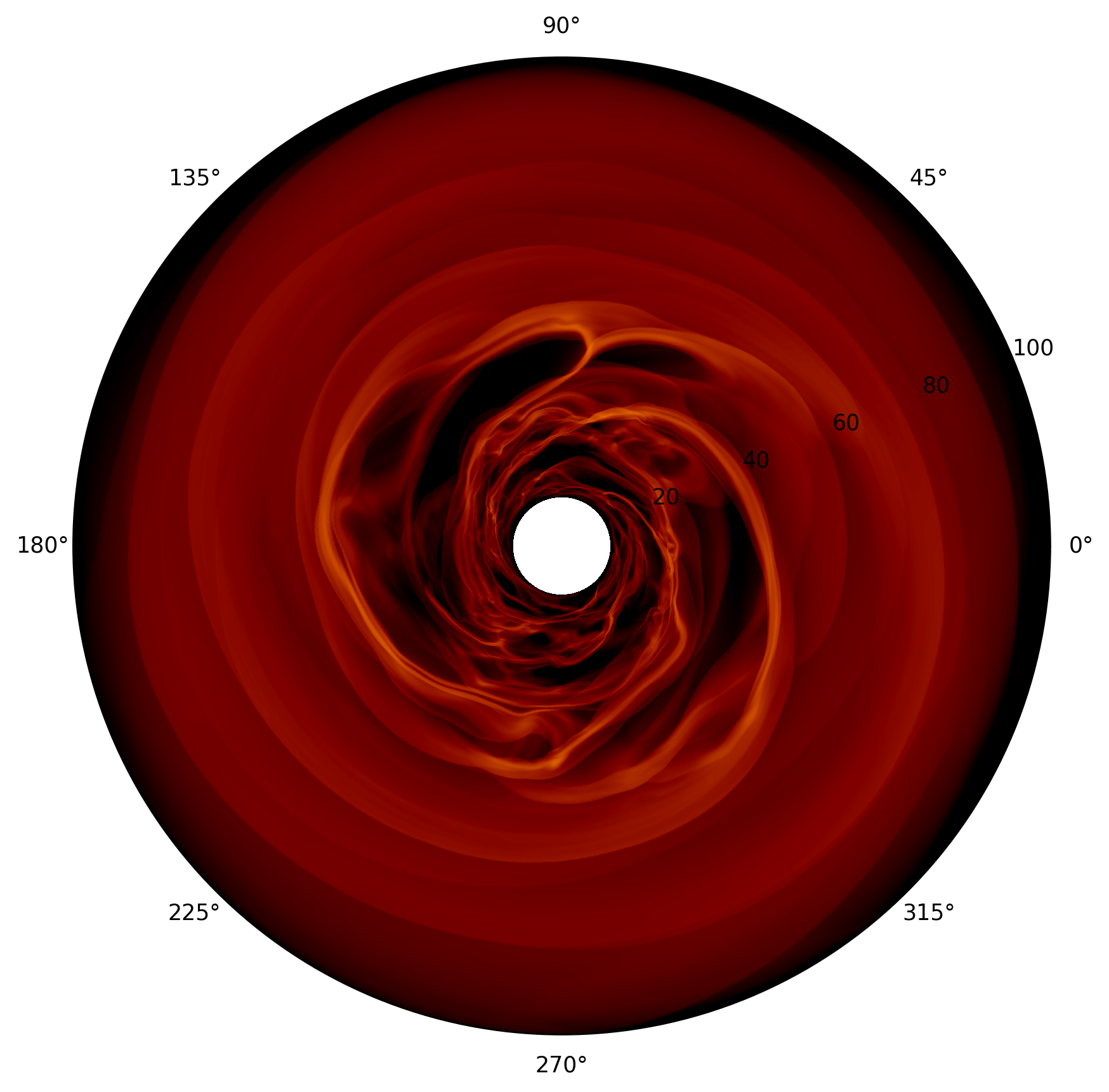} &
\includegraphics[width=\linewidth]{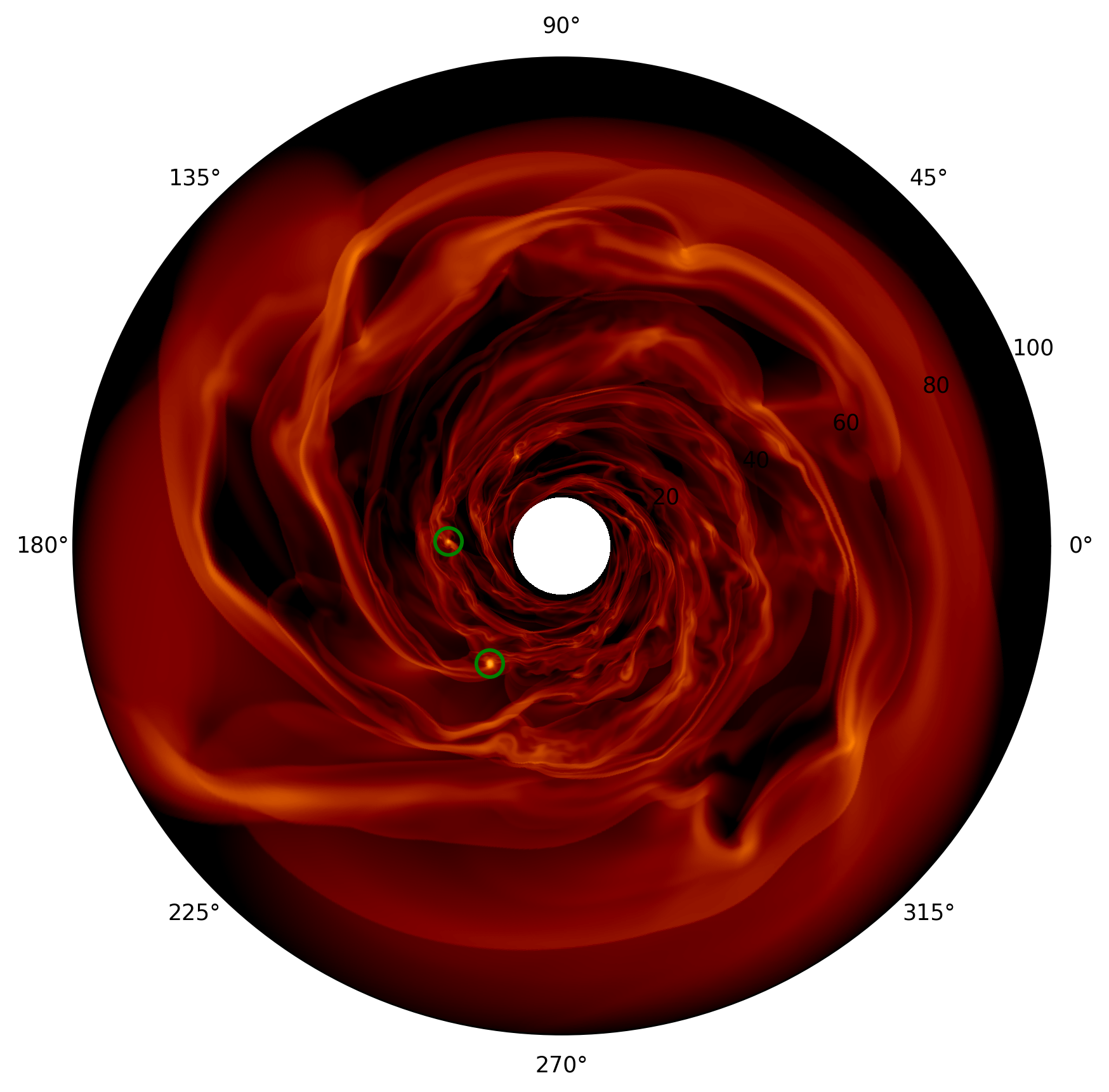} &
\includegraphics[width=\linewidth]{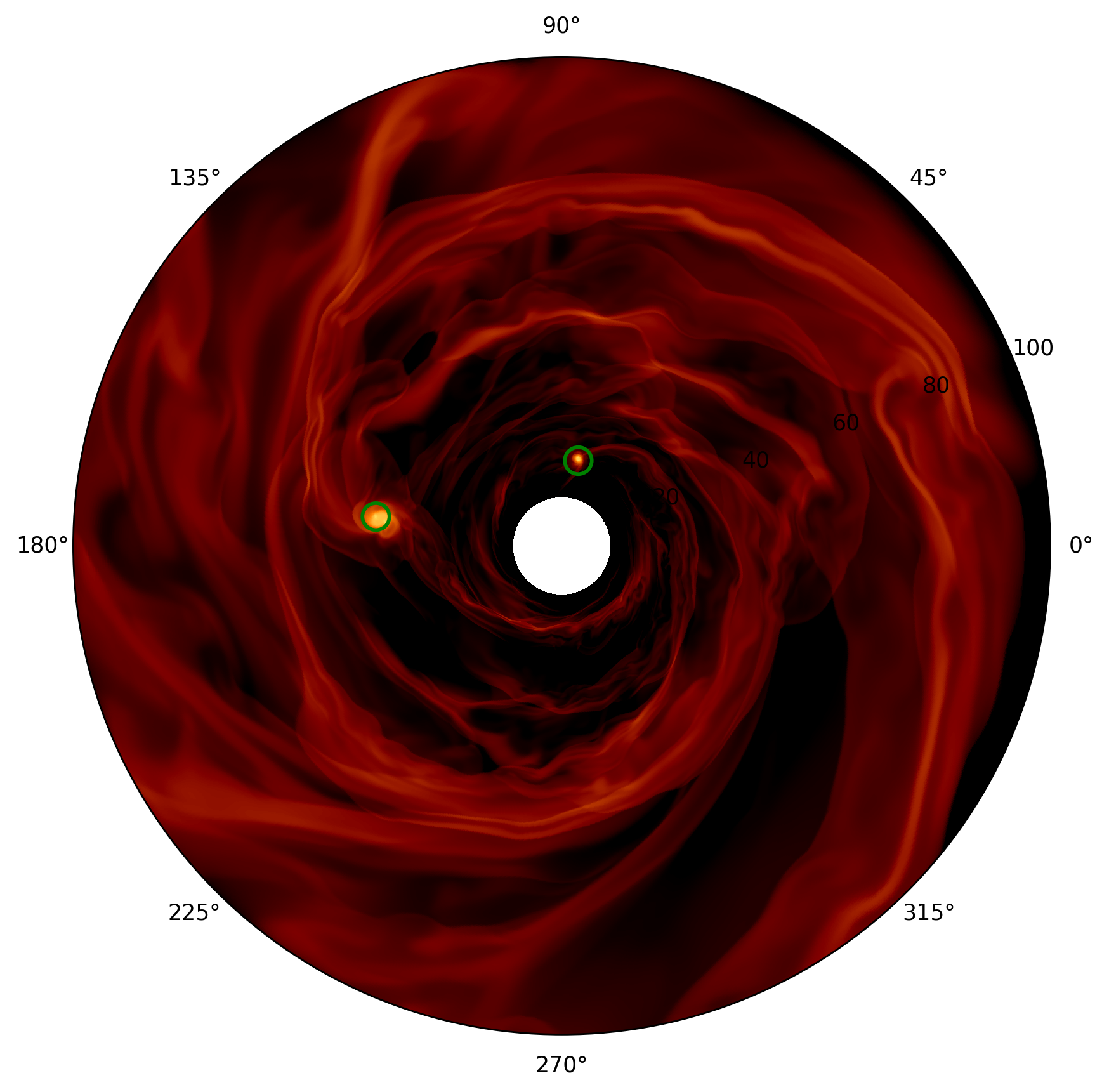} \\
\rotatebox{90}{\hspace{4.8em} \textbf{$\displaystyle \frac{\epsilon}{H_{\rm rms}}=0.0$}} &
\includegraphics[width=\linewidth]{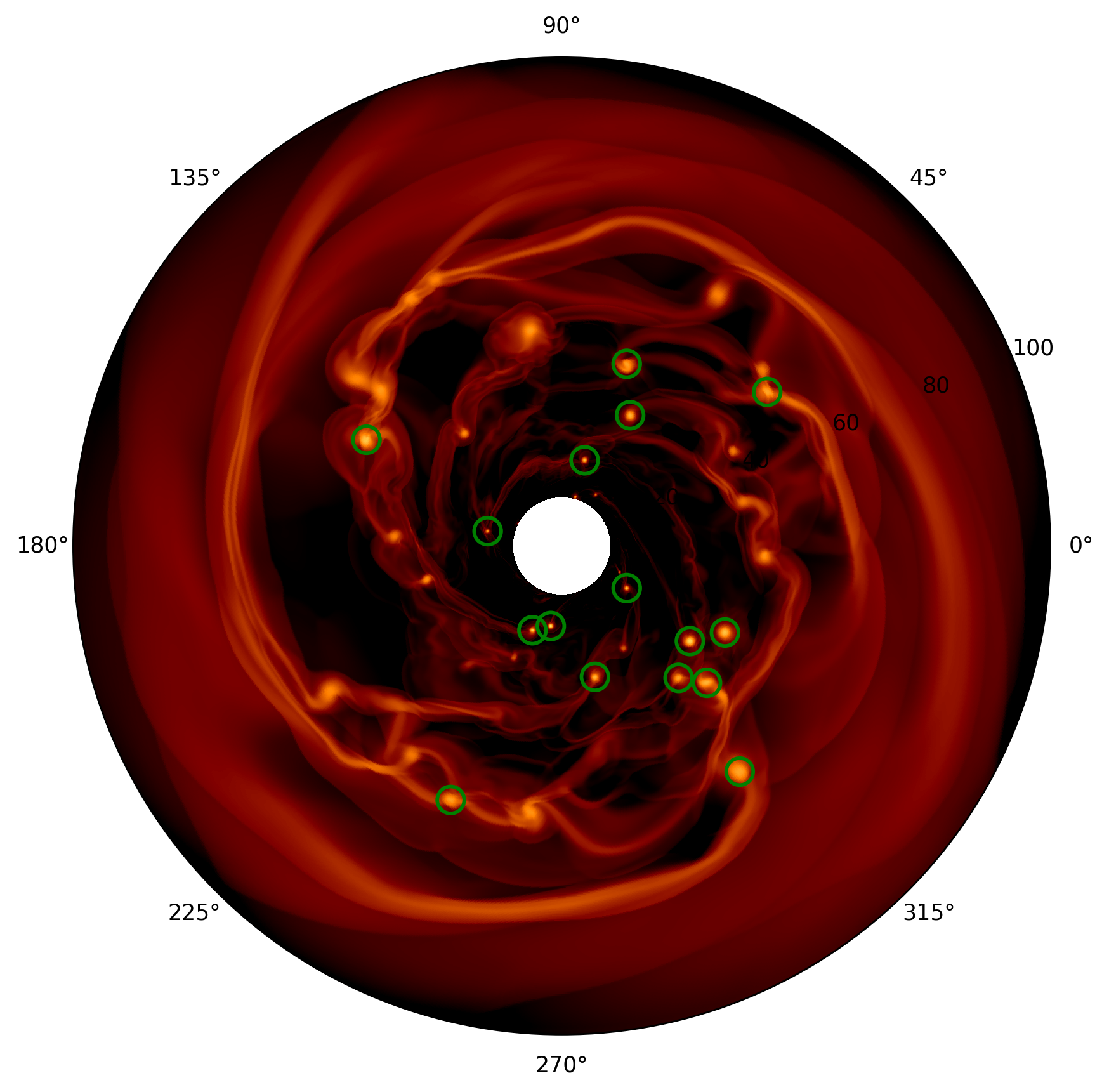} &
\includegraphics[width=\linewidth]{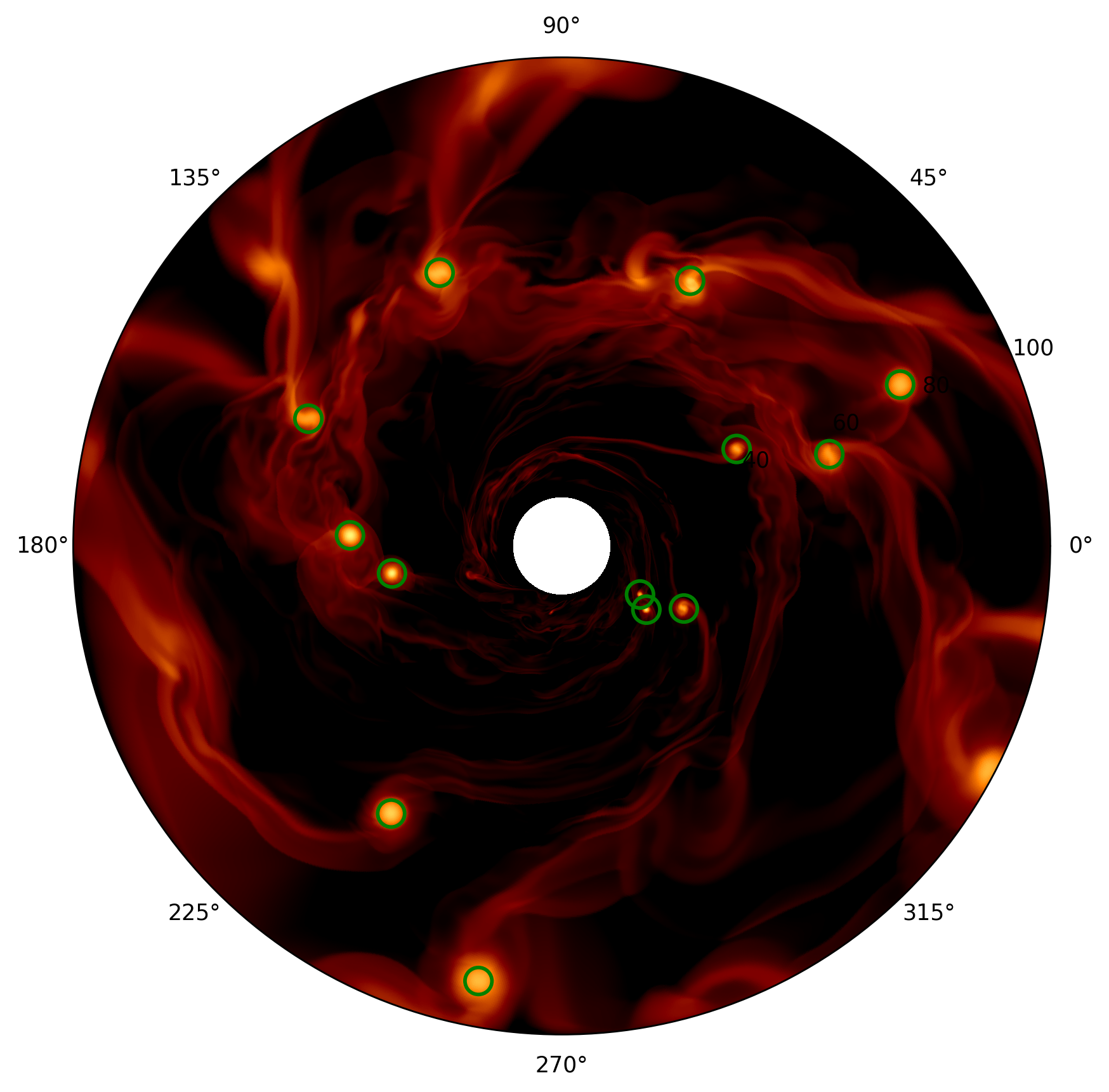}  &
\includegraphics[width=\linewidth]{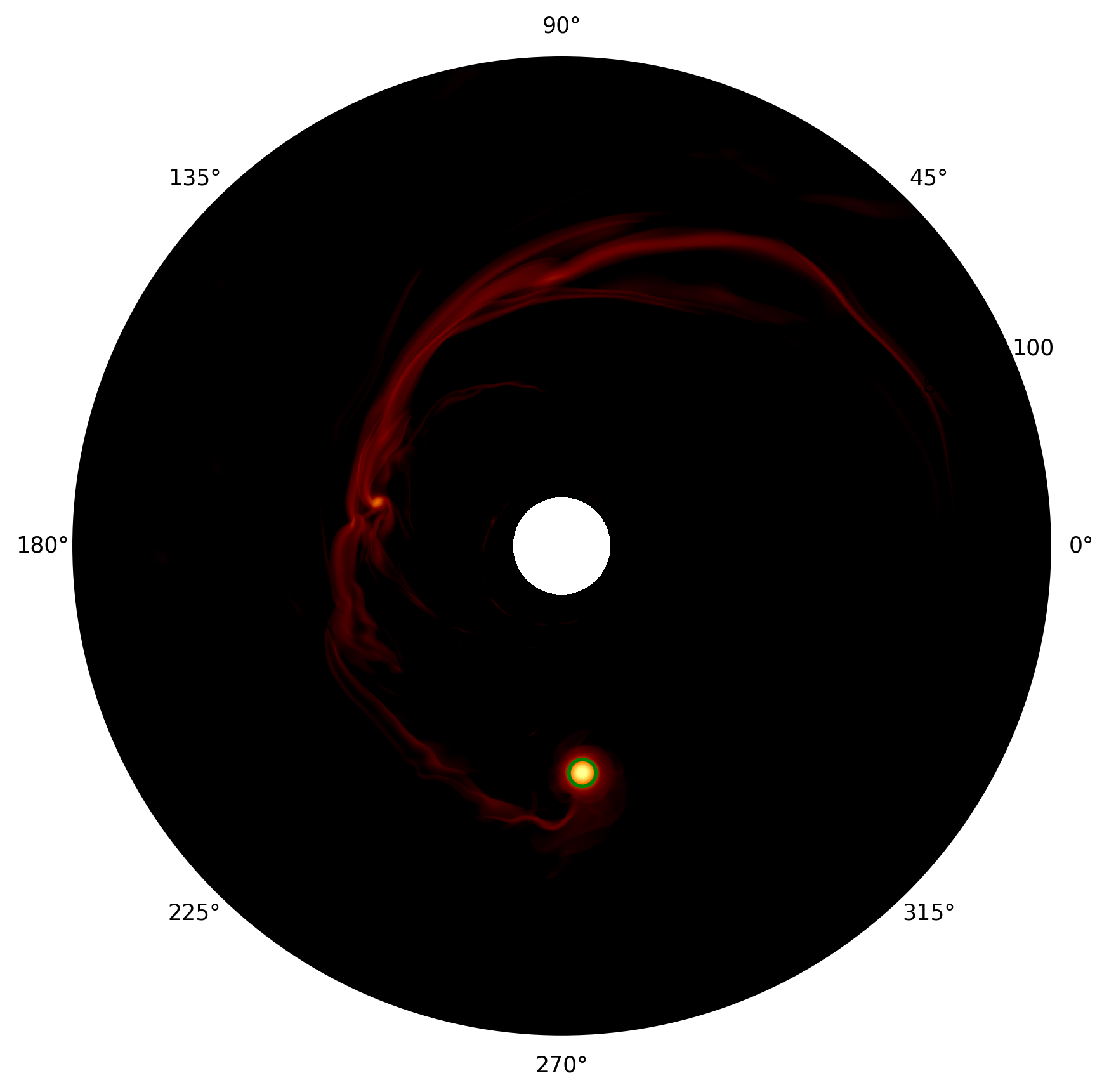}  \\
\rotatebox{90}{\hspace{4.8em} \textbf{$\displaystyle \frac{\epsilon}{H_{\rm rms}}=0.3$}} &
\includegraphics[width=\linewidth]{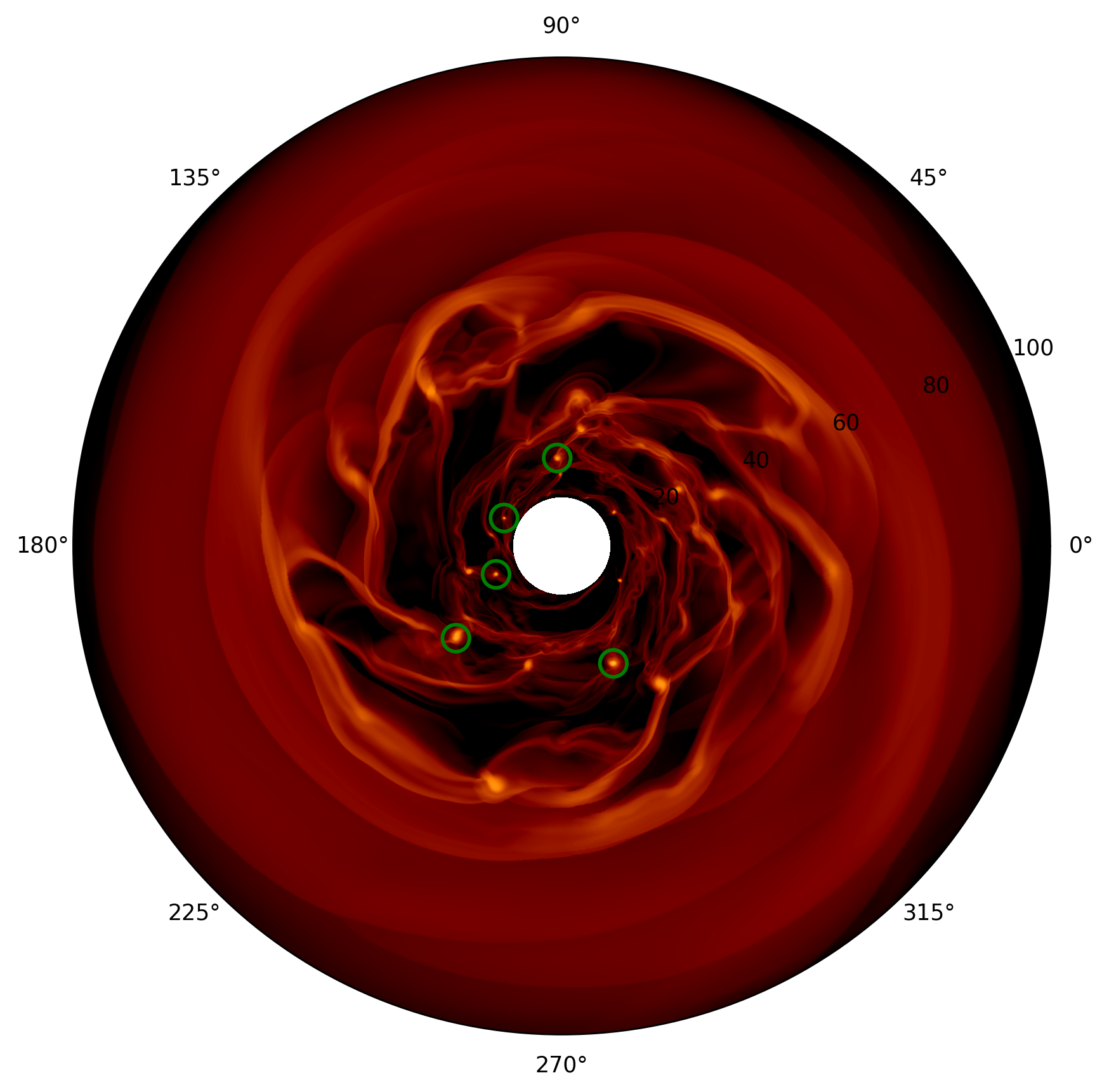} &
\includegraphics[width=\linewidth]{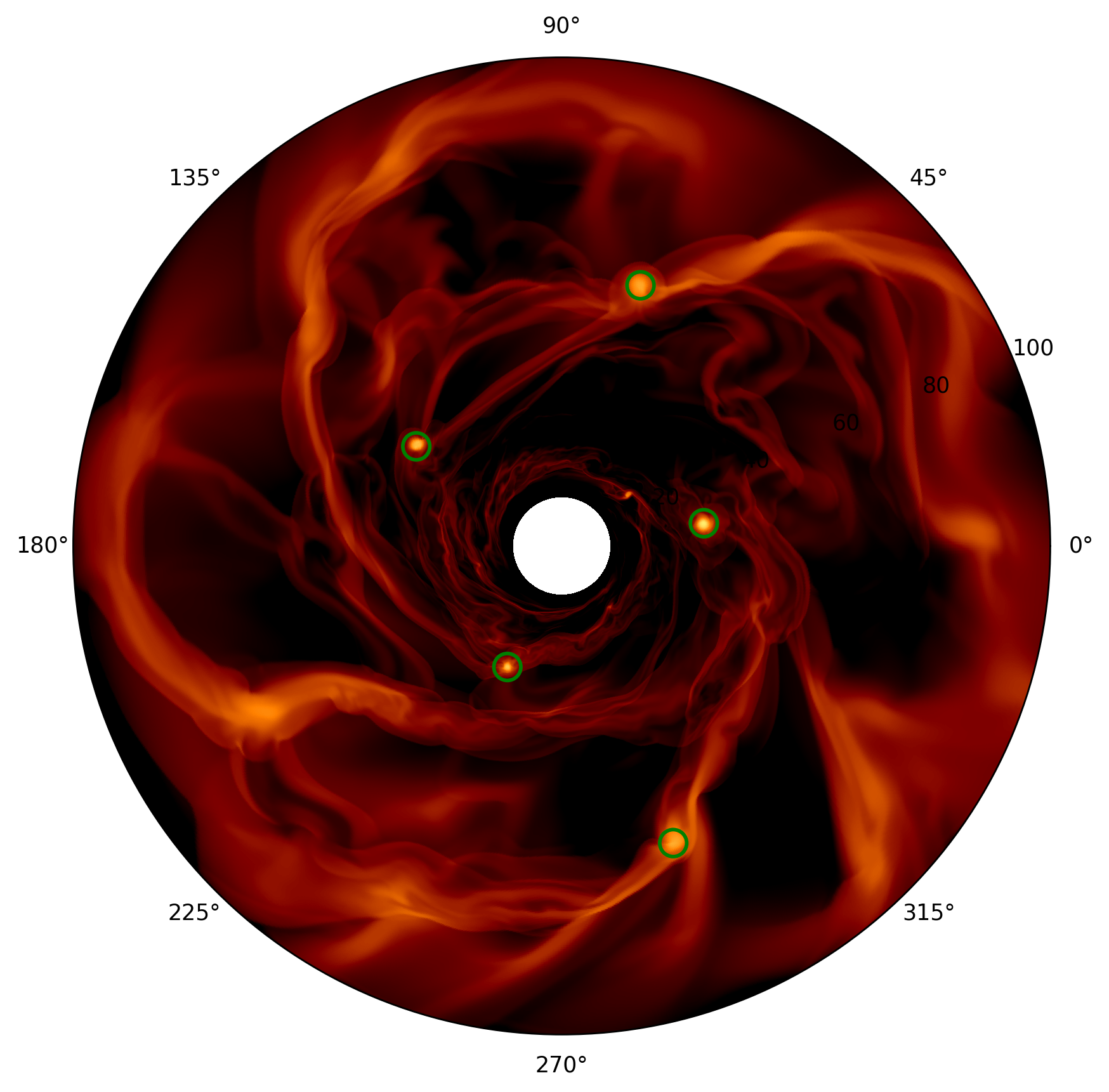}  &
\includegraphics[width=\linewidth]{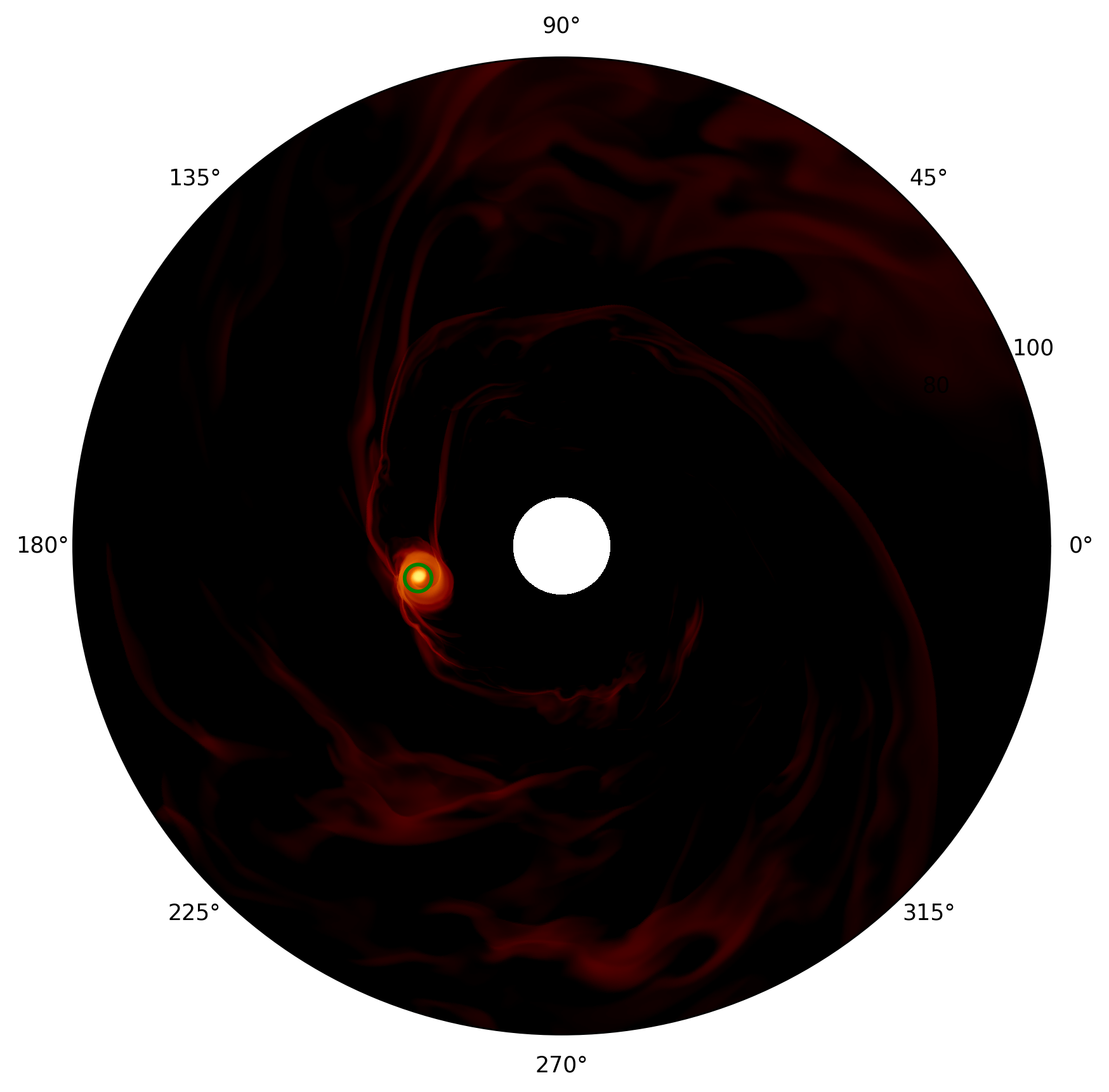}  \\
\rotatebox{90}{\hspace{4.8em} \textbf{$\displaystyle \frac{\epsilon}{H_{\rm rms}}=0.6$}} &
\includegraphics[width=\linewidth]{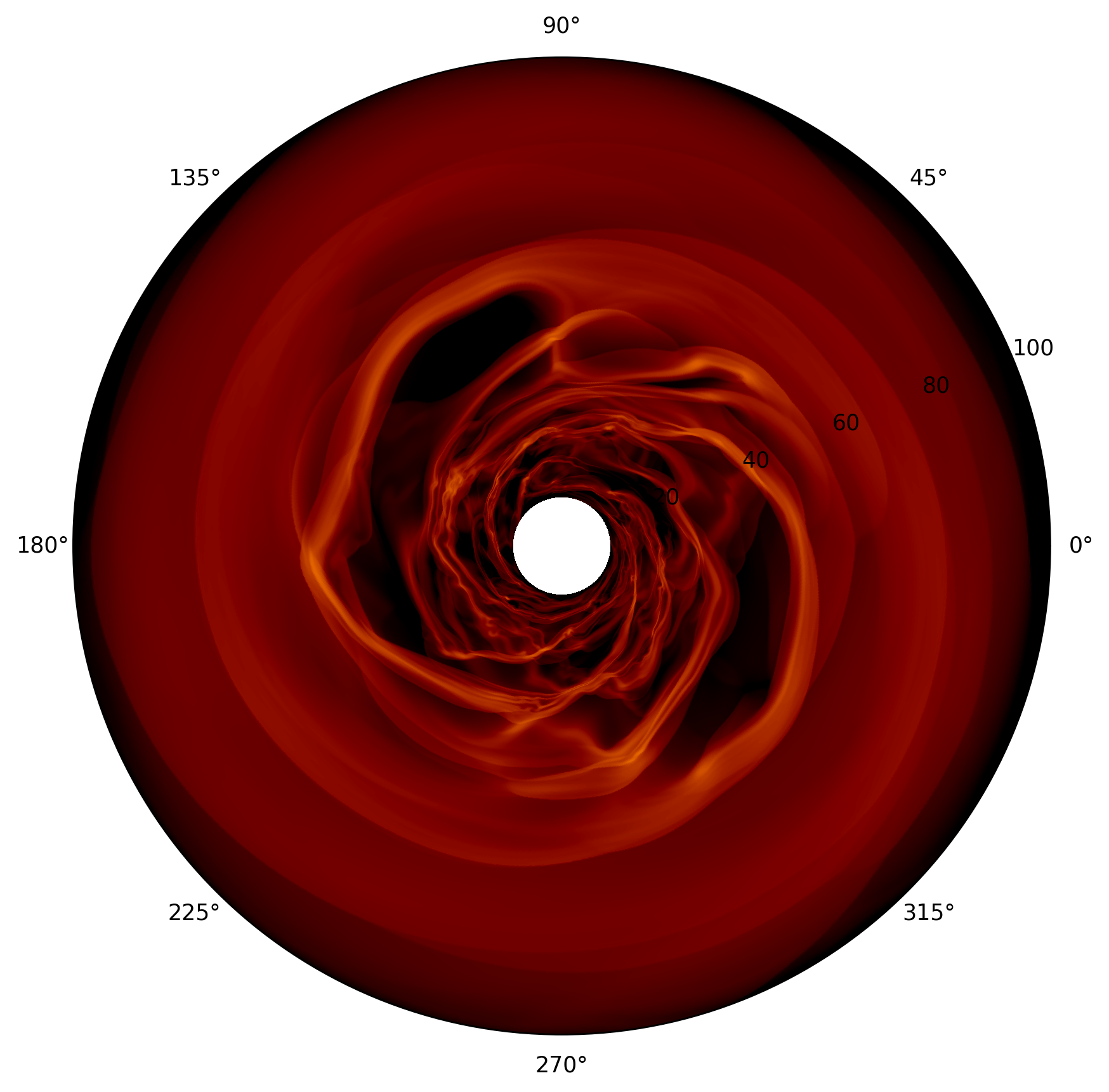} &
\includegraphics[width=\linewidth]{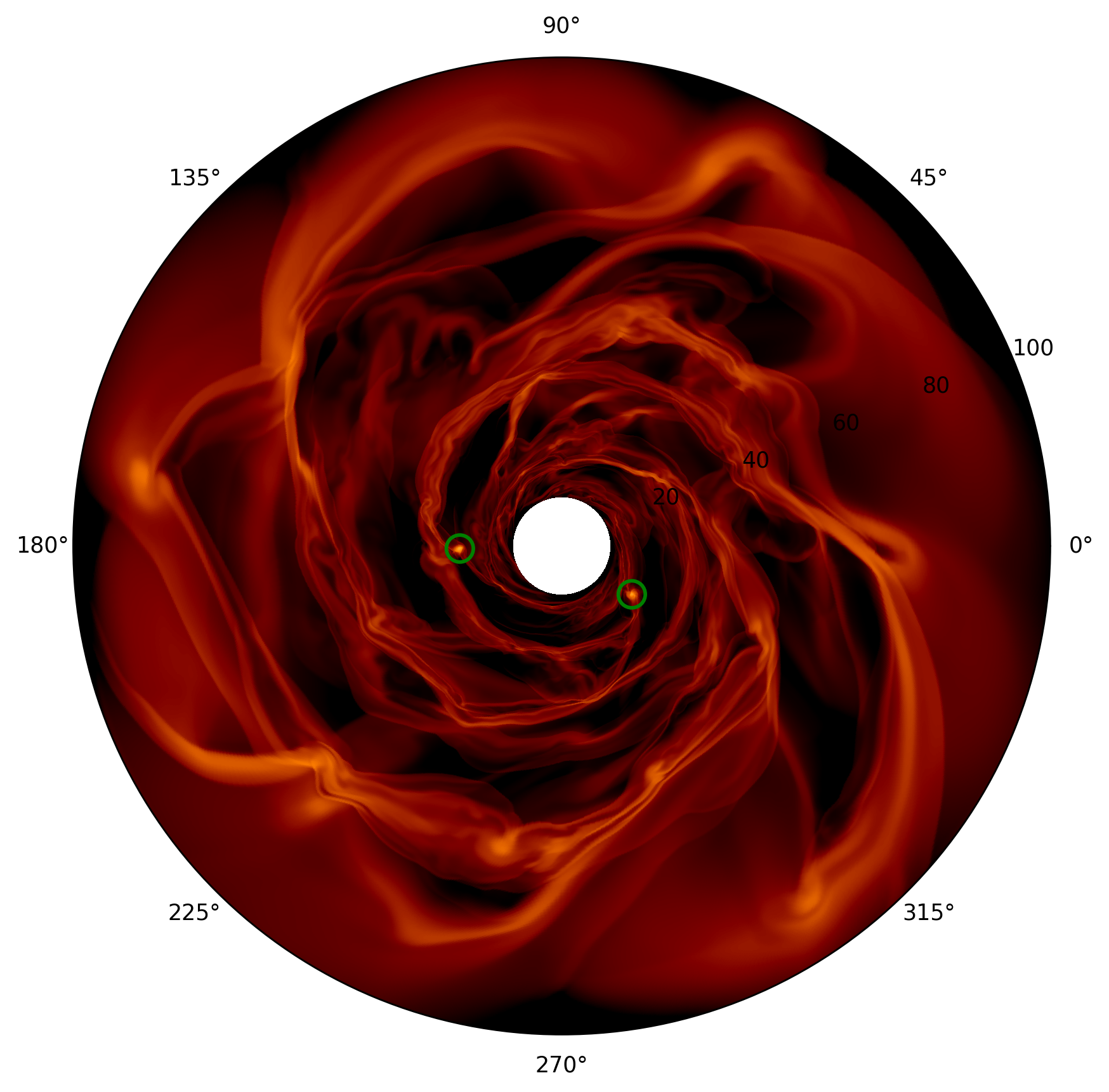}  &
\includegraphics[width=\linewidth]{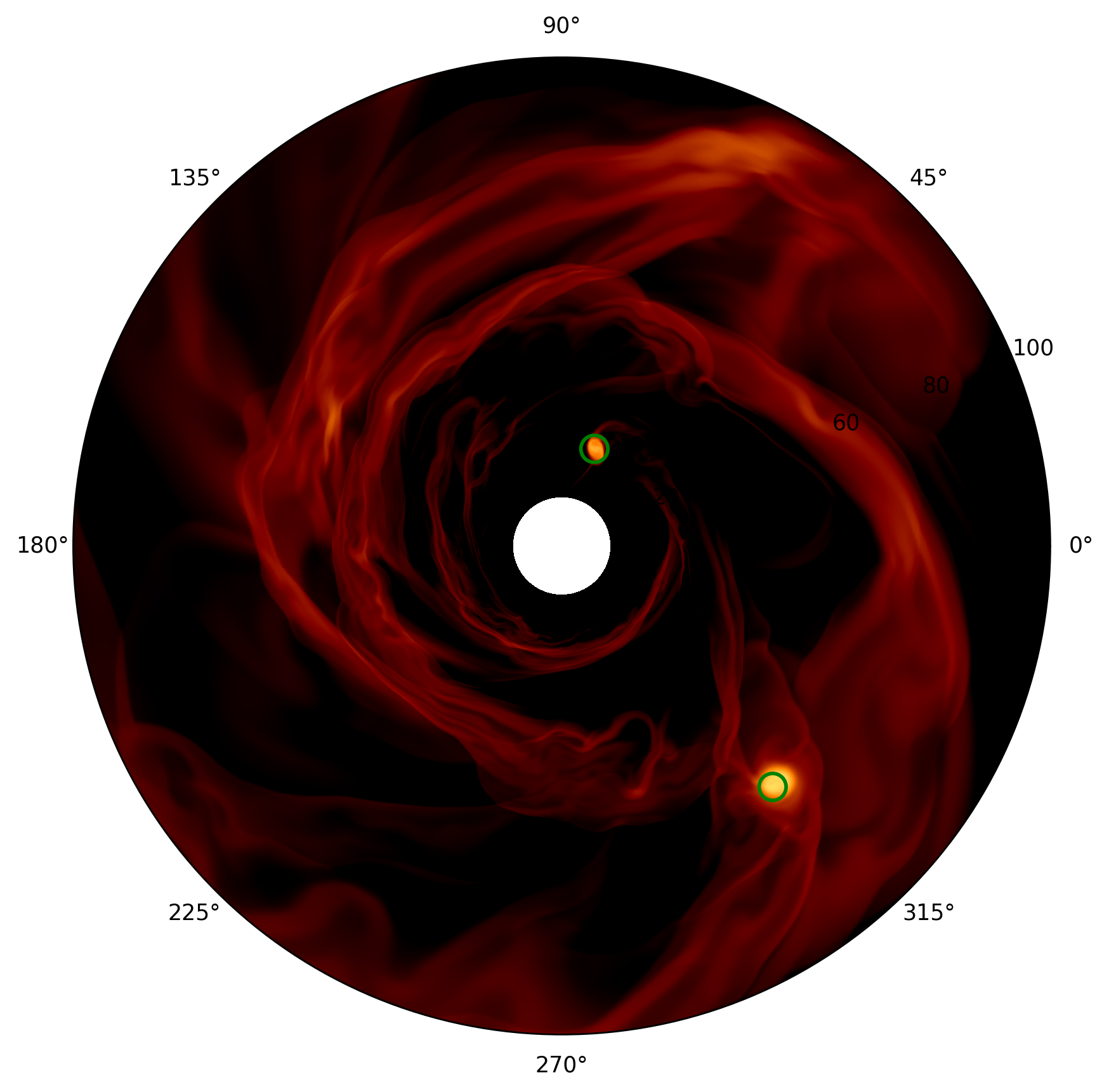}  \\
\end{tabular}

\includegraphics[width=0.9\linewidth]{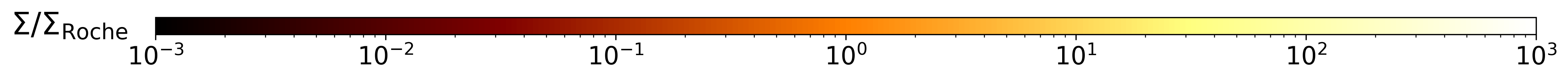}

\caption{Comparison of the fragmentation regime ($\beta=2$) for the different gravity prescriptions at different times.
Fragments are highlighted with green circles. 
For $\epsilon/H_{\rm rms}=1.2$ (not shown here), no fragments form.
}
\label{fig: comparison fragmentation for different kernels}
\end{figure*}

\begin{figure}
\centering
\includegraphics[width=\hsize]{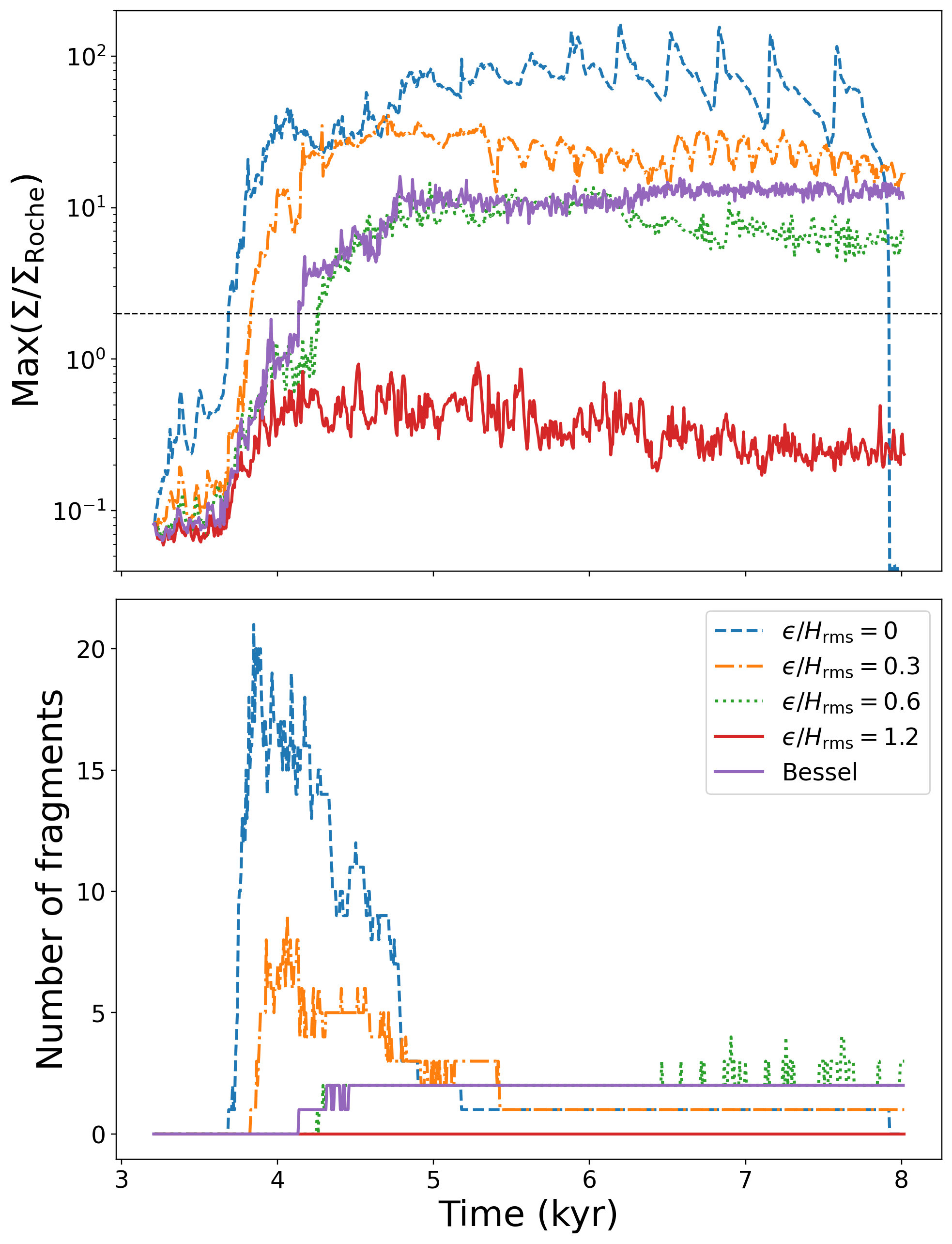}

\caption{Properties of the fragmentation regime for the different gravity prescriptions. 
The black dash line indicate the fragmentation threshold.
} 
\label{fig: properties of fragmentation for different kernels}
\end{figure}

\begin{figure}
\centering
\includegraphics[width=\hsize]{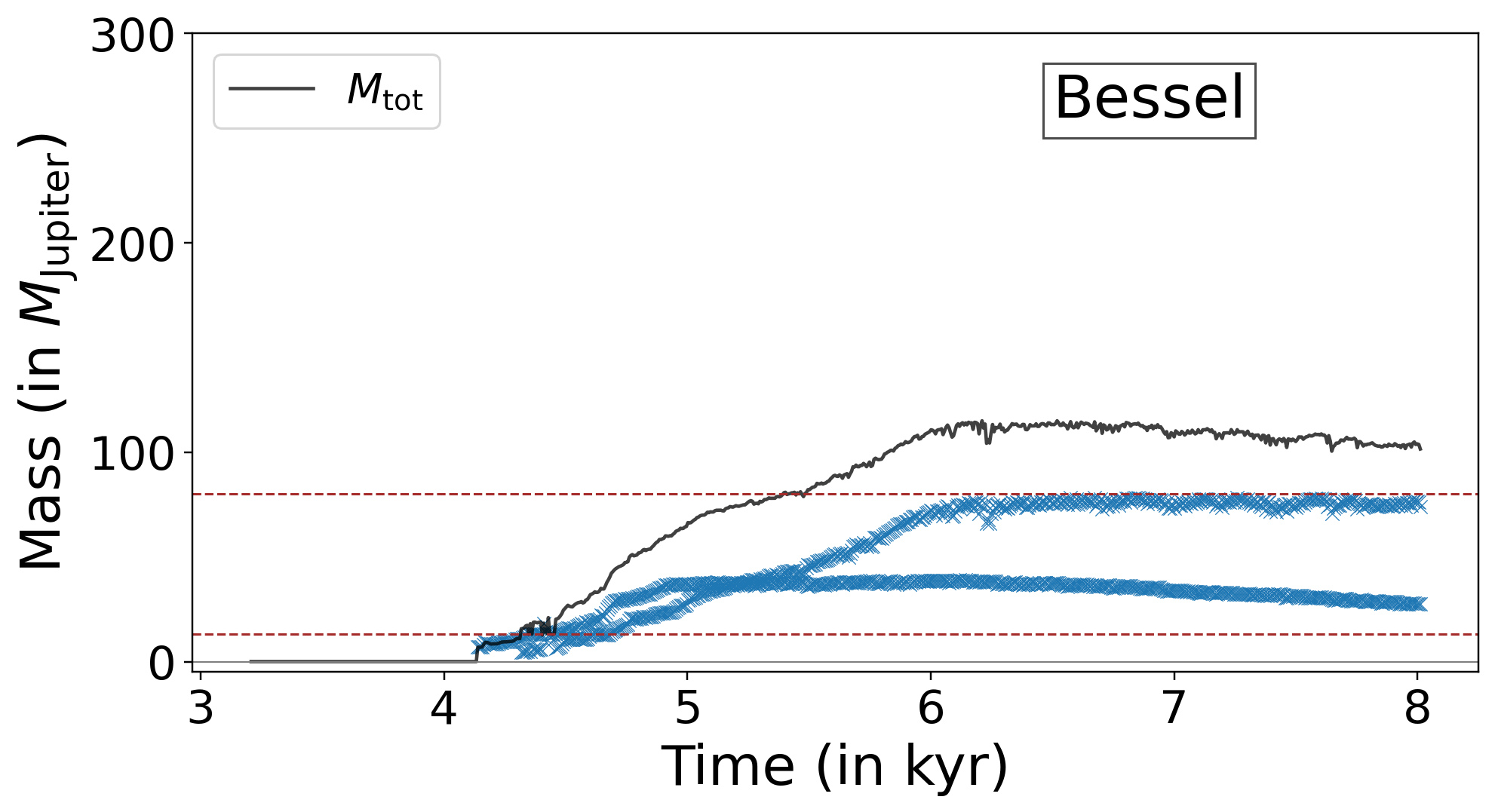}
\includegraphics[width=\hsize]{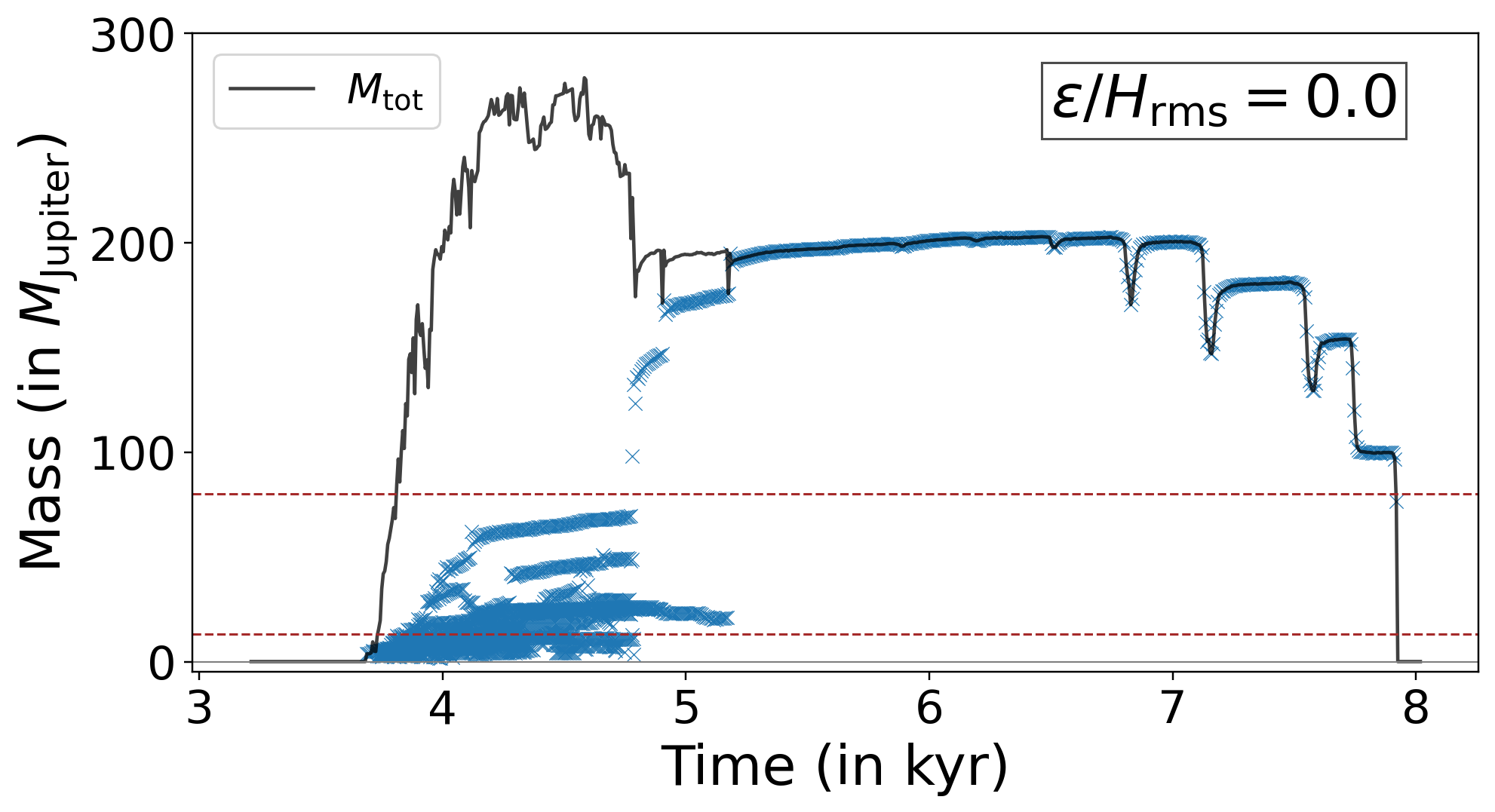}
\includegraphics[width=\hsize]{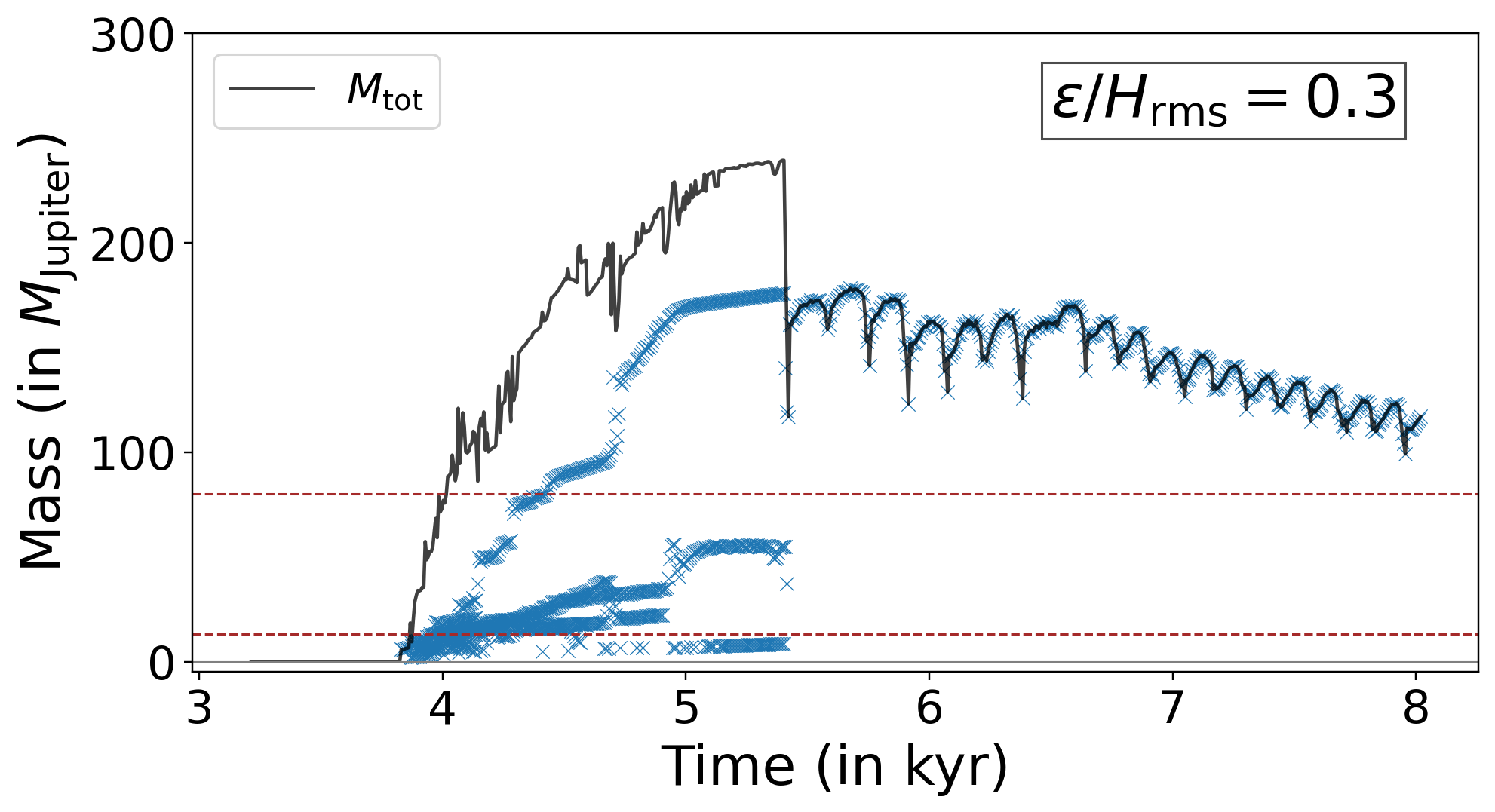}
\includegraphics[width=\hsize]{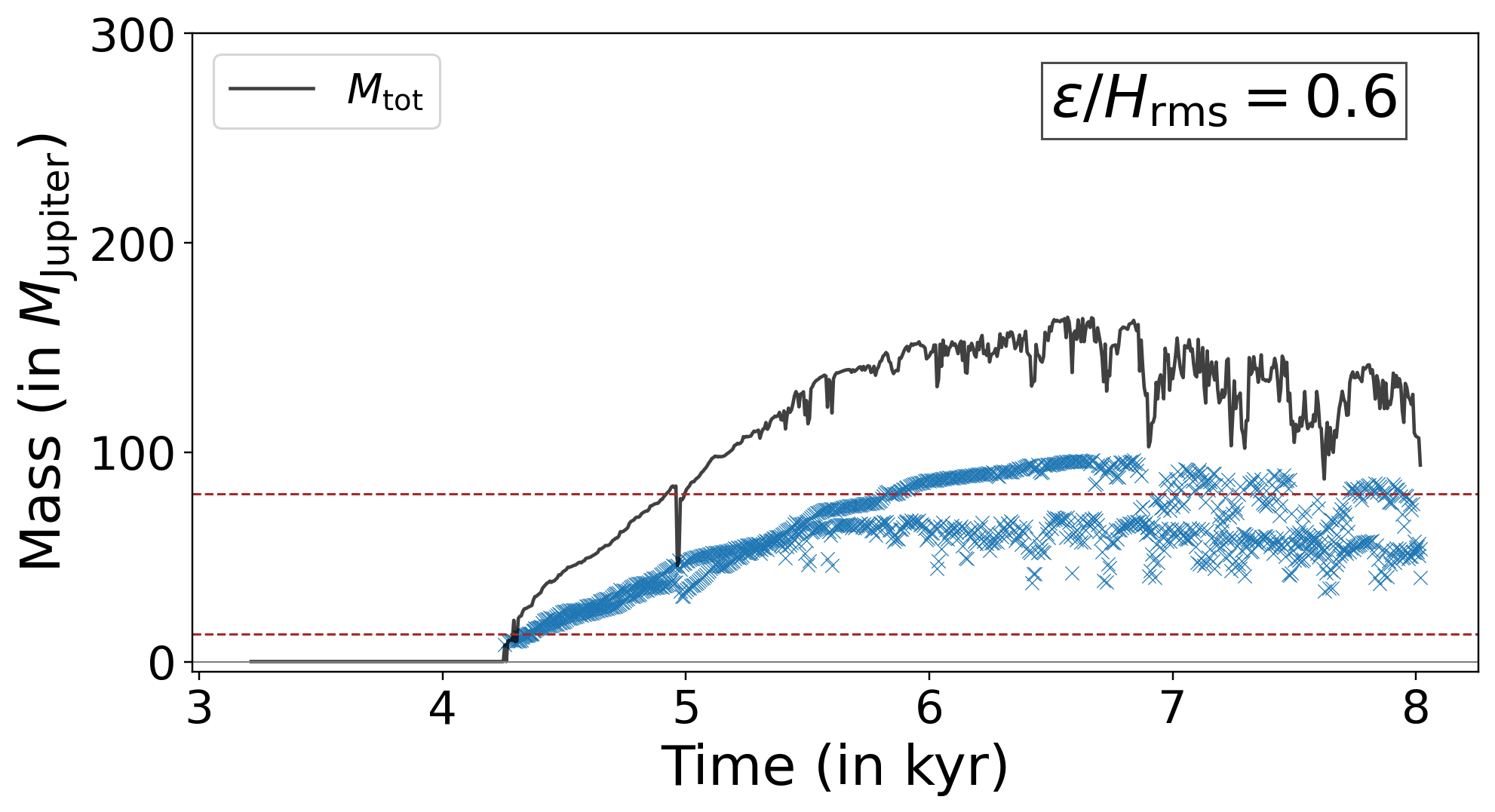}

\caption{Evolution of gravitationally bound objects under different gravity prescriptions for $\beta=2$.
The solid line represents the total mass, while cross markers denote individual fragments. 
The brown dashed lines indicate the lower and upper mass limits for brown dwarf formation.
For $\epsilon/H_{\rm rms}=0.0$, no fragments form, and for $\epsilon/H_{\rm rms}=0.6$, the total mass may be overestimated beyond $\sim 6.7$ kyr  (see Sect.~\ref{subsec:characterisation of GI different kernels - beta 2}).
} 
\label{fig: mass fragments for different gravity prescriptions}
\end{figure}

\begin{figure}
\centering
\begin{tabular}{cc}
\textbf{Bessel} & \textbf{$\epsilon/H_{\rm rms}=0.6$} \\
\includegraphics[width=0.46\hsize]{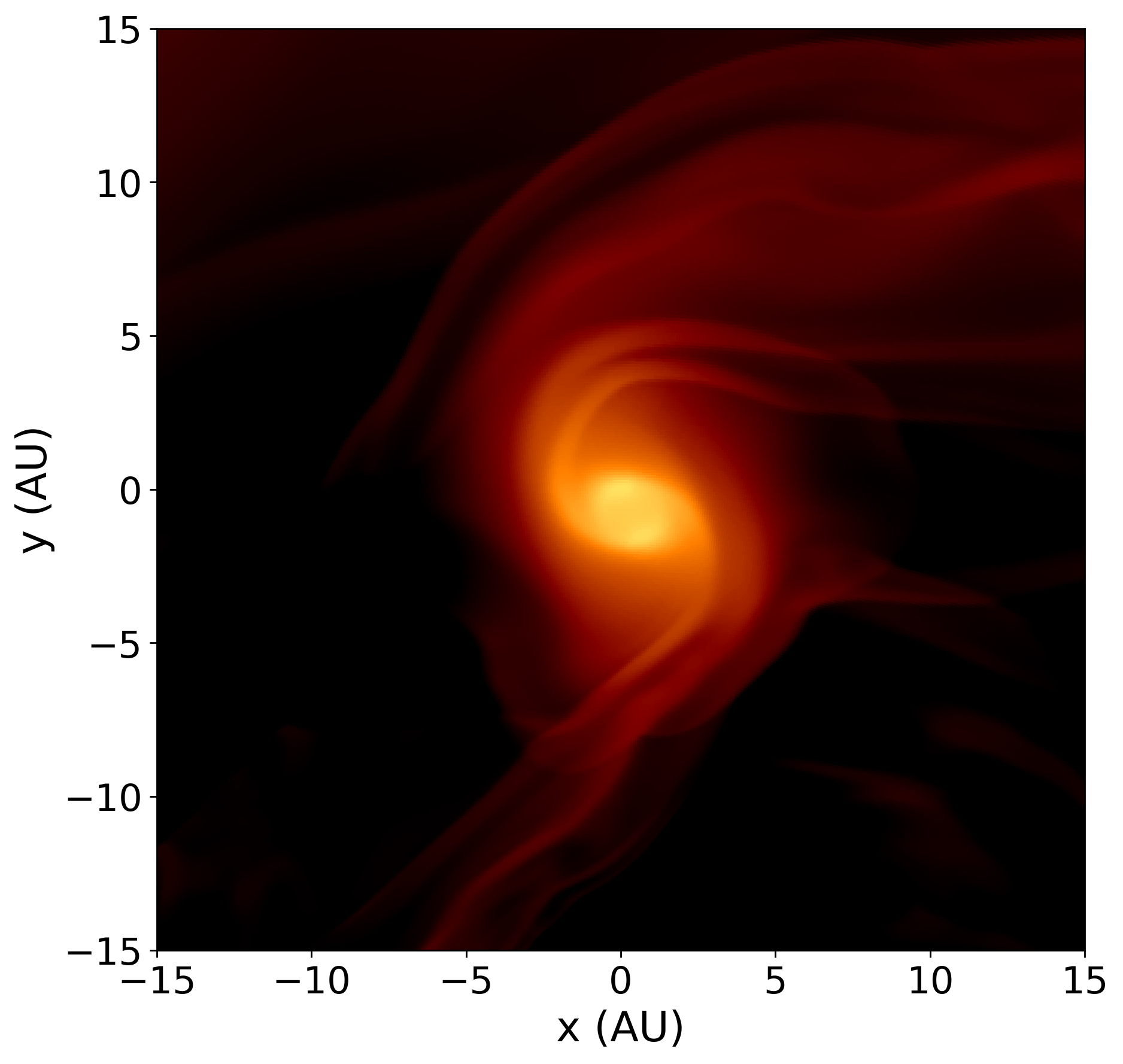} &
\includegraphics[width=0.46\hsize]{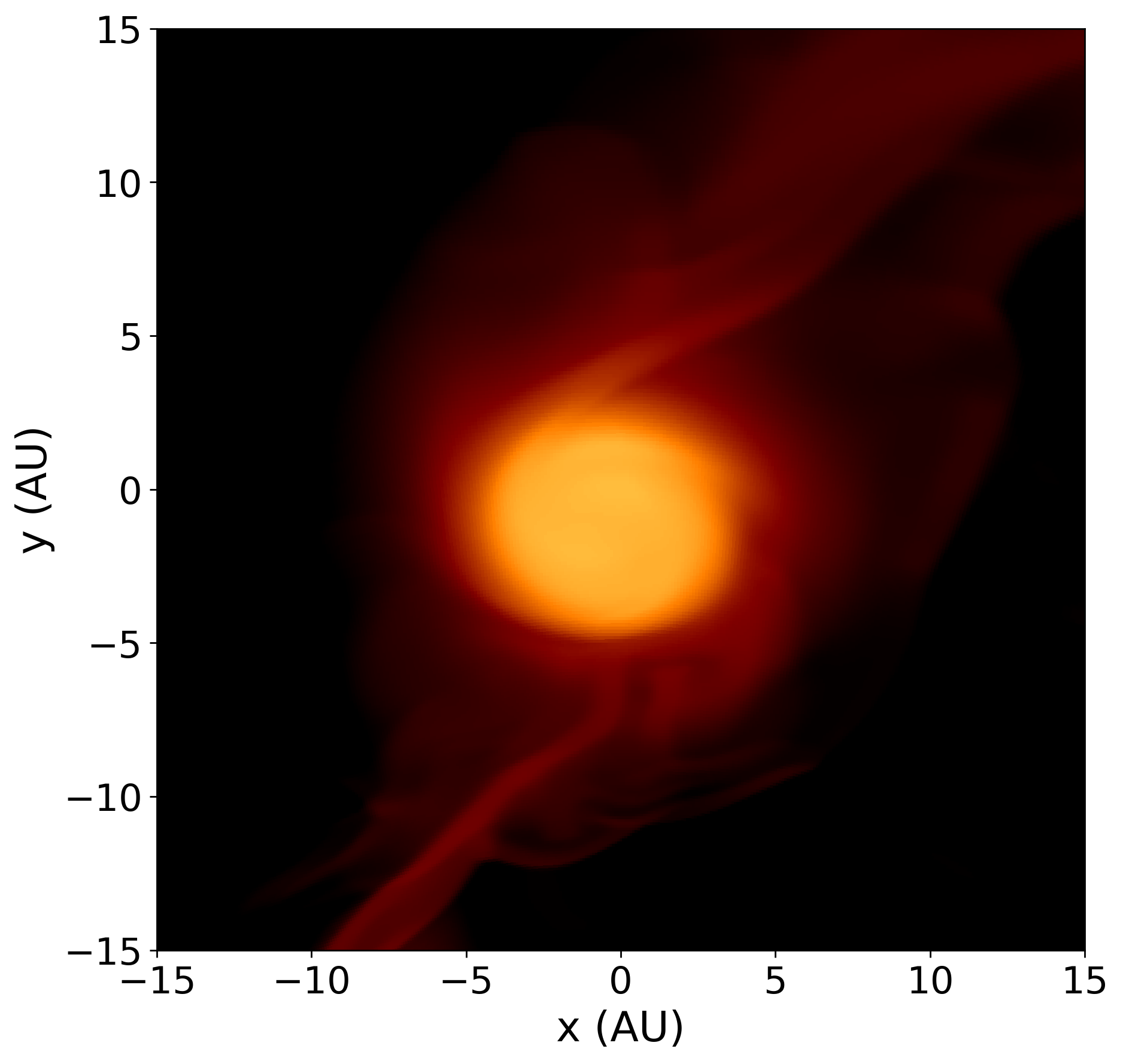} \\
\includegraphics[width=0.46\hsize]{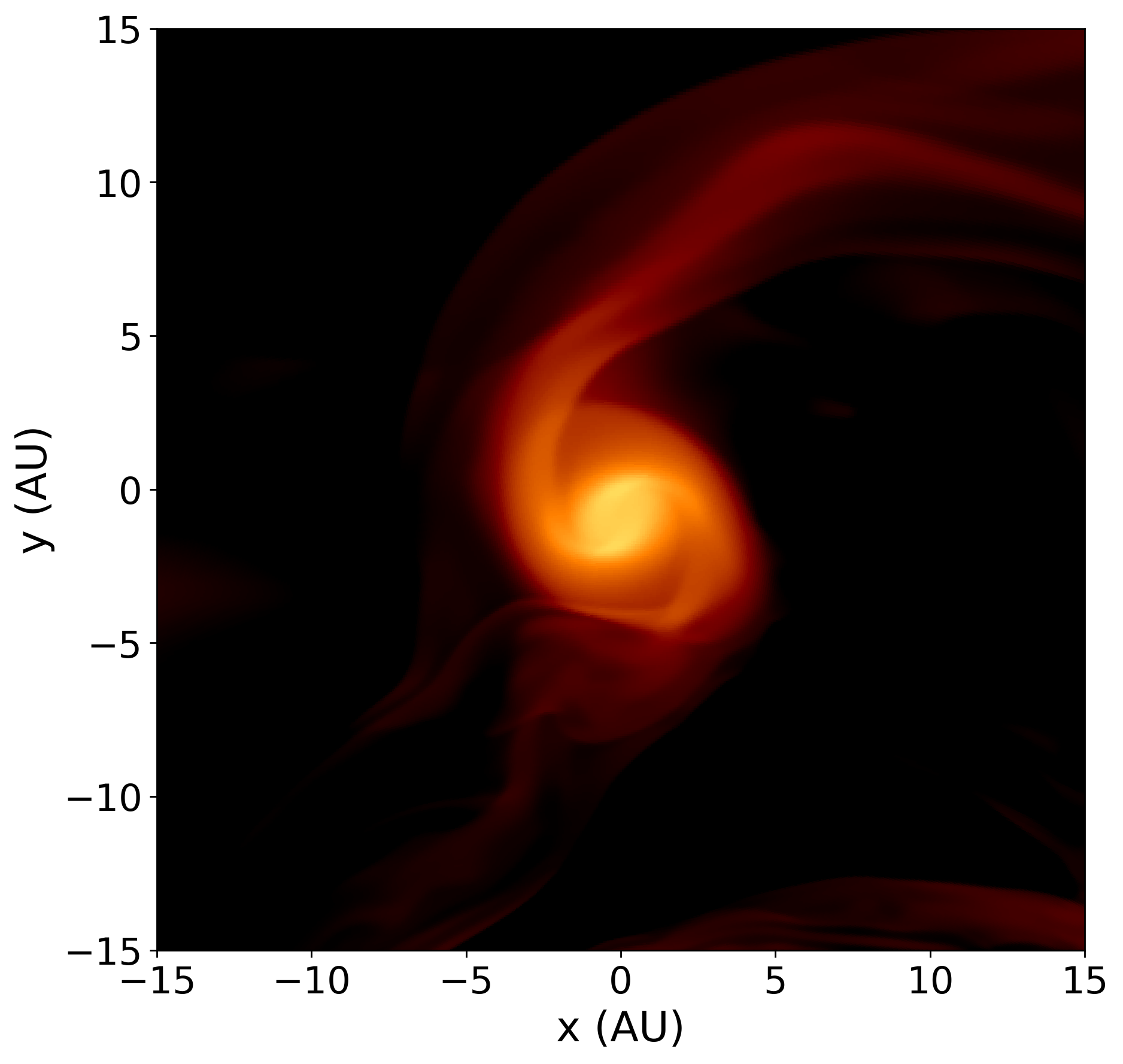} &
\includegraphics[width=0.46\hsize]{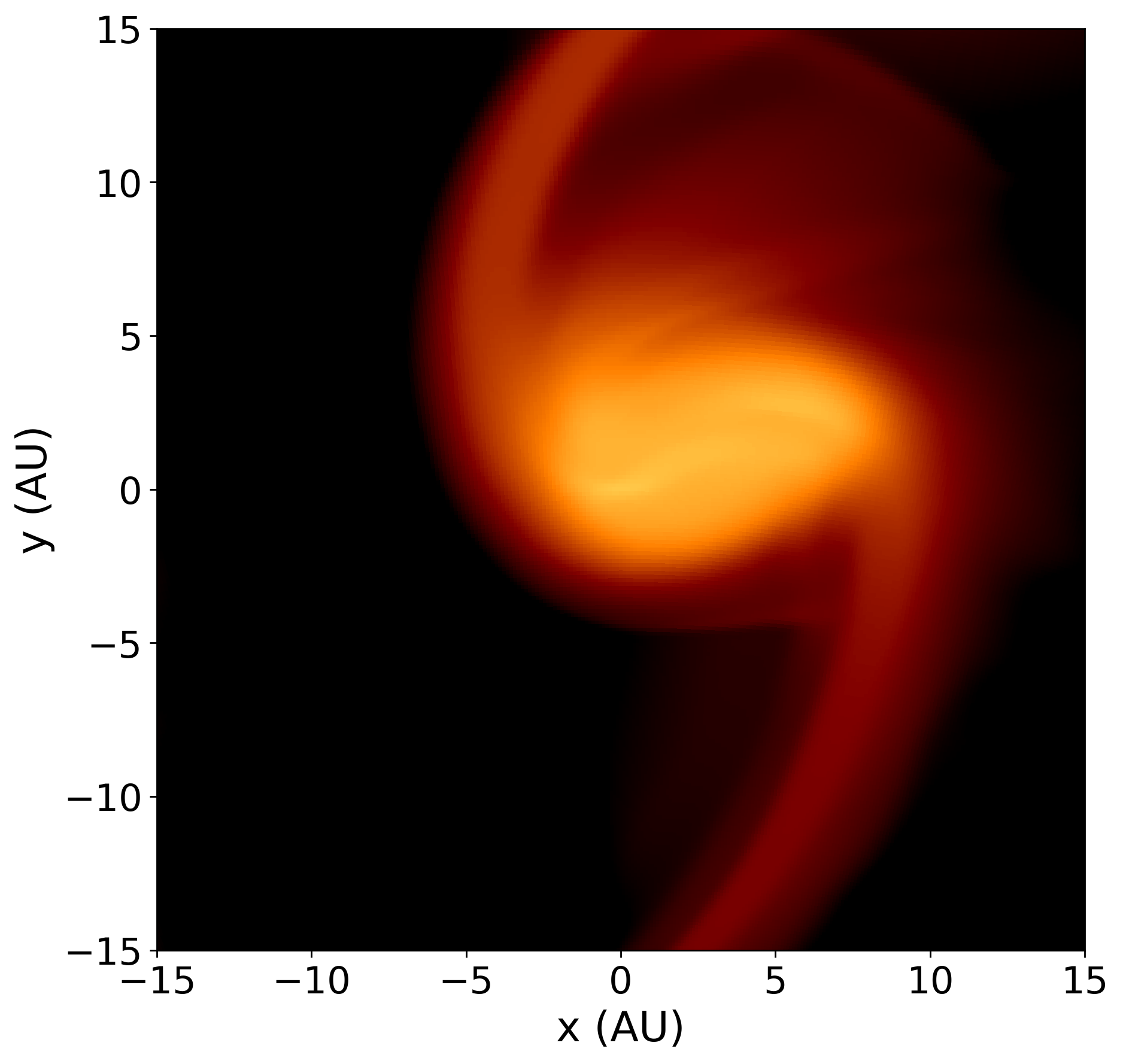} \\
\includegraphics[width=0.46\hsize]{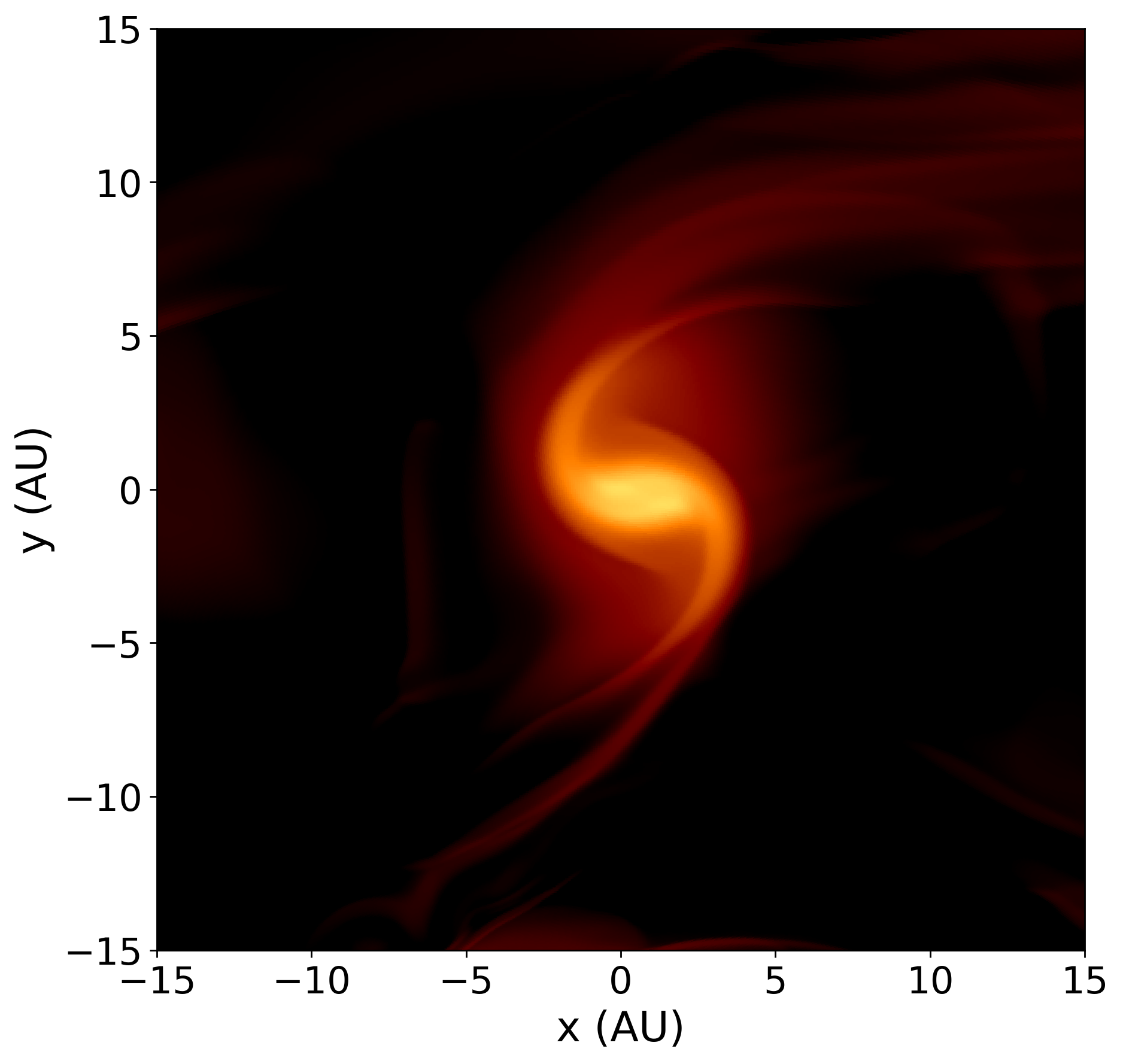} &
\includegraphics[width=0.46\hsize]{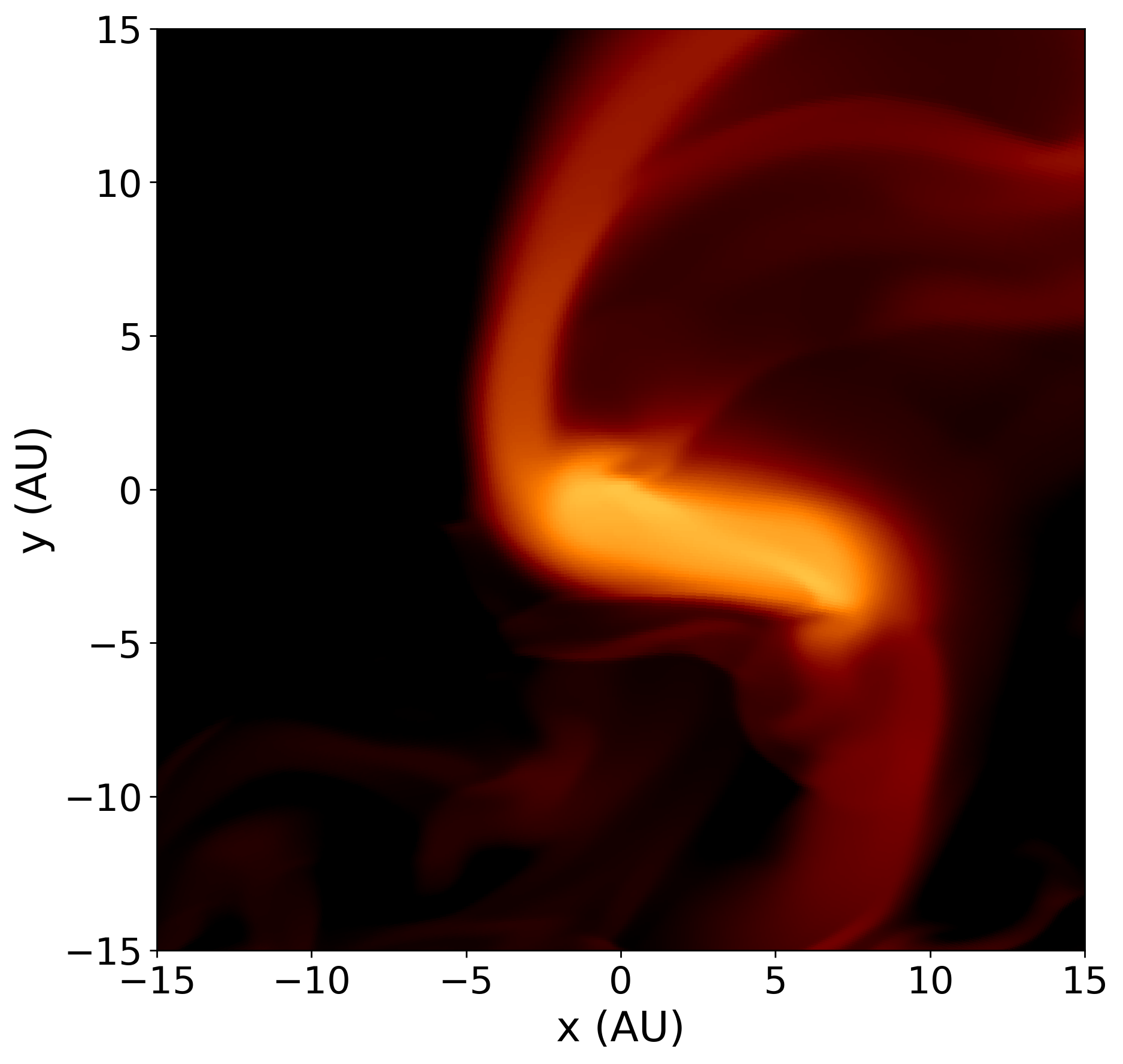} \\
\includegraphics[width=0.46\hsize]{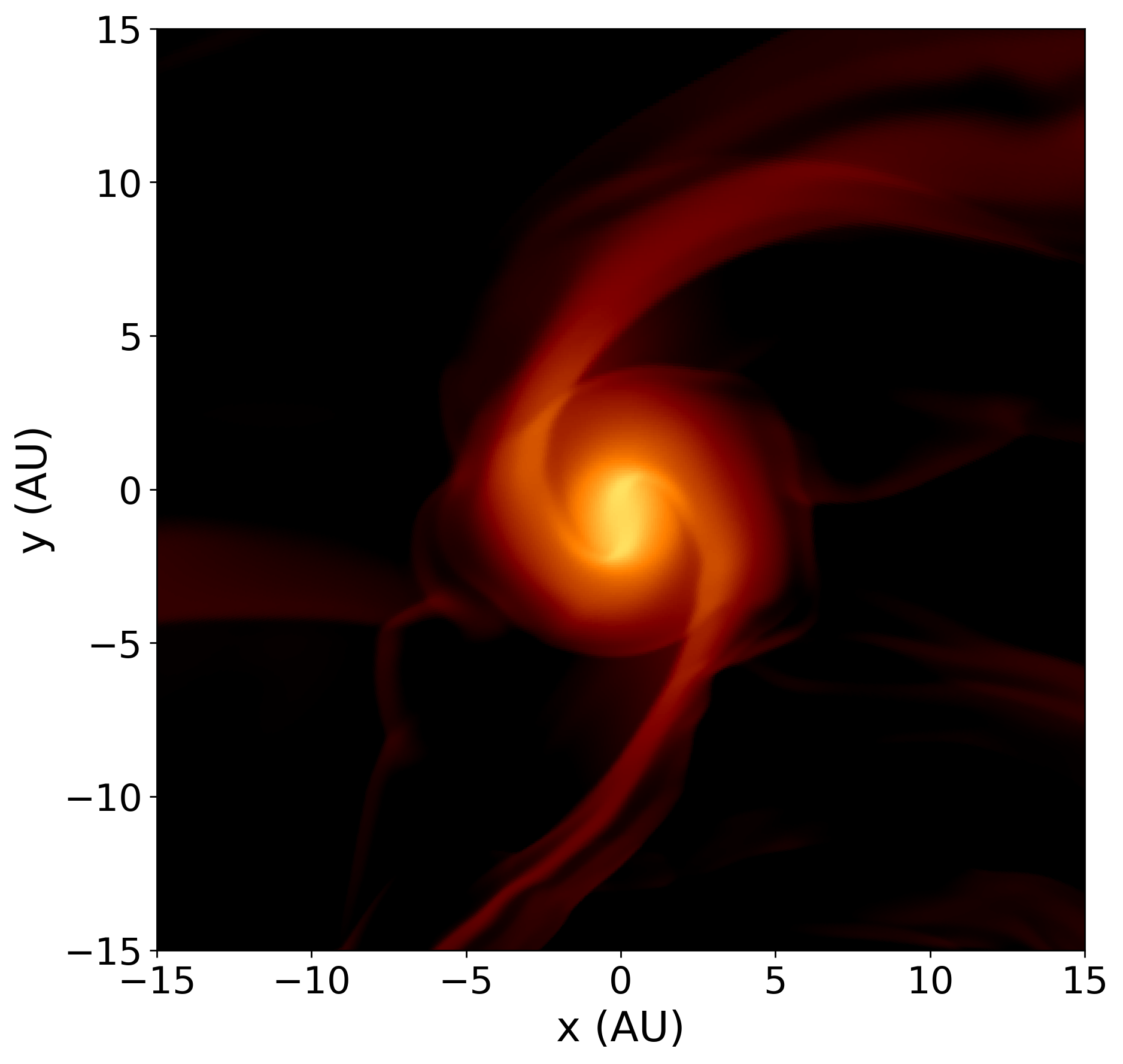} &
\includegraphics[width=0.46\hsize]{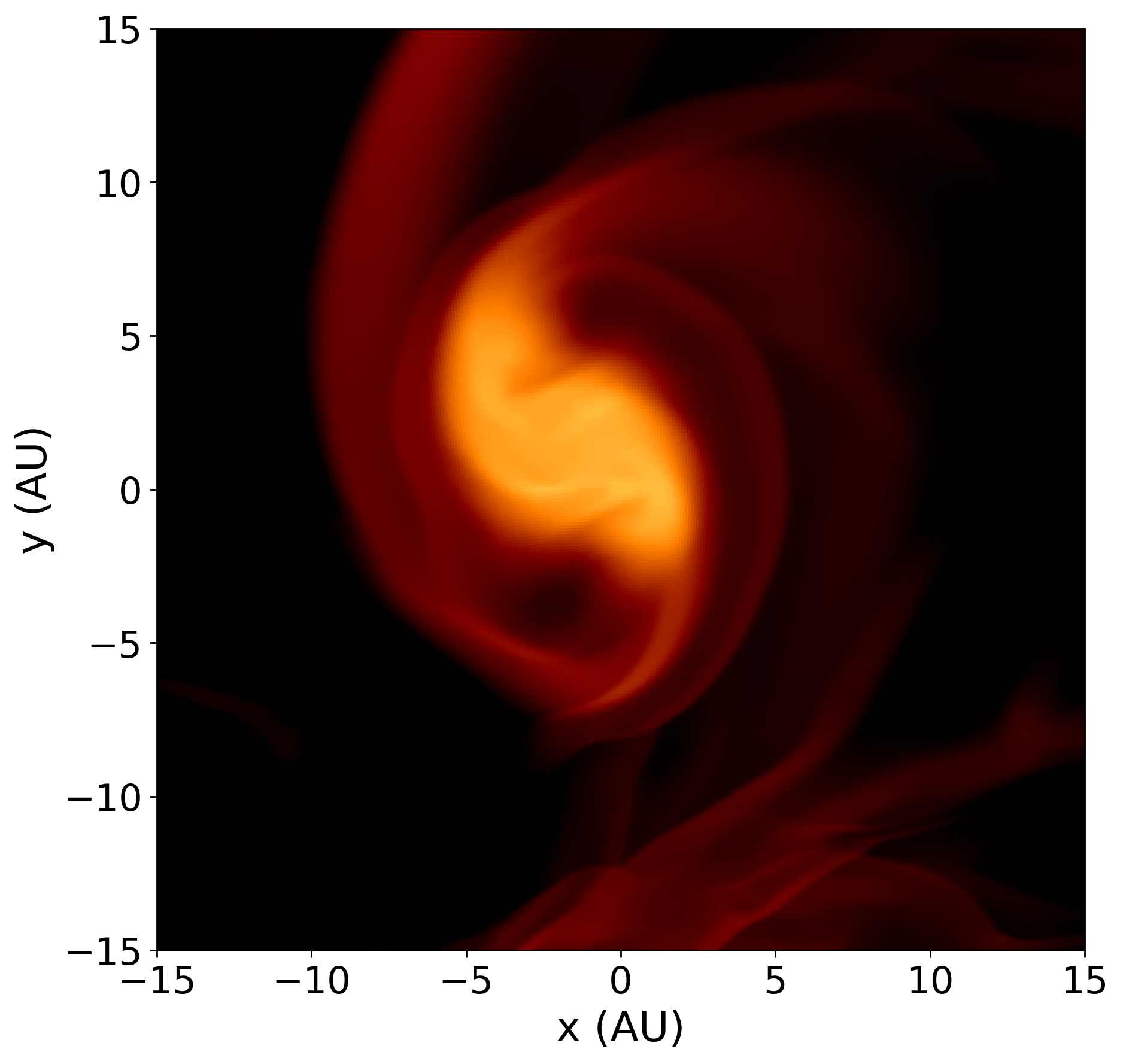} \\
\includegraphics[width=0.46\hsize]{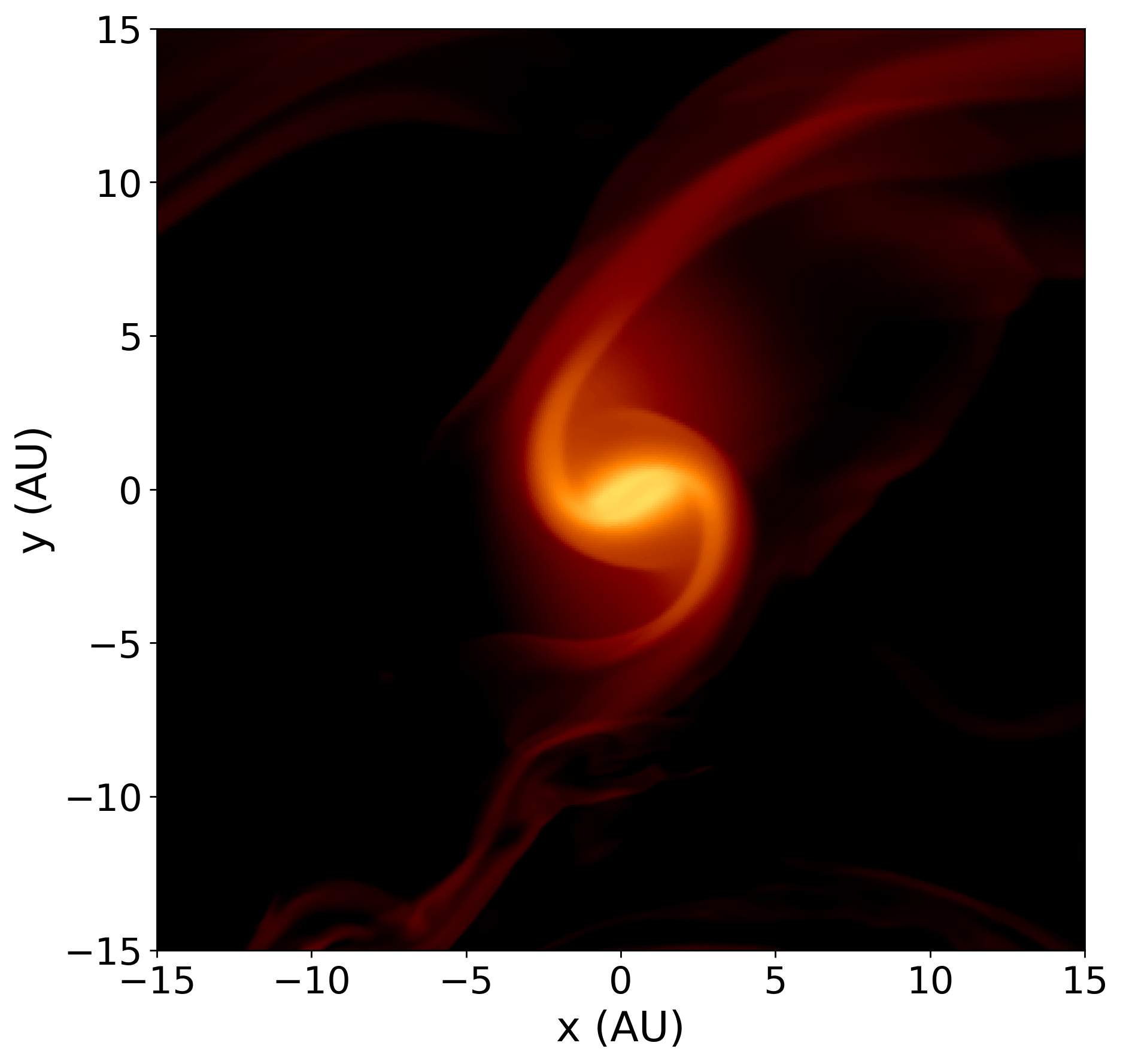} &
\includegraphics[width=0.46\hsize]{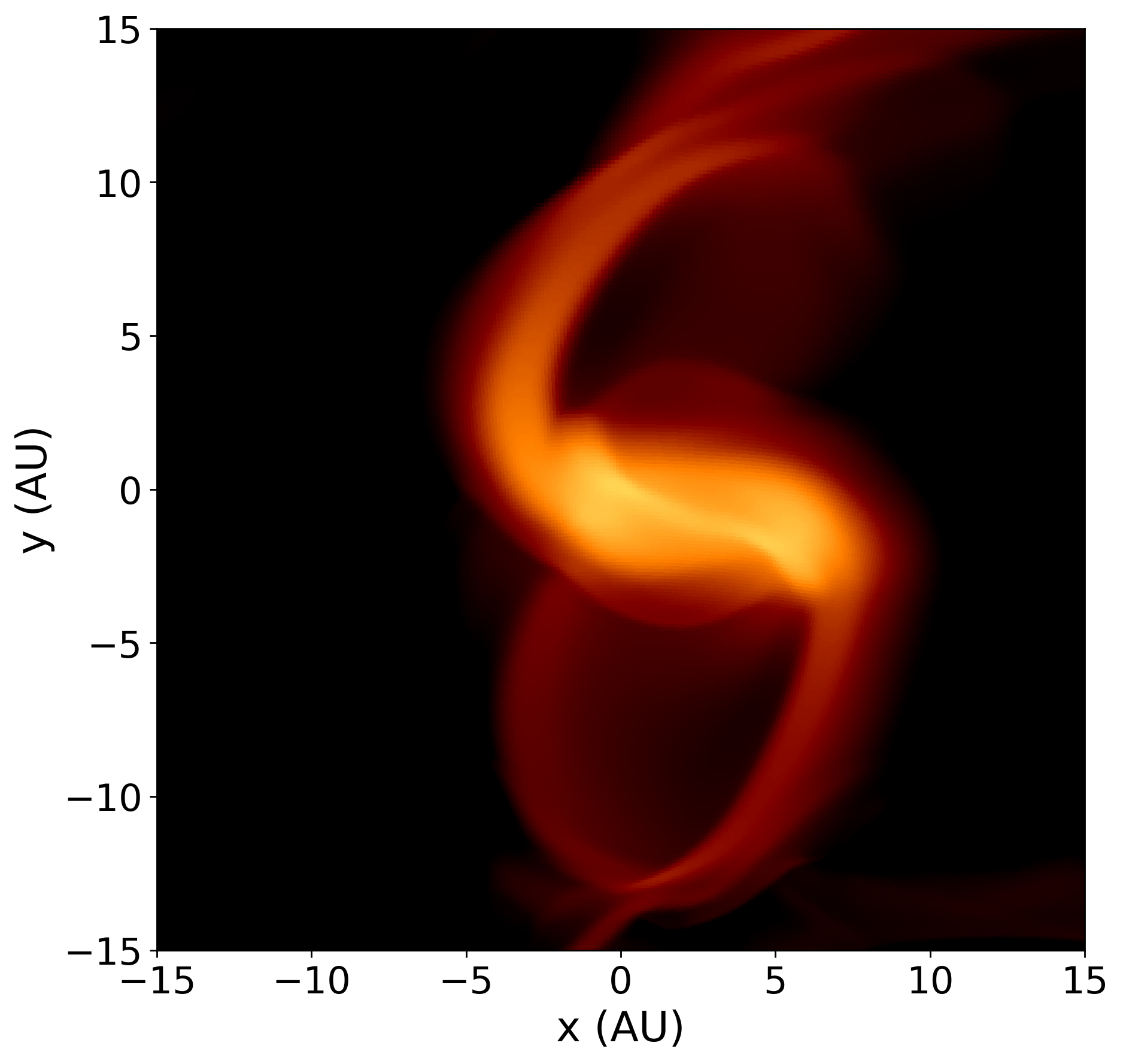} \\
\end{tabular}

\includegraphics[width=0.98 \hsize]{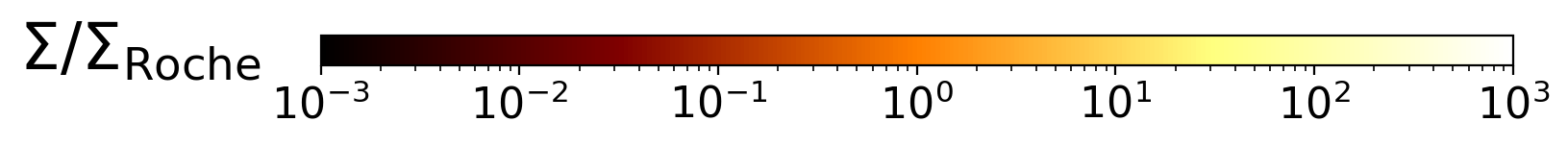}

\caption{Detailed comparison of fragment evolution using the Bessel kernel (left) and the Plummer potential paradigm with $\epsilon/H_{\rm rms} = 0.6$ (right).
The snapshots, from top to bottom, correspond to $t=6.7, 7.7, 8.05, 8.1$ and $8.6$ kyr.
All plots are centered at the peak density.
}
\label{fig: detailed comparison bessel and SL}
\end{figure}

In Fig.~\ref{fig: comparison fragmentation for different kernels}, I present three snapshots from GI simulations for $\beta=2$, comparing the outcomes for each gravity kernel.
Fragments are highlighted with green circles for clarity.
Fig.~\ref{fig: properties of fragmentation for different kernels} illustrates the properties of the fragmentation regime—specifically, the maximum density and the number of fragments—throughout the entire simulation.
Fig.~\ref{fig: mass fragments for different gravity prescriptions} shows the individual masses of the formed fragments, as well as their cumulative mass distribution.
For the latter Figure, it is important to highlight that, in the simulation with $\epsilon/H_{\rm rms}=0.6$, the total mass may be overestimated beyond $\sim 6.7$ kyr.
In this particular setup, fragments tend to undergo disruption, resulting in two closely situated density peaks.
While these peaks are accurately identified by the detection algorithm, their respective areas—used for mass calculation—may overlap.
This potentially leads to double counting, thereby inflating the estimated total mass.

I begin by describing the results obtained using the Bessel kernel, which serves as the reference case. 
Once the fragmentation criterion (Eq.~\ref{Eq: fragmentation criterion}) is satisfied (at $t \sim 4.2$ kyr), the density rapidly increases, reaching values of approximately $\sim 10 \Sigma_{\rm Roche}$.
Only two fragments emerge and evolve into brown dwarfs: one with an intermediate mass of $\sim 40 M_J$ and a second, more massive $\sim 75 M_J$.
Both fragments persist until the end of the simulation, appearing to settle into a stable orbital configuration.
Notably, the brown dwarf's mass piles up just below the lower mass limit required for a main-sequence star.

In analysing the results obtained using the smoothing length paradigm, I observe that simulations employing small smoothing lengths ($\epsilon/H_{\rm rms} \leq 0.3$) systematically overestimate both the density and number of first fragments by a factor of 3 to 10 compared to those using the Bessel prescription. 
Furthermore the fragments form at $t\sim3.7$ kyr, namely 12\% faster compared to the simulations employing the Bessel prescription.
At later stages of these simulations, several fragments are ejected from the numerical box early on, consistent with the recent findings of \citet{2026_calovic, 2026_nayakshin}.
The fragments that remain within the box subsequently undergo successive mergers, ultimately forming a single, extremely massive stellar-type object with a mass in the range of 0.15-0.2 $M_{\odot}$, which is more than 50\% of the initial mass of the disc.
By the end of the simulation, most of the gas has either been accreted by this object or wiped out.
Notably, the fragment settles into a wide eccentric orbit when $\epsilon/H_{\rm rms}=0$, whereas for $\epsilon/H_{\rm rms}=0.3$, it adopts a tight eccentric orbit.
When approaching the outer or inner boundary, they experience periodic mass loss, ultimately leading to the ejection of the fragment in the unsoftened simulation.
In stark contrast, simulations with large smoothing lengths ($\epsilon/H_{\rm rms} \geq 1.2$) fail to produce any fragments. 
In this case, the density never exceeds the Roche density, occasionally reaching peak values of up to $0.9 \Sigma_{\rm Roche}$.

The most compelling case arises when $\epsilon/H_{\rm rms}=0.6$, where the results closely resemble those obtained using the Bessel kernel.
Specifically, the initial number of fragments. the peak density and the initial time of fragmentation are nearly identical to those observed with the Bessel prescription.
The masses of the fragments in the softened simulations reach 66 $M_J$ and 96 $M_J$, surpassing the threshold required for stellar object formation. 
These values are 25\% higher than the masses obtained with the Bessel kernel, when comparing both the least massive and the most massive objects formed in each case.
Despite this overestimation, one might initially assume that no fundamental differences exist between the two approaches. 
However, a key distinction emerges after 6.5 kyr: while the maximum density in the Bessel case remains constant, it begins to decline in the $\epsilon/H_{\rm rms}=0.6$ simulation. 
Furthermore, a closer examination of the fragment masses reveals a notable contrast: in the Bessel kernel simulations, the mass evolution of the objects is smooth and devoid of oscillations, unlike the behavior observed in the $\epsilon/H_{\rm rms}=0.6$ case.
This phenomenon, expected for $\epsilon/H_{\rm rms} \leq 0.6$ (see Fig. 1 and Sect. 3.2 of \citet{2025c_rendon}), arises due to the shielding of gravity at distances smaller than one scale height, which disrupts the fragments.
Conversely, gravity is overestimated at distances between one and two scale heights.
This discrepancy triggers a cyclic process of disruption and collapse: the core mass of the clump, unable to be sustained by gravity, is released.
However, as it moves away from the clump's center, it is reaccreted due to the overestimation of gravity in the surrounding region. 
Importantly, during the release phase, a fraction of the mass is permanently lost due to the tidal forces exerted by the central object.
This mechanism is more clearly illustrated in Fig. \ref{fig: detailed comparison bessel and SL}, which presents a series of snapshots comparing the outer fragment under the Bessel formalism and the $\epsilon/H_{\rm rms}=0.6$ prescription.
Notably, for the softened simulation at $t=8.05$ kyr and $t=8.6$ kyr, the fragment exhibits pronounced shearing.
An additional morphological feature of interest is observed in all snapshots: the fragment appears more extended, likely as a consequence of the shielded gravity.
In contrast, the fragment in the simulation utilizing the Bessel kernel maintains a consistent morphology throughout the entire duration of the run.
The cyclic behaviour described above does not occur with the Bessel kernel, as it inherently preserves the nature of Newtonian gravity at small scales.
Once an object forms, it retains its mass because gravity is not shielded.
Moreover, at small distances, gravity scales as $\propto r^{-1}$—unlike the $\epsilon/H_{\rm rms} =0$ case, where it scales as $\propto r^{-2}$, resulting in an overestimation of self-gravity.
This overestimation leads to more violent runaway collapse and increased fragment merging.
A paradox arises when attempting to mitigate the disruption-collapse cycle in the softened gravity prescription: while reducing the smoothing length could prevent disruption, this adjustment would simultaneously amplify the already overestimated mass of formed objects and exacerbate artificial fragmentation. 
Consequently, this approach risks producing excessively massive stellar-like objects and potentially unrealistic outcomes.

\subsection{Gravito-turbulence: $\beta=8$}

\begin{figure}
\centering
\includegraphics[width=\hsize]{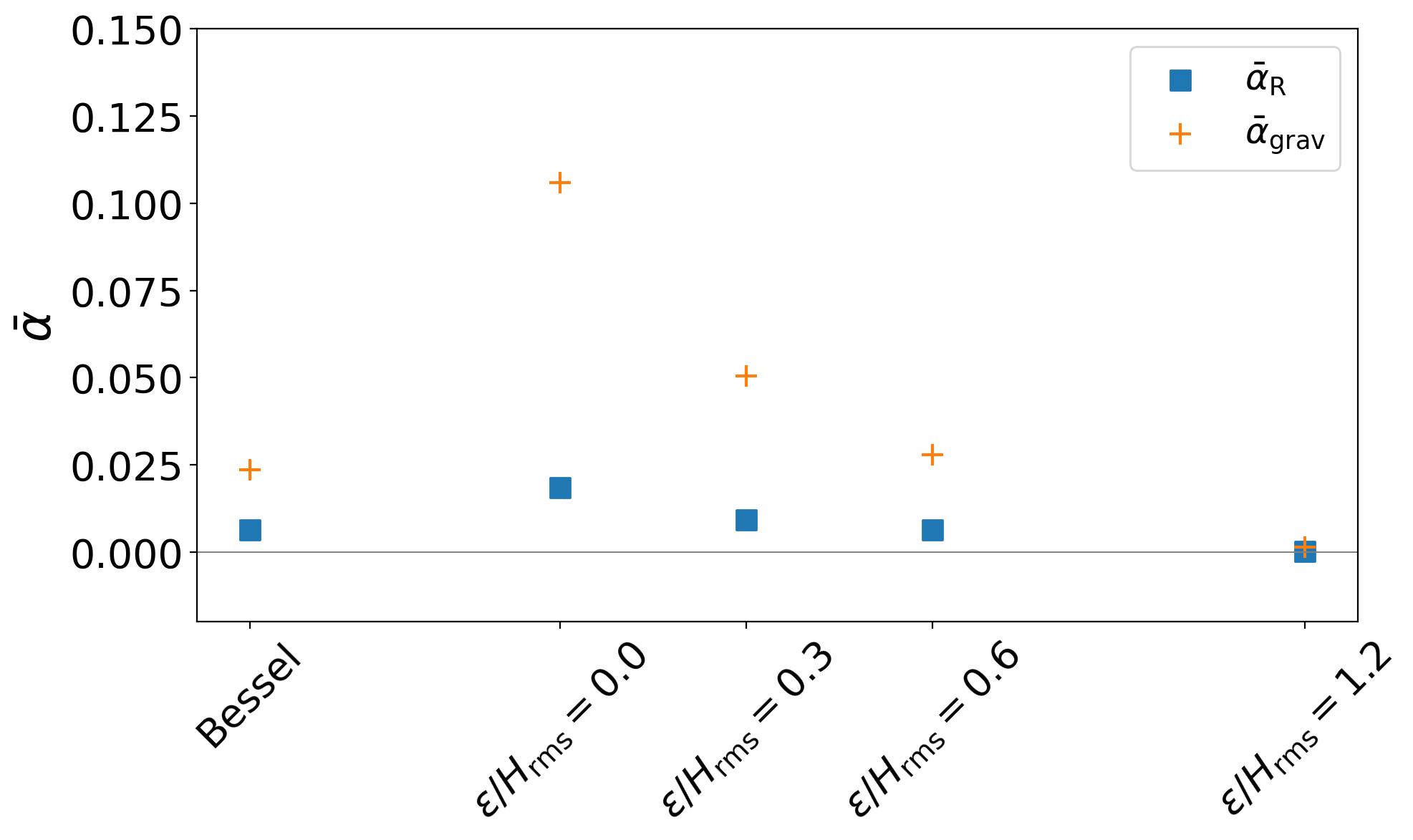}

\caption{Comparison of the gravito-turbulent regime ($\beta=8$) for the different gravity prescriptions.
} 
\label{fig: stresses different gravity prescriptions}
\end{figure}

\begin{figure}
\centering
\includegraphics[width=\hsize]{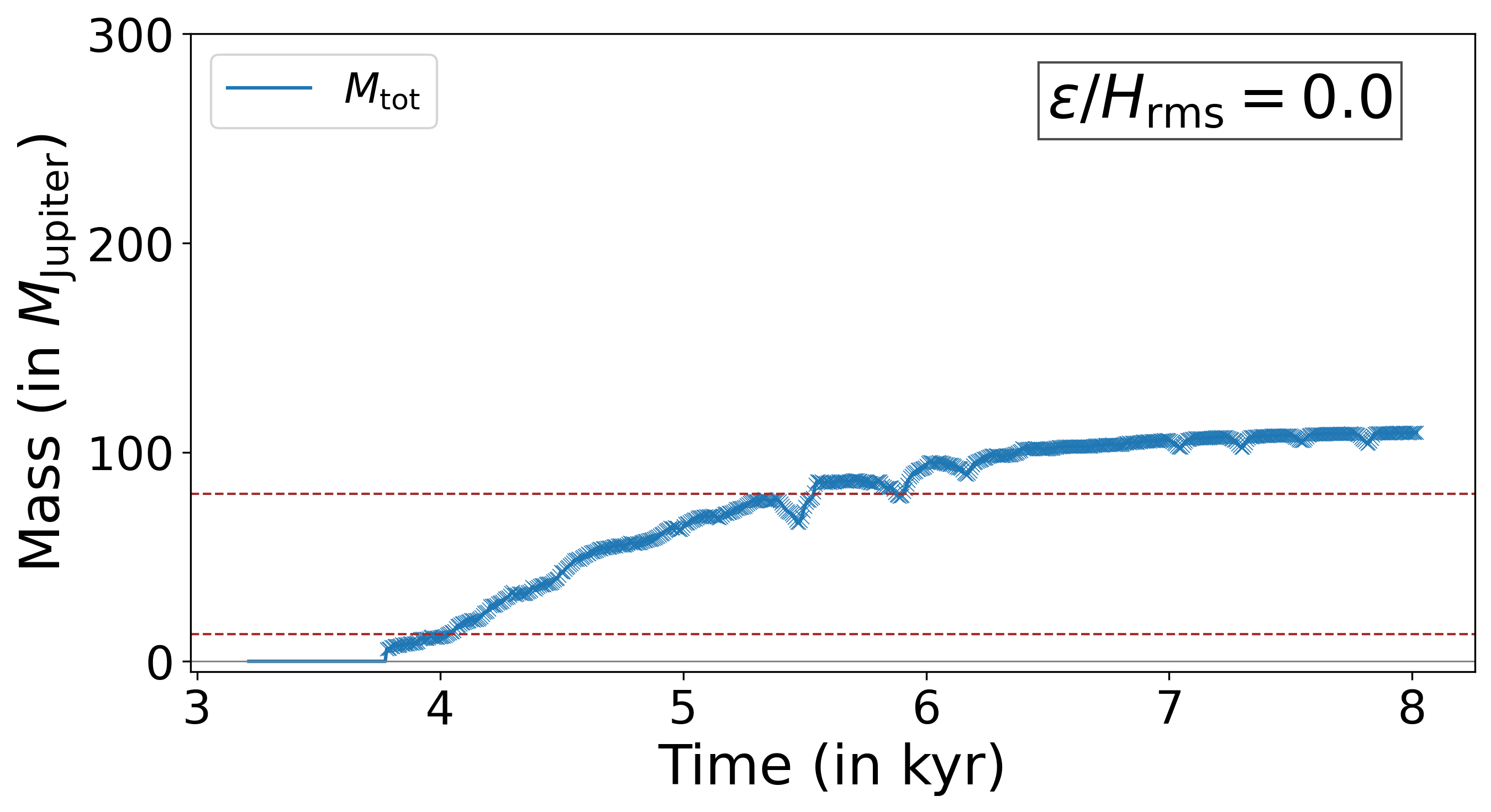}

\caption{Evolution of the gravitationally bound objects for $\epsilon/H_{rm rms}=0$ and $\beta=8$.
} 
\label{fig: mass for beta=8}
\end{figure}

In Fig. \ref{fig: comparison gravito-turbulence different kernels} I present three snapshots of GI simulations for $\beta=8$ comparing the outcomes of each gravity kernel.
Fig.\ref{fig: stresses different gravity prescriptions} illustrates the time- and space-averaged Reynolds and gravitational stresses for all the gravity prescriptions under consideration. 
The mass of the formed fragment in the unsoftened simulation is shown in Fig.~\ref{fig: mass for beta=8}.
Morphologically, all kernels yield a similar turbulent state and spiral structures, with one notable exception: the simulation employing unsoftened gravity. 
In this case, the system fragments into a single object, which continuously accretes mass throughout the simulation, ultimately reaching a final mass of $\sim 110 M_J$.
This mass exceeds the lower limit required for the formation of a main-sequence star.
For $\beta=4$, fragments also form when $\epsilon/H_{\rm rms} = 0.3$.
This suggests that, for small smoothing lengths ($\epsilon/H_{\rm rms} \leq 0.3$), the critical cooling parameter shifts toward higher values.

When comparing the $\alpha$ stresses, a clear trend emerges: the smaller the smoothing length, the higher the stresses.
Specifically, for $\epsilon/H_{\rm rms} \leq 0.3$, the gravitational stress is two to four times greater than that observed with the Bessel kernel.
For $\epsilon/H_{\rm rms} = 0.6$, the stress closely resemble those obtained with the Bessel prescription.
Conversely, for the highest softening, both the Reynolds and gravitational stresses are highly reduced.
This last observation aligns with expectations, as the most unstable wavelength of GI are on the order of $H$.
These wavelengths are inherently unresolved when the smoothing length exceeds $H$.

\subsection{Summary}

In this section, a comprehensive comparison was conducted between the outcomes of GI using the recently rediscovered Bessel kernel and the Plummer potential with various smoothing lengths used in the literature. 
The key findings are as follows:
For large smoothing lengths ($\epsilon/H_{\rm rms} \geq 1.2$), gravito-turbulence is weakened, and fragmentation is inherently suppressed. 
Conversely, for small smoothing lengths ($\epsilon/H_{\rm rms} \leq 0.3$), gravity is significantly overestimated, resulting in elevated Reynolds and gravitational stresses in the gravito-turbulent regime. 
More concerning is the observation that, in the fragmentation regime, the final masses of the formed objects are overestimated by a factor of 2 to 3. This leads to the formation of stellar objects with masses of $0.15-0.2 M_{\odot}$.
Two distinct pathways contribute to the formation of these overestimated massive objects when $\epsilon/H_{\rm rms} \leq 0.3$: under low cooling conditions, a single object forms and efficiently accretes gas material during its orbit due to the overestimated gravity.
Under efficient cooling, multiple Jupiter-mass fragments form and rapidly merge, again driven by the overestimated gravity that enhances encounters, ultimately resulting in a single massive fragment.

For $\epsilon/H_{\rm rms} = 0.6$, the outcomes in both the gravito-turbulent and fragmentation regimes closely resemble those obtained with the Bessel prescription.
However, it is important to note that the masses of the objects formed with this smoothing length are approximately 25\% larger than those formed using the Bessel kernel.
A closer examination of the fragments reveals a cyclic disruption/collapse mechanism.
This phenomenon arises because gravity is shielded within the core of the clump, causing material to be released.
As the material moves away from the clump, it is reaccreted due to the overestimated gravity in its vicinity.
During the release phase, a fraction of the mass is permanently lost due to tidal forces.
It is also worth noting that the initial formation of fragments in this simulation occurs later compared to setups with $\epsilon/H_{\rm rms} \leq 0.3$, which likely explains the preference for smaller smoothing lengths in the literature. 
In this context, I speculate that the destruction of the clump during its inward migration to a tight orbit, as discussed in Section 4 of \citet{2026_zhang}, may be attributable to the shielding of gravity mentioned in this paragraph.
In conclusion, while this option may appear suitable at first glance, it raises concerns regarding the physical consistency of self-gravity.

\section{Discussion and perspectives}\label{sec: discussion and perspectives}

\subsection{Consistency of fragments evolution when using 2D setups}

Regardless of the gravity prescription employed in 2D simulations, once fragments form, the question arises of how to model the newly formed object using sink particles \citep{2010_federrath}.
In the literature, it is common practice to employ sink particles that also utilize a smoothing length for planet-disc interactions.
The choice of the used softening length is typically tailored to the specific focus of the study. 
For example, the recent work of \citet{2026_nayakshin, 2026_zhang} adopt a value of $0.5 H$ for this purpose.

While my simulations do not incorporate sink particles following fragmentation, they nevertheless provide meaningful insights into the implications of using softening in planet-disc interactions. 
Specifically, the shortcomings identified in Sect.~\ref{subsec:characterisation of GI different kernels - beta 2} for a softened gravity are likely to manifest in some form and potentially influence both the accretion processes and the torques experienced by the planets, thereby impacting their migration history.
Given these considerations, I advocate for the use of the Bessel potential in modeling planet-disc interactions \citep{muller_kley_2012,2024_brown_joshua,2025_cordwell}. 
Notably, the Bessel potential for planet-disc interactions represents a specific case of the Bessel kernel utilized in this study, as demonstrated in Sect.~4 of \citep{2025c_rendon}.

\subsection{Initial mass of giant gas planets}

Our simulations indicate that the choice of gravity prescription is not critical for triggering GI, as the most unstable wavelengths are on the order of $H$ \citep{2016_Kratter_Lodato}—a result that also holds when using the Bessel prescription (work in preparation).
However, the prescription of self-gravity becomes pivotal during the non-linear evolution of clumps, particularly in determining their mass and, more importantly, their ability to merge with other fragments.
This study suggests that the masses of objects formed by GI may be overestimated by a factor of 2–3, given that the typical softening lengths used in 2D fragmentation studies satisfy $\epsilon/H_{\rm rms} \leq 0.3$. 
In this context, I anticipate that combining the Bessel kernel with a realistic cooling prescription will further support the conclusion that GI can produce giant gas planets in the mass range of 0.3-10 $M_J$, as recently demonstrated by \citet{2025_ni}.

A key advantage of 2D simulations over their 3D counterparts—despite the latter being computationally accessible today—lies in their ability to achieve higher in-plane resolutions, facilitate long-term evolutionary studies, and explore a broader parameter space. 
The substantial numerical cost and complex data analysis required for 3D often limit such studies to only a few simulations.

\subsection{Numerical aspects}

The numerical computation of the gravitational potential using the Bessel prescription follows the same approach as that employing a smoothing length, as both methods rely on Fast Fourier Transforms (FFTs), which are renowned for their accuracy and computational efficiency, scaling as $\mathcal{O} (N \log(N))$.
The primary distinction lies in the use of special functions for evaluating the Bessel kernel, which increases its computational time by a factor of 20 compared to the smoothing length approach.
However, this additional cost is mitigated by precomputing the kernel once and reusing it at every time step.
Given that this prescription introduces no performance penalty during runtime and enhances the physical consistency of the gravitational calculation, I recommend its adoption.

A major physical limitation inherent in my computational method—and also present in the smoothing length paradigm—is the constraint $H_{sg}/r=const.$.
This condition "freezes" the temperature to its initial profile for the calculation of self-gravity.
It is important to note, however, that this constraint is purely numerical; the framework proposed by \citet{2025c_rendon} is more general and accommodates arbitrary temporal and spatial variations in the scale height.
To fully leverage the Bessel kernel while maintaining computational efficiency comparable to FFTs, a more general approach is required.
One promising avenue is the use of Hankel transforms, which represent a weighted sum of a radial function with Bessel functions of the first kind \citep{2011_baddour,2014_debnath}.
Conceptually, Hankel transforms can be viewed as an extension of Fourier transforms to radial polar coordinates. 
They naturally emerge in polar coordinate systems, particularly in the context of solving the Poisson equation in cylindrical coordinates, and may offer a viable alternative.
Recent theoretical and numerical studies have demonstrated that Hankel transforms can be computed efficiently on non-uniform grids \citep{2024_beckman}, although a parallel algorithm has yet to be developed.
An intriguing numerical perspective for future work would therefore be to assess the feasibility of adapting our gravitational framework to Fast Hankel Transforms and to develop an efficient parallel algorithm, which remains an open challenge.

\subsection{Future research perspectives}

Our simulations demonstrate that, when employing the appropriate prescription for two-dimensional gravity, GI simulations in 2D grid-based codes converge at a critical cooling parameter of $\beta_c \sim 3$.
For cooling times above this threshold, the disc exists in a state of gravito-turbulence, whereas fragmentation occurs for values below it.
I expect that this outcome will hold with 2D SPH simulations provided the gravity prescription is appropriately adjusted.
Our results also align with established findings indicating that, for cases involving zero smoothing lengths, the threshold cooling value $\beta_c$ shifts toward higher values as gravitational forces increase with resolution \citep{2015_young_clarke}.
However, above simple picture is complicated by the work of \citet{2012_paardekooper}, who—through high-resolution, long-term 2D shearing box simulations—demonstrated that fragmentation is inherently a stochastic process: if simulations are run for long enough, the disc will fragment no matter the cooling.
Notably, his simulations exhibited fragmentation for cooling values as high as 20, well beyond the expected critical threshold.

In light of my findings—and given that solving the 2D Poisson equation is equivalent to adopting a zero smoothing length—I hypothesize that stochastic fragmentation may not arise when using a more physically accurate gravitational prescription, such as the Bessel formulation.
This assertion, however, is already challenged by the softened simulations conducted by \citet{2012_paardekooper}, which still exhibited stochastic fragmentation. 
A plausible explanation for this discrepancy lies in its use of the \textsc{SuperBee} slope limiter, which is known to induce oversteepening and thereby artificially enhance fragmentation \citep{2017_klee}. 
Indeed, \citet{2017_klee} demonstrated that employing more dissipative limiters suppresses stochastic fragmentation.

Based on these observations, I speculate that stochastic fragmentation may result from a combination of numerical oversteepening effects and an artificial overestimation of self-gravity due to the gravitational prescription.
To further explore this hypothesis, I intend to investigate the stochastic nature of fragmentation in a future study.
As a preliminary step, I will introduce a Bessel-based prescription tailored for shearing box simulations.
Specifically, this prescription will be formulated as an analytical multiplication in Fourier space, enabling both efficient numerical computation via FFTs and facilitating an analytical treatment of the linear perturbation analysis in self-gravitating systems.

\section{Conclusion}\label{sec: conclusion}

In this study, I apply the 2D Bessel formalism of gravity in a realistic simulation of thin discs, specifically within the framework of GI.
Our findings demonstrate that the straightforward application of this first-principles-derived approach naturally resolves the convergence issues identified in 2D SPH and 2D grid-based codes.
Notably, I observe a resolution-independent distinction between gravito-turbulence and fragmentation at a threshold of 44 cells/H. 
Furthermore, my analysis reveals that, for the investigation of pure gravito-turbulence at $\beta \geq 8$, a resolution of 22 cells/H is sufficient. 
Additionally, I find that fragments in low-resolution simulations are disrupted, whereas they remain intact at higher resolutions, suggesting a potential influence of grid diffusion effects.
We anticipate that all these results will extend to two-dimensional SPH simulations of GI, provided that the Bessel prescription is used.

We further conducted an in-depth comparison of GI outcomes using the widely adopted Plummer potential prescription with varying smoothing lengths.
Our results indicate the following:
\begin{itemize}
\item The Bessel formalism leads to the formation of two brown dwarfs in a stable orbit within the fragmentation regime. For $\beta=8$, I measured Reynolds stresses of $\reynolds \sim 10^{-3}$ and gravitational stresses of $\grav \sim 10^{-2}$.
\item For $\epsilon/H \leq 0.3$, the disc undergoes early fragmentation, producing an excess number of fragments whose masses are probably systematically overestimated. In this case, the final products of fragmentation reach masses of $0.15-0.2 M_{\odot}$. Additionally, gravitational stresses are overestimated in the gravito-turbulent regime.
\item When $\epsilon/H = 1.2$, no fragments form, and the stresses are underestimated.
\item For $\epsilon/H = 0.6$, the outcomes closely resemble those obtained with the Bessel prescription, except that fragments do not remain gravitationally bound over the long term. Instead, they undergo cycles of disruption and collapse.
\end{itemize}
In conclusion, while a smoothing length of $\epsilon/H=0.6$ appears to be the optimal choice-with other values not being suitable- the loss of the Newtonian character of gravity may rise concerns regarding the consistency of formed fragments and their evolution.

The remarkable agreement between the results obtained using the Bessel kernel and their three-dimensional counterparts (as illustrated in Fig. 5 and Table 3 of \citep{2025c_rendon}), combined with the resolution of the long-standing convergence problem of GI in 2D—achieved without the need for outcome-dependent fine-tuned parameters— and the gravitational binding of fragments upon formation provides compelling evidence that this prescription is the most suitable approach for modelling gravity in two-dimensional discs.
These findings strongly advocate for the adoption of this prescription in future studies.

\begin{acknowledgements}

Funded by the European Union (ERC, Epoch-of-Taurus, 101043302). Views and opinions expressed are however those of the author(s) only and do not necessarily reflect those of the European Union or the European Research Council. Neither the European Union nor the granting authority can be held responsible for them.
I would like to warmly thank O. Gressel, A. Mandal, M. Van den Bossche, C. Baruteau and T. Rometsch for their insightful discussions.
The authors gratefully acknowledge the computing time made available to them on the high-performance computer at the NHR Center of TU Dresden. This center is jointly supported by the Federal Ministry of Research, Technology and Space of Germany and the state governments participating in the NHR (www.nhr-verein.de/unsere-partner).
\end{acknowledgements}

\bibliographystyle{aa}
\bibliography{bibliography}

\begin{appendix}

\section{Additional material}

\begin{figure*}
\centering

\begin{tabular}{cp{0.3\hsize}p{0.3\hsize}p{0.3\hsize}}
& \centering{\textbf{time = $3.9$ kyr}} & \centering{\textbf{time = $4.3$ kyr}} & \textbf{time = $6.1$ kyr} \\
\rotatebox{90}{\hspace{6.4em} \textbf{Bessel}} &
\includegraphics[width=\linewidth]{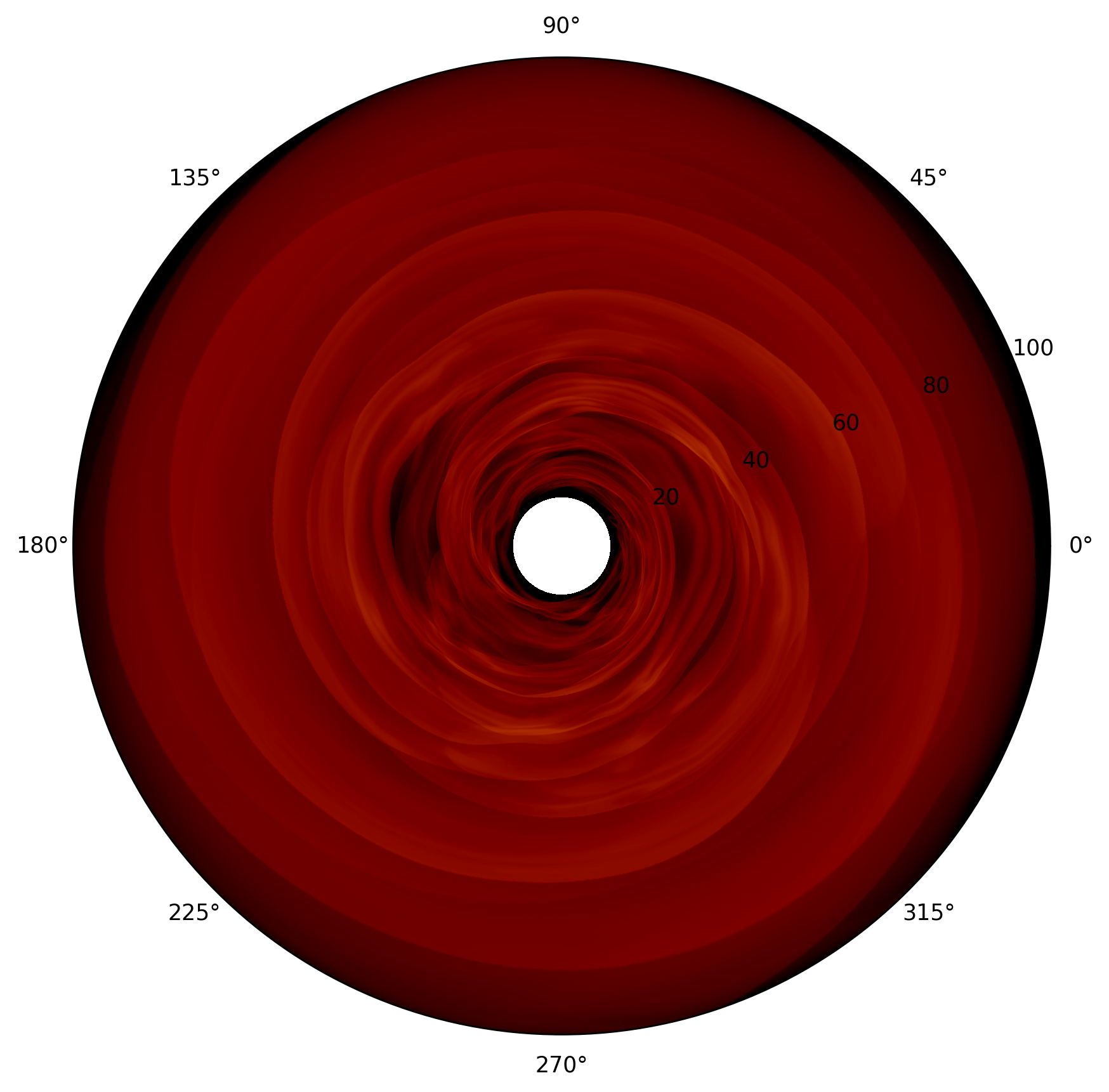} &
\includegraphics[width=\linewidth]{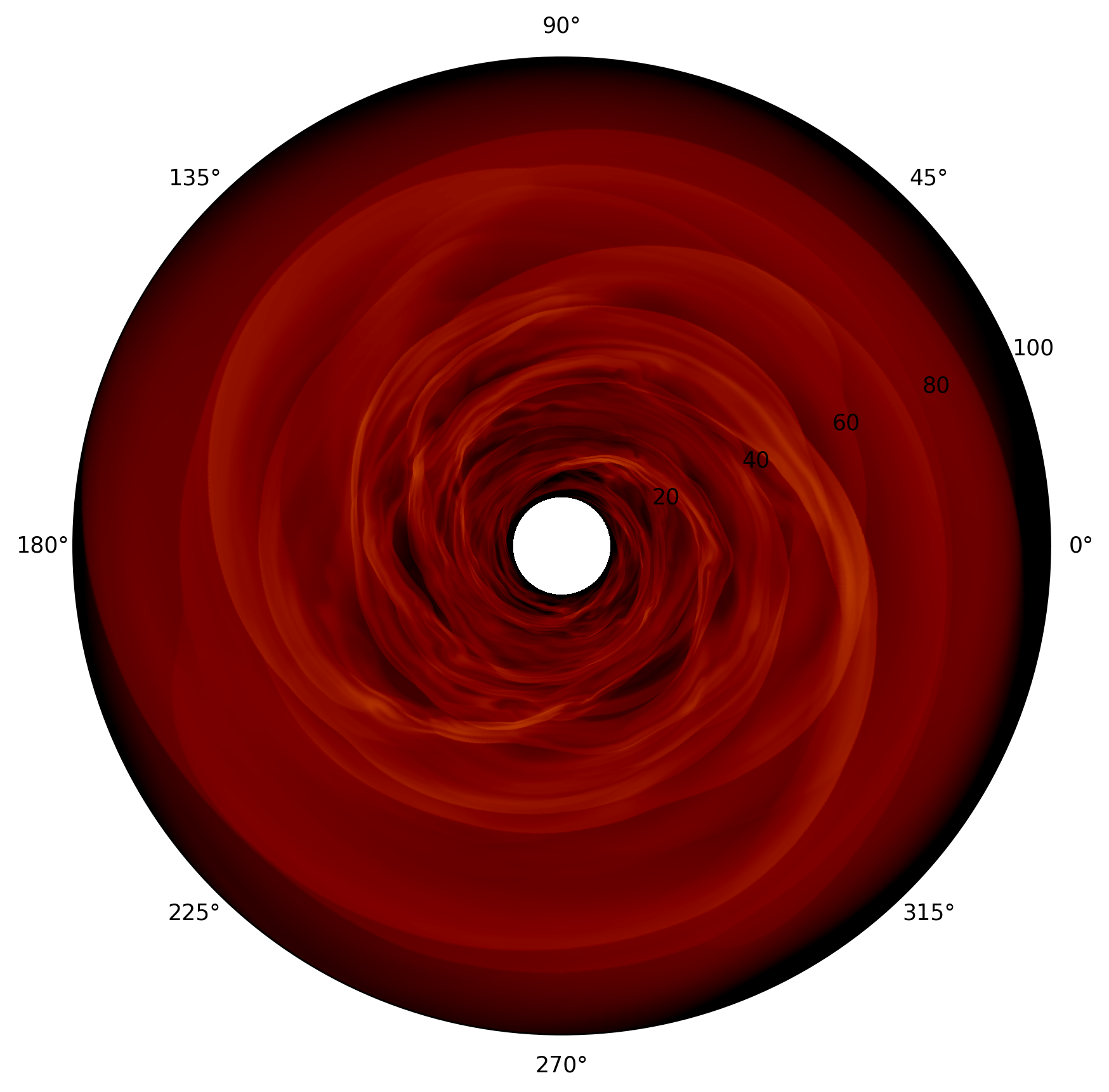} &
\includegraphics[width=\linewidth]{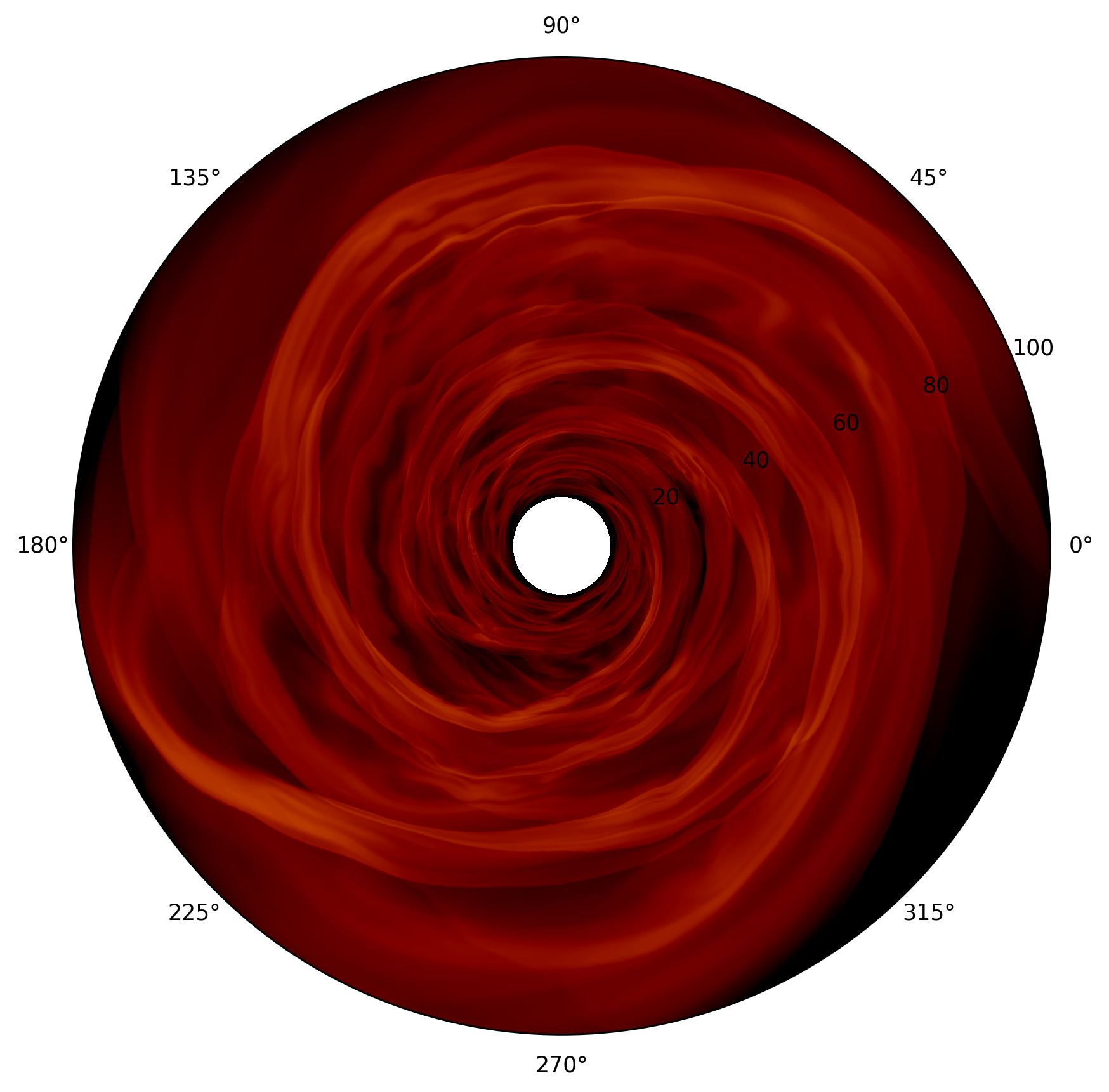} \\
\rotatebox{90}{\hspace{4.8em} \textbf{$\displaystyle \frac{\epsilon}{H_{\rm rms}}=0.0$}} &
\includegraphics[width=\linewidth]{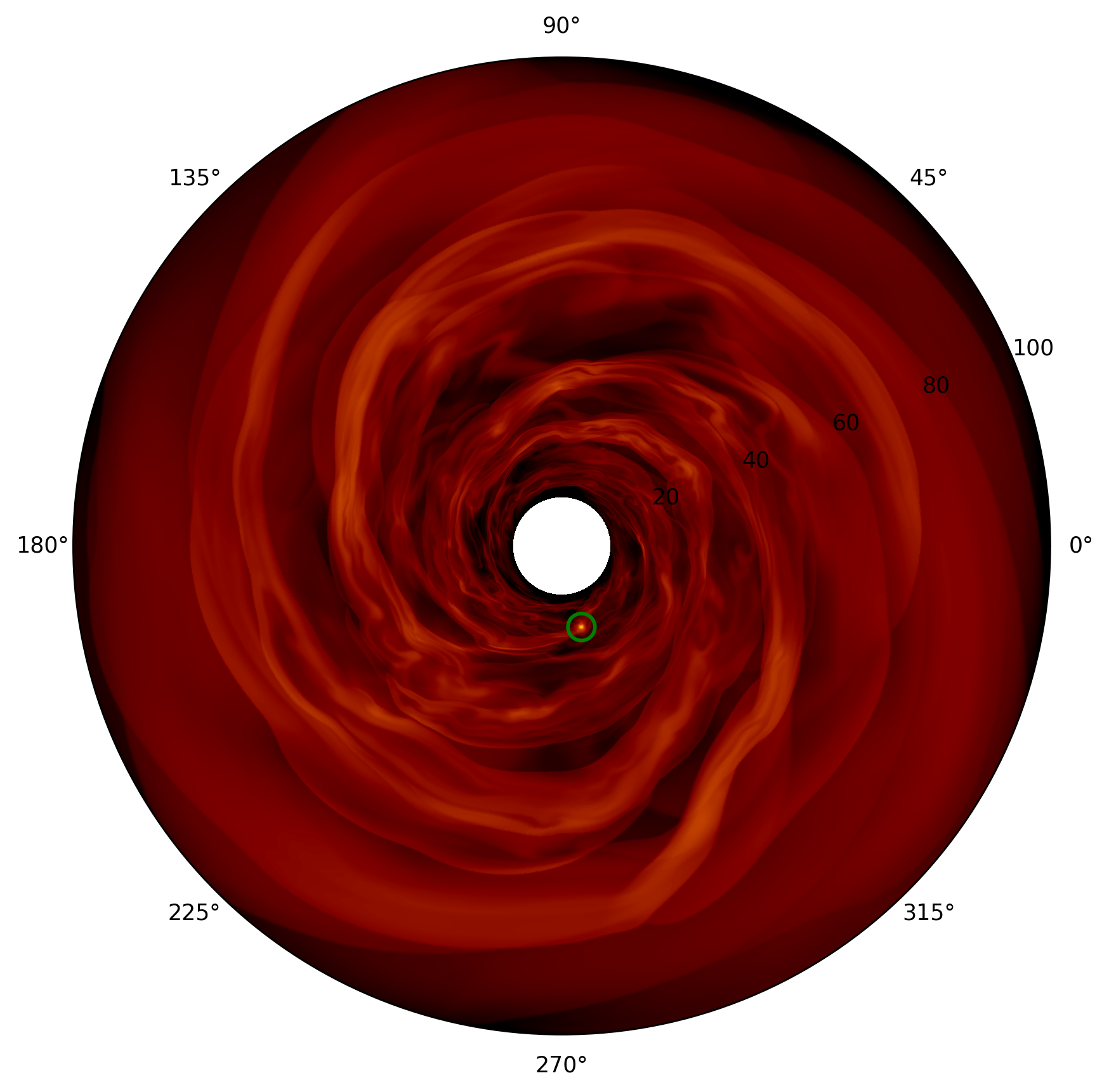} &
\includegraphics[width=\linewidth]{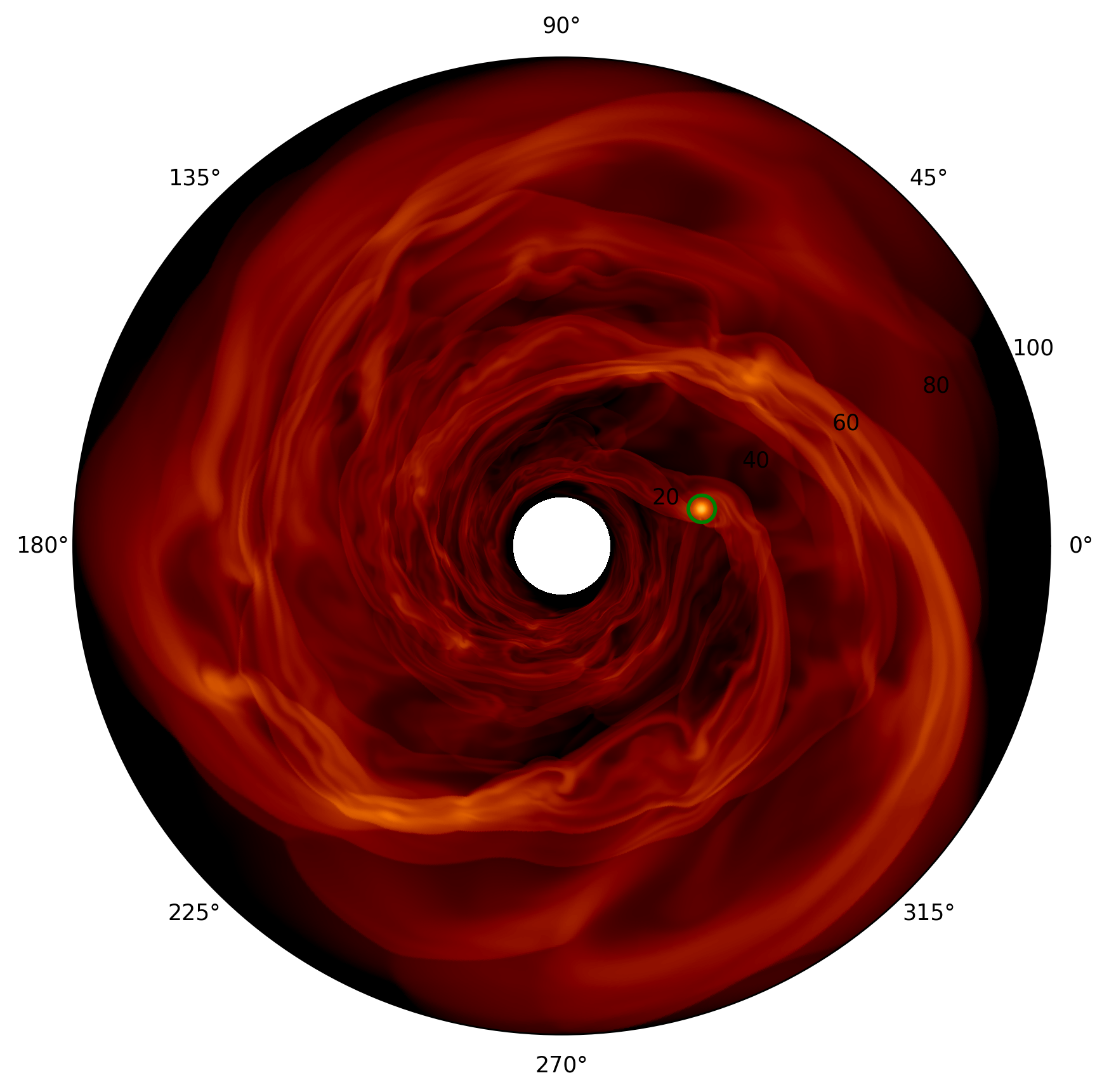}  &
\includegraphics[width=\linewidth]{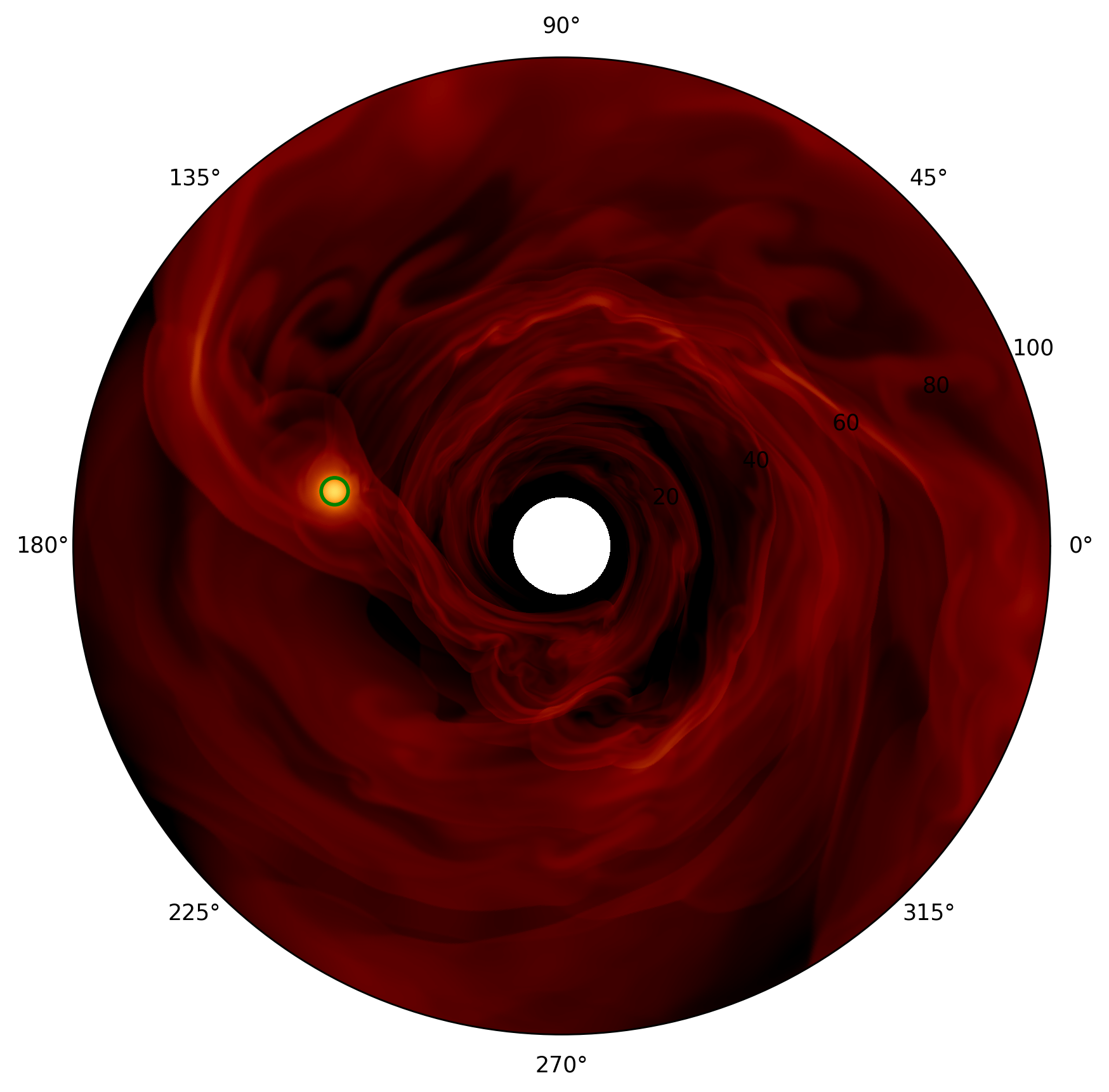}  \\
\rotatebox{90}{\hspace{4.8em} \textbf{$\displaystyle \frac{\epsilon}{H_{\rm rms}}=0.3$}} &
\includegraphics[width=\linewidth]{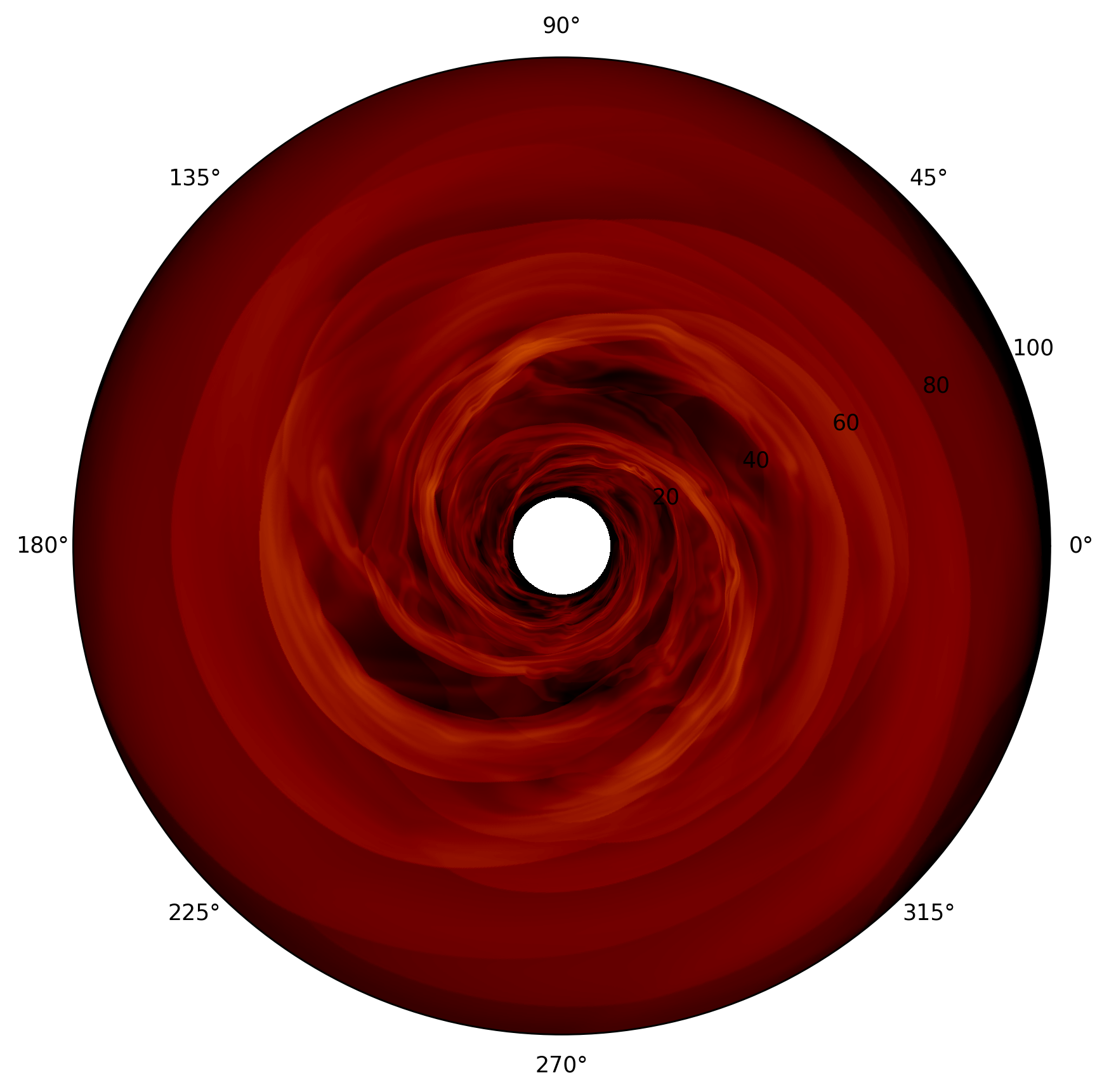} &
\includegraphics[width=\linewidth]{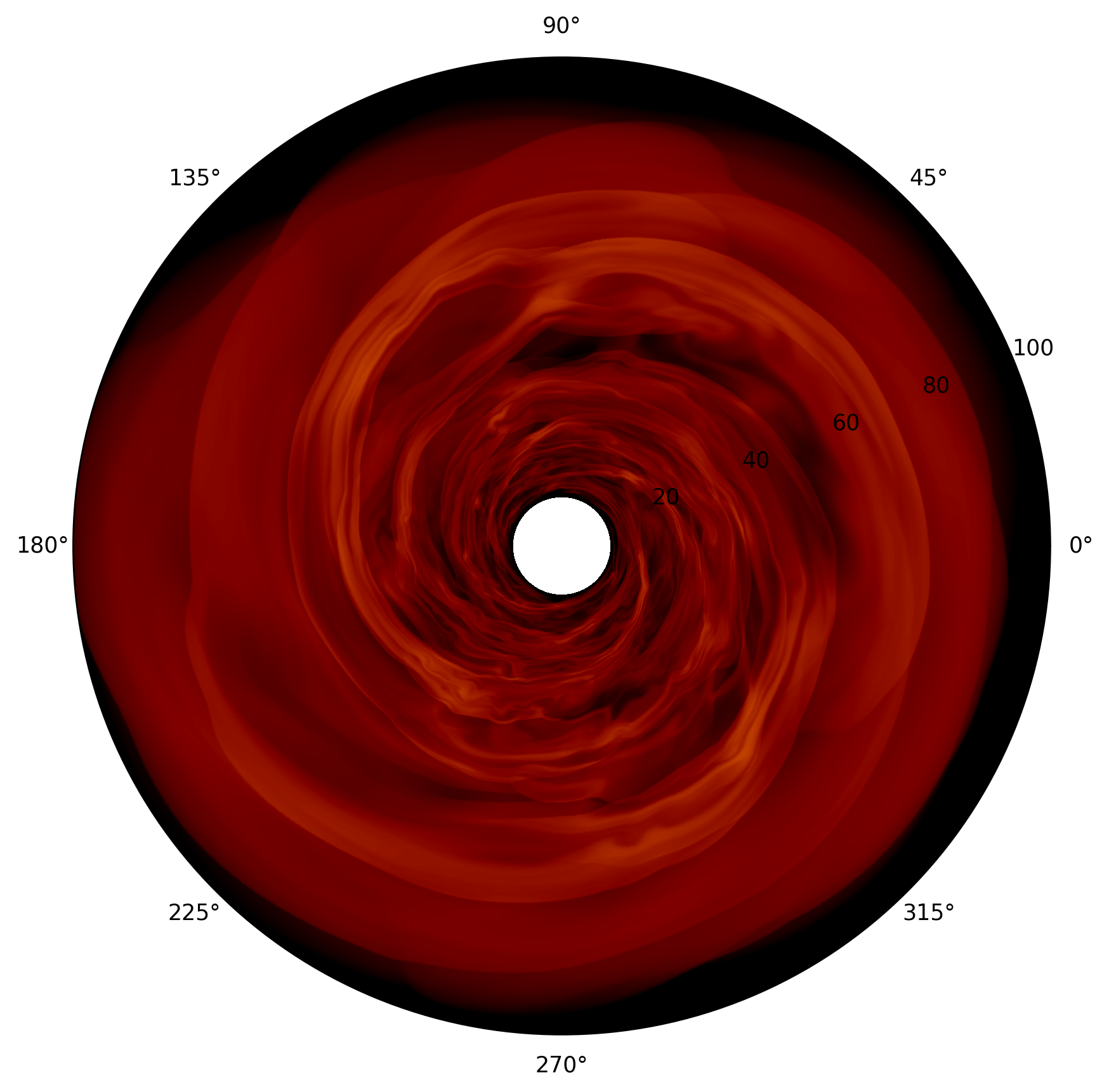}  &
\includegraphics[width=\linewidth]{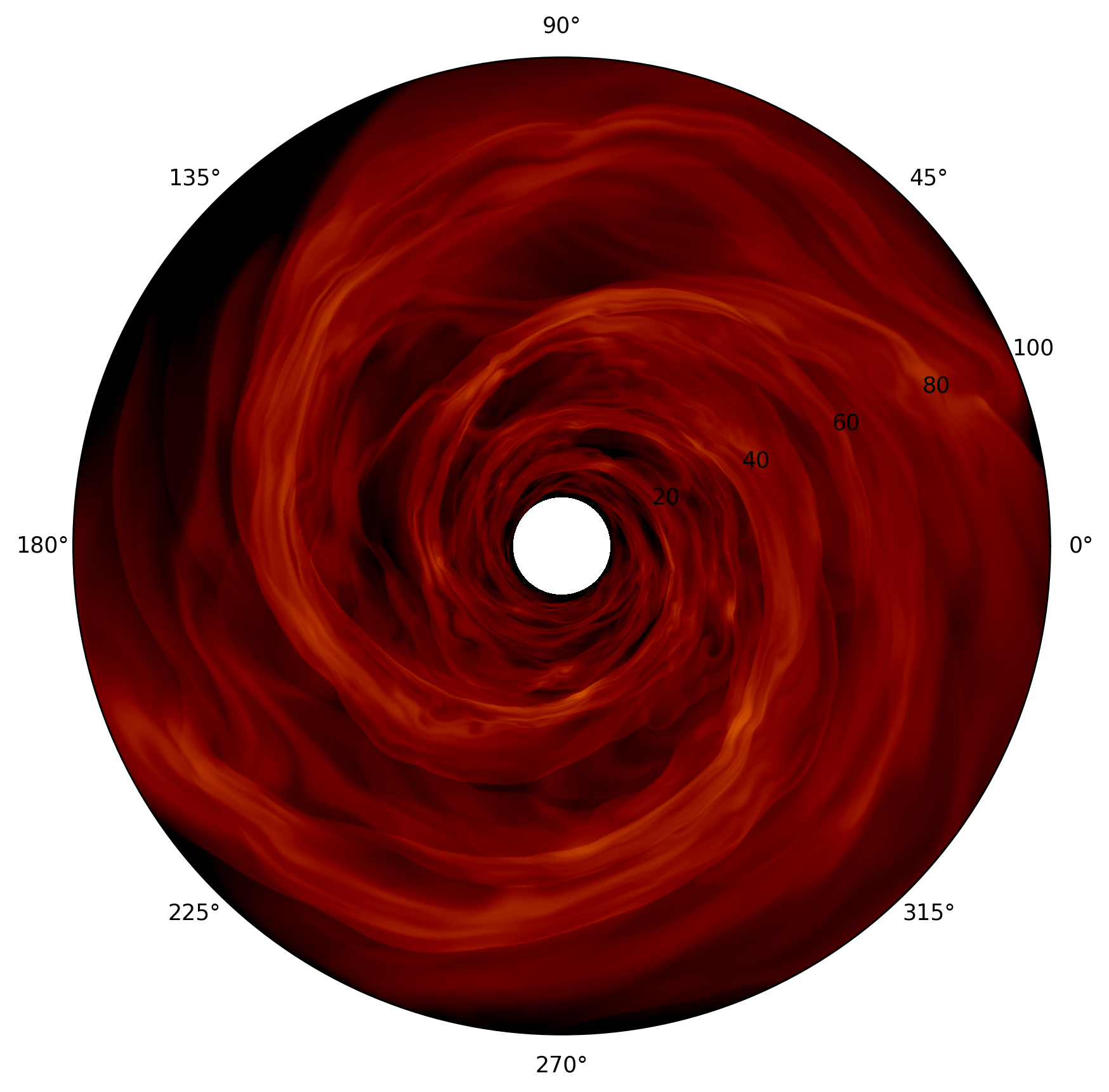}  \\
\rotatebox{90}{\hspace{4.8em} \textbf{$\displaystyle \frac{\epsilon}{H_{\rm rms}}=0.6$}} &
\includegraphics[width=\linewidth]{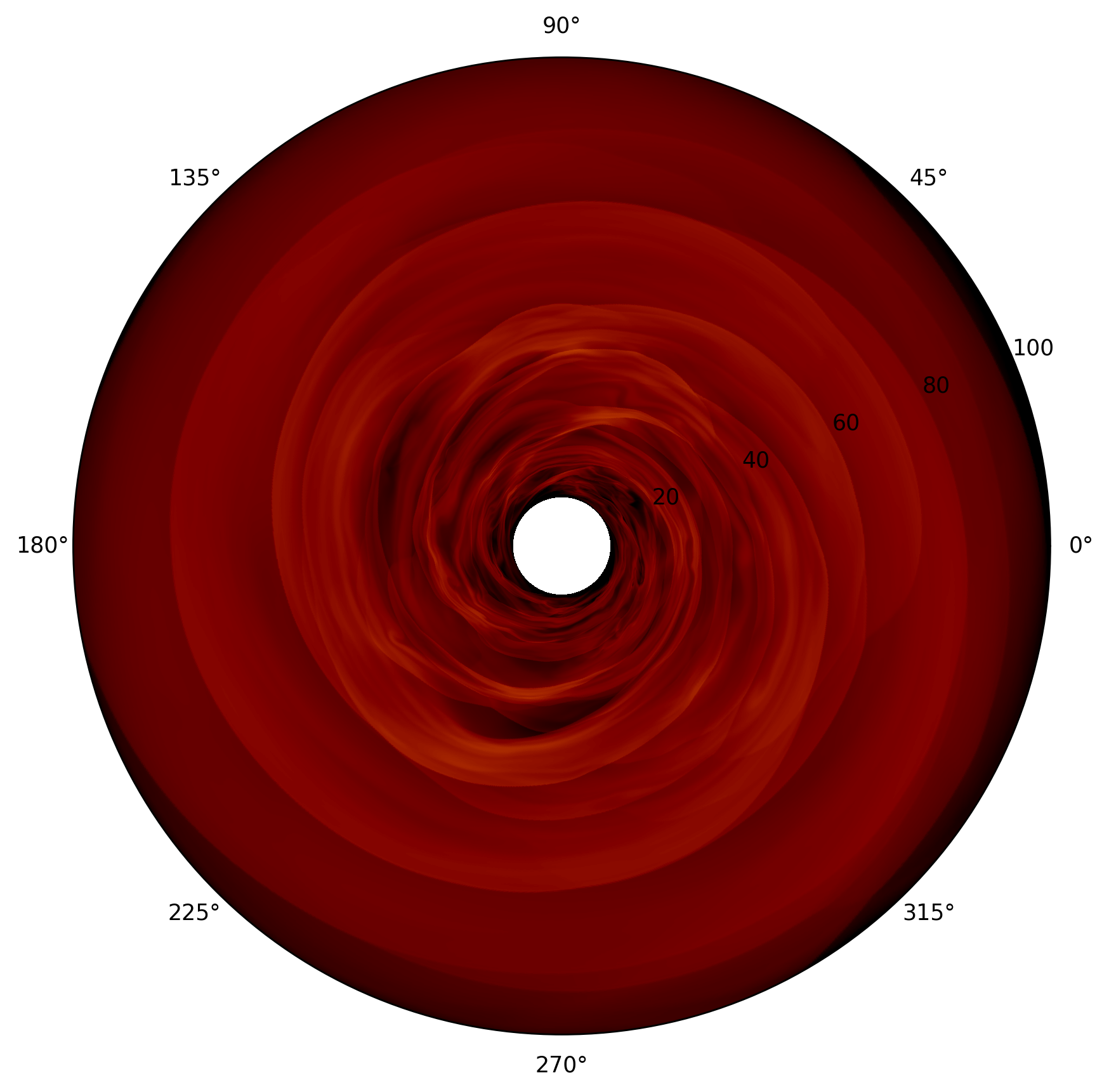} &
\includegraphics[width=\linewidth]{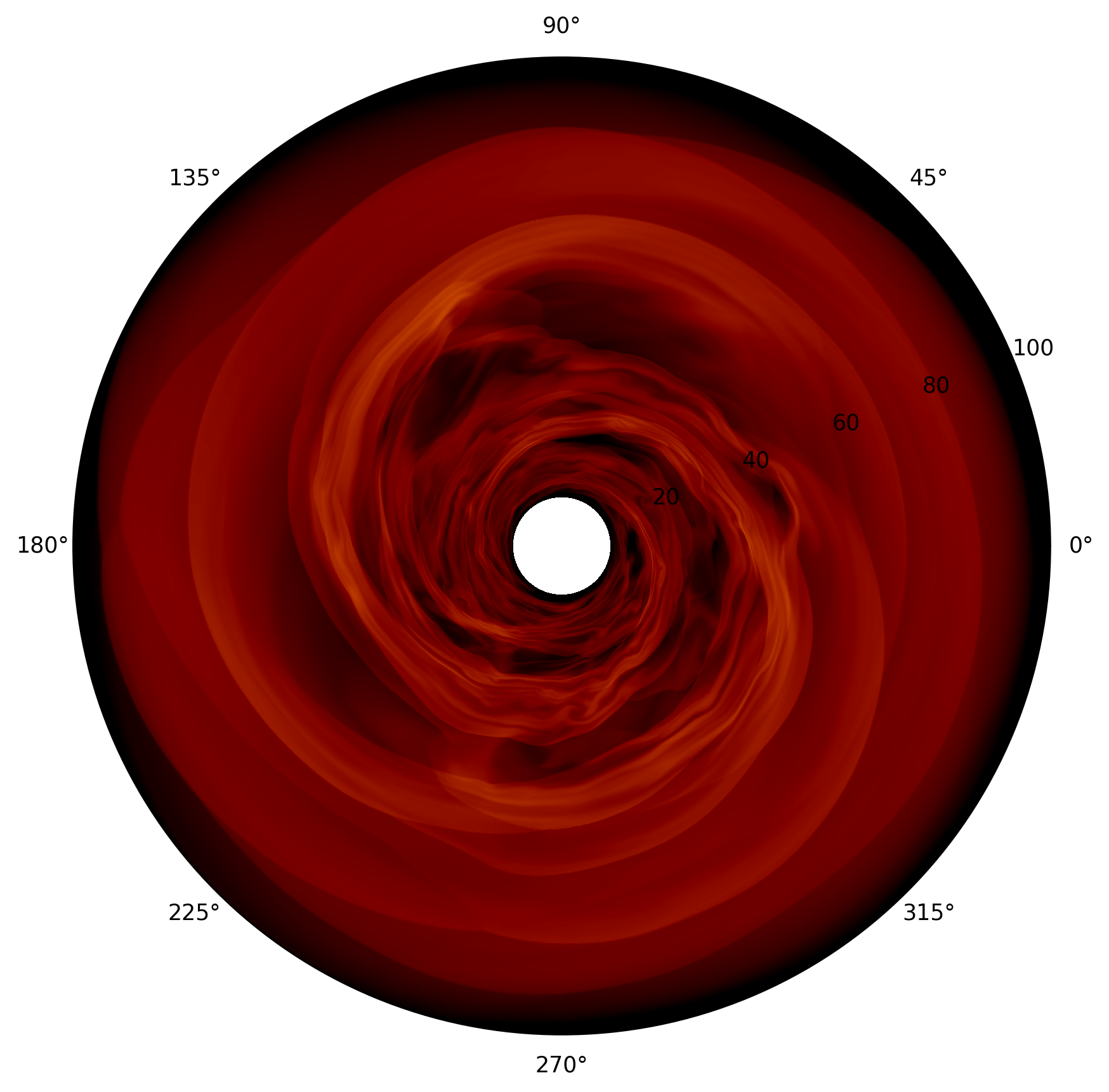}  &
\includegraphics[width=\linewidth]{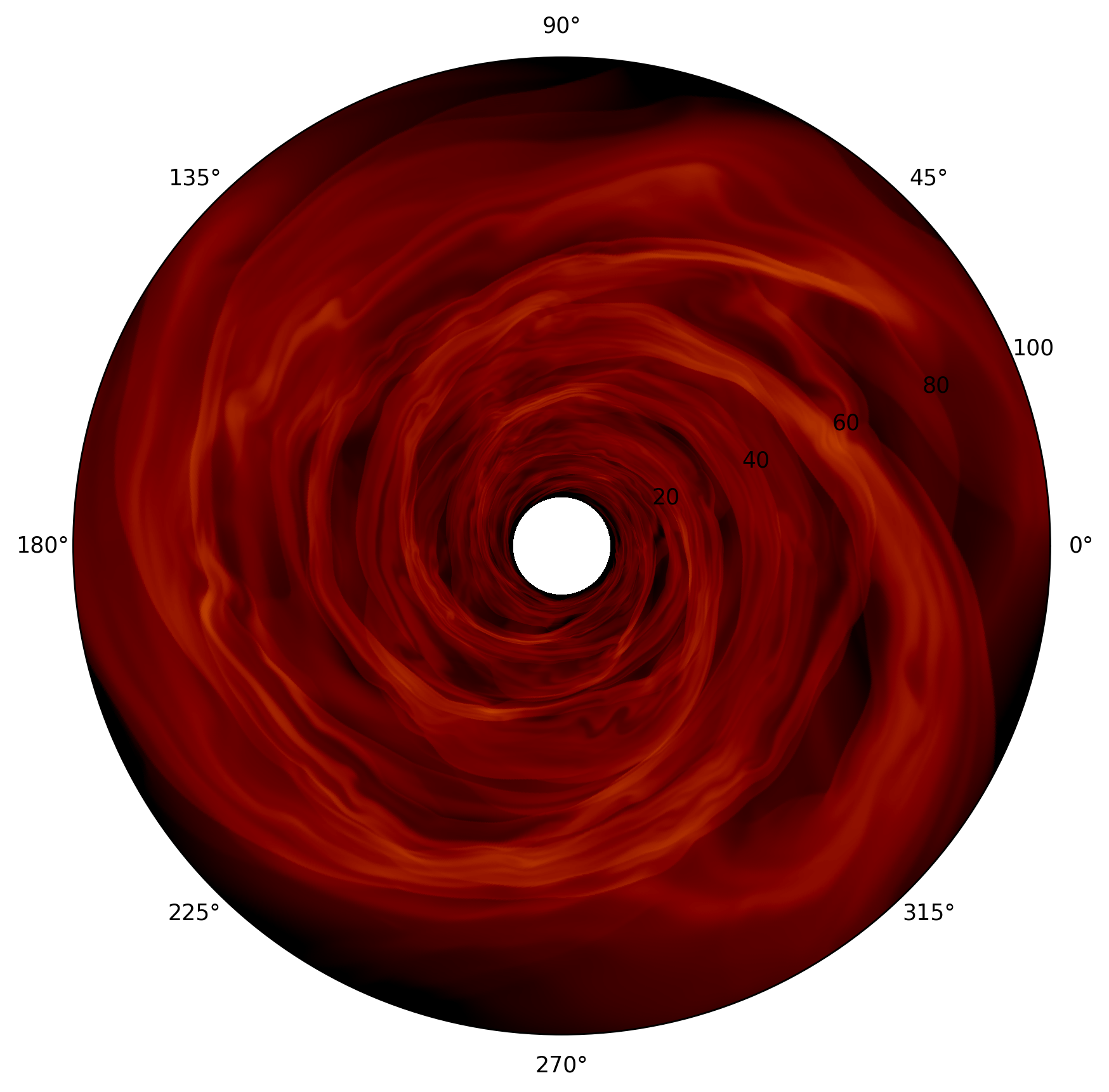}  \\
\end{tabular}

\includegraphics[width=0.9\linewidth]{images/colorbar_Sigma_polar.png}

\caption{Comparison of the gravito-turbulent regime ($\beta=8$) for the different gravity prescriptions at different times.
Fragments are highlighted with green circles. 
}

\label{fig: comparison gravito-turbulence different kernels}
\end{figure*}

\end{appendix}

\label{LastPage}

\end{document}